\documentclass[
 longbibliography,
 twocolumn,
 superscriptaddress,
 amsmath,
 amssymb,
 amsthm,
 amsfonts,
 prl,
 floatfix,
]{revtex4-2}

\usepackage[colorlinks=true,urlcolor=blue,citecolor=blue,linkcolor=blue]{hyperref} 
\usepackage{graphicx}
\usepackage{dcolumn}
\usepackage{kotex}
\usepackage{natbib}
\usepackage{color}
\usepackage{amsfonts}
\usepackage{hyperref}
\usepackage{bm}
\usepackage{mathtools}
\usepackage{float}

\usepackage{tabularx}
\usepackage{multirow}
\usepackage{booktabs}
\usepackage{siunitx}

\newcommand{\matbf}[1]{\boldsymbol{\mathsf{#1}}}

\begin{document}

\title{Exact Identity Linking Entropy Production and Mutual Information}

\author{Doohyeong Cho}
\email{lokie@kaist.ac.kr}
\affiliation{Department of Physics, Korea Advanced Institute of Science and Technology, Daejeon 34141, Rep. of Korea}

\author{Hawoong Jeong}
\email{hjeong@kaist.edu}
\affiliation{Department of Physics, Korea Advanced Institute of Science and Technology, Daejeon 34141, Rep. of Korea}
\affiliation{Center for Complex Systems, Korea Advanced Institute of Science and Technology, Daejeon 34141, Rep. of Korea}

%\date{\today}

\begin{abstract}
We establish an exact identity for overdamped Langevin dynamics: the total entropy production rate equals four times the mutual information rate between an infinitesimal displacement and its actual temporal midpoint, plus a mean-flow term. This provides a forward information-theoretic representation of irreversibility and makes the Shannon chain rule act directly on dissipation. For additive bipartitions, the chain rule reveals nonnegative self and interaction channels, with the self channel coinciding with apparent entropy production.
This channel structure sharpens the learning-rate bound by showing that learning is constrained solely by the interaction channel. The same structure is also shown to resolve a recently reported coarse-graining reversal in hydrodynamically coupled colloids. Together, these results exhibit that, within this channel decomposition, learning and coarse-graining probe complementary parts of dissipation in overdamped Langevin systems.

\end{abstract}

\maketitle

%Entropy production (EP), a central measure of nonequilibrium irreversibility, admits equivalent representations that expose different structures. Its path-space form is the Kullback--Leibler divergence between forward and backward path measures, which exposes arrow-of-time distinguishability and fluctuation relations~\cite{Crooks1999Entropy, Maes2003time-reversal, Seifert2005Entropy, parrondo2009entropy, seifert2012stochastic}. Related information-theoretic measures of time asymmetry likewise compare forward and reversed trajectory ensembles~\cite{feng2008length}. For overdamped diffusions, the equivalent current-quadratic form instead exposes instantaneous geometry and projections~\cite{seifert2012stochastic,nakazato2021geometrical,dechant2022geometricPRR,dechant2022geometric,frishman2020learning}. Ordinary mutual information (MI) asks a different question: how variables depend on other variables within one joint ensemble. MI is central to information thermodynamics~\cite{Sagawa2010generalized,sagawa2013role,ito2013Informatioin,horowitz2014second,hartich2014stochastic,parrondo2015thermodynamics}, and direct relations between MI and EP links exist in special settings~\cite{diana2014mutual, dabelow2019irreversibility,louwerse2022information}.Yet, to our knowledge, a general total EP identity built from such within-ensemble MI has been lacking.
%horowitz2015multipartite,Chetrite2019information,belenchia2022informaional

Entropy production (EP), a central measure of nonequilibrium irreversibility, admits equivalent representations that expose different structures. Its path-space form is the Kullback–Leibler divergence between forward and backward path measures, revealing arrow-of-time distinguishability and fluctuation relations~\cite{Crooks1999Entropy, Maes2003time-reversal, Seifert2005Entropy, parrondo2009entropy, seifert2012stochastic}. Related information-theoretic measures of time asymmetry likewise compare forward and reversed trajectory ensembles~\cite{feng2008length}. For overdamped diffusions, the equivalent current-quadratic form instead exposes instantaneous geometry and projections~\cite{seifert2012stochastic,nakazato2021geometrical,dechant2022geometricPRR,dechant2022geometric,frishman2020learning}. Mutual information (MI), a central measure of statistical dependence in information theory, asks a different question: how variables depend on one another within a single joint ensemble. MI has long played an important role in information thermodynamics, entering generalized second laws and relations involving feedback or information exchange~\cite{Sagawa2010generalized,sagawa2013role,ito2013Informatioin,horowitz2014second,hartich2014stochastic,parrondo2015thermodynamics}. Direct relations between MI and EP are also known in special settings~\cite{diana2014mutual, dabelow2019irreversibility,louwerse2022information}. Yet it is not obvious that the total EP rate itself admits an exact representation in terms of ordinary MI. To our knowledge, a general identity of this kind has been lacking.

\begin{figure}[!t]
    \centering
    \includegraphics[width=\linewidth]{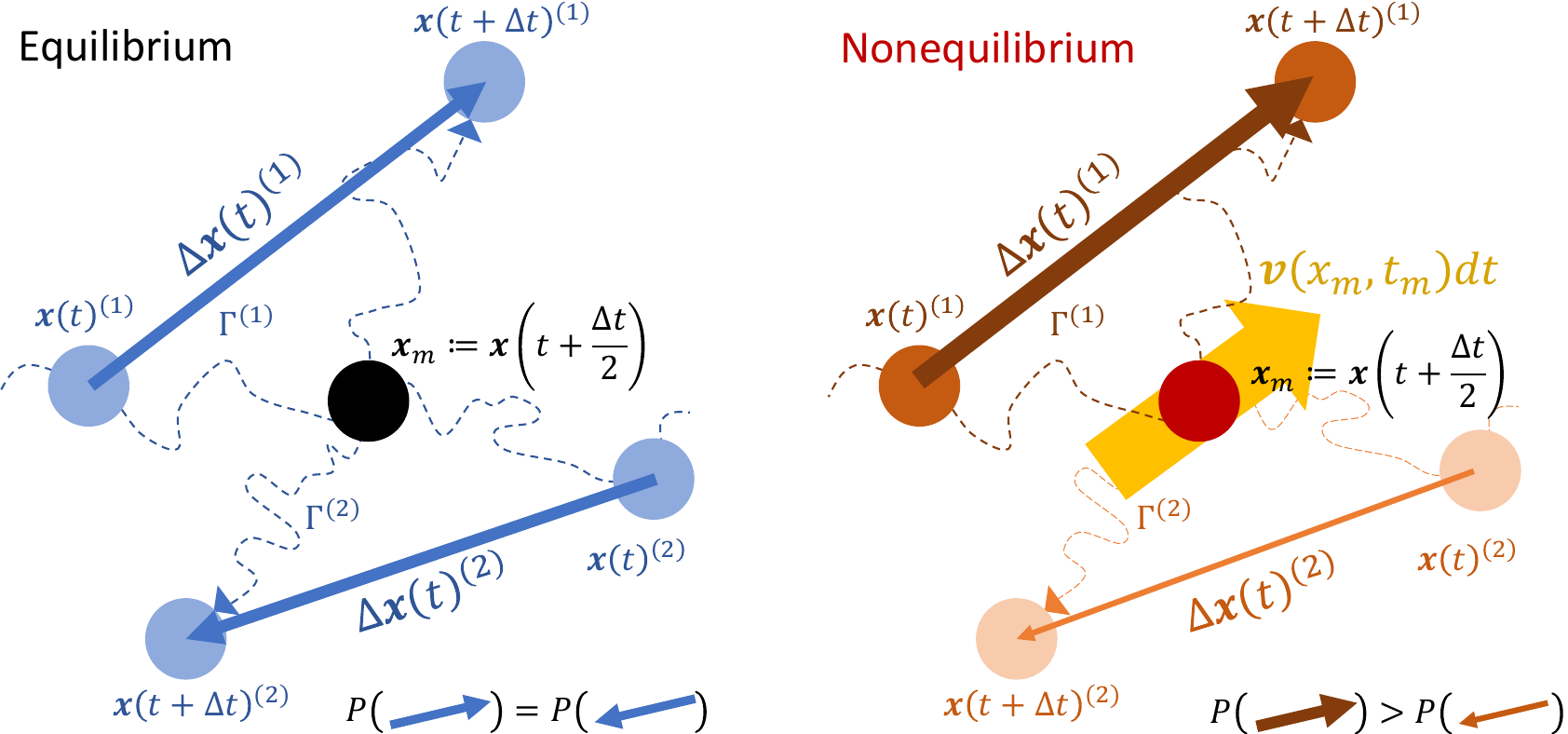}
    \vskip -0.1in
    \caption{Schematic illustration of displacement $\Delta\bm{x}$ conditioned on $\bm{x}_m$. (Left) In equilibrium, detailed balance implies that $\bm{x}_m$ contains no information about $\Delta\bm{x}$. (Right) In nonequilibrium, a finite current velocity $\bm{v}_t(\bm{x}_m)$ breaks detailed balance and biases $\Delta\bm{x}$ conditioned on $\bm{x}_m$, rendering the midpoint informative. In the infinitesimal-time limit $\mathrm{d}t\to 0$, $\Delta\bm{x}$ corresponds to $\mathrm{d}\bm{x}$ in the main text.
    }\label{fig1}
    \vskip -0.2in
\end{figure}

Here we establish such an identity for overdamped Langevin dynamics. The total EP rate equals four times the MI rate between an infinitesimal displacement and its actual temporal midpoint, plus a mean-flow term.
Both variables belong to the same forward trajectory segment; no time-direction label or separate reversed ensemble enters the final MI.
The midpoint is the natural time-reversal-symmetric reference and converts irreversible current into displacement--midpoint dependence (Fig.~\ref{fig1}), while the mean-flow term accounts for uniform translation invisible to MI.
The result is therefore a forward information-theoretic representation.

In this representation, the physical degrees of freedom appear explicitly as MI arguments, allowing the Shannon chain rule to act directly on EP. For additive bipartitions, this maps onto the marginal-current projection~\cite{nakazato2021geometrical,frishman2020learning} and identifies nonnegative self and interaction channels. We use this structure to sharpen the learning-rate bound~\cite{matsumoto2025learningrate} and to resolve a reported coarse-graining reversal in hydrodynamically coupled colloids~\cite{das2025irreversibility,das2025uraveling}.

\begin{figure*}[!t]
    \centering
    \includegraphics[width=\textwidth]{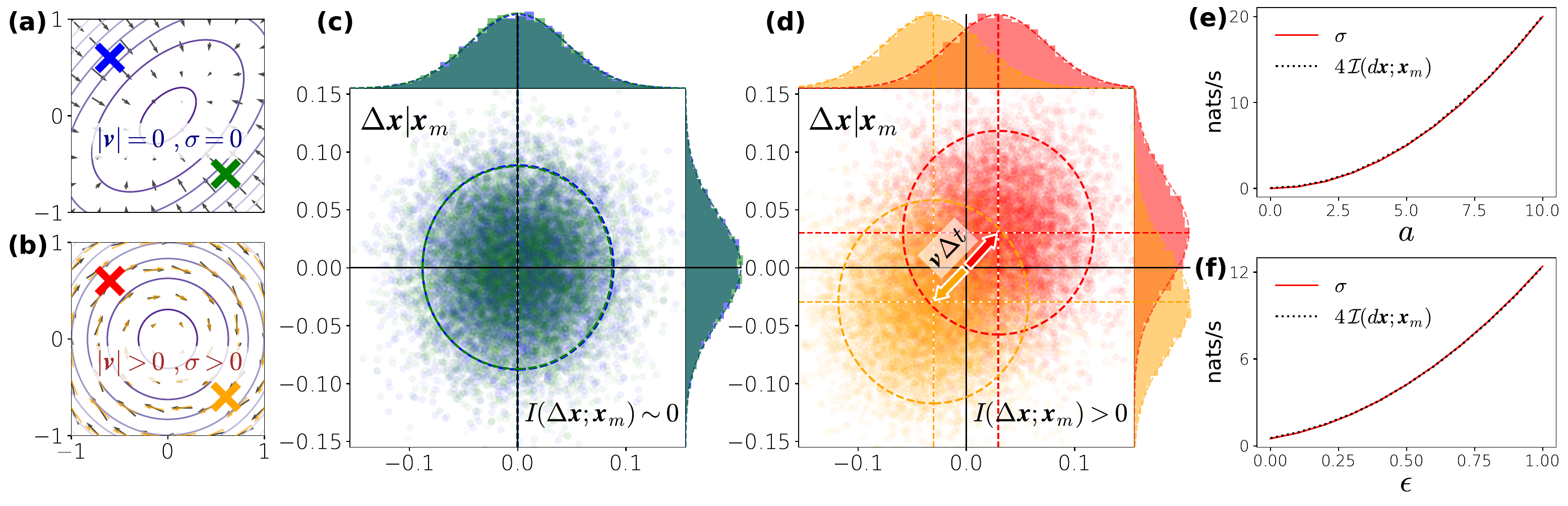}
    \vskip -0.1in
    \setlength{\abovecaptionskip}{0pt}
    \caption{Physical mechanism and numerical validation of Eq.~\eqref{eq:main_identity}. 
    \textbf{(a–d)} Linear model defined by $\dot{x}=-kx+(s+a)y+\sqrt 2\xi_{x}$ and $\dot{y}=-ky+(s-a)x+\sqrt 2\xi_{y}$ with $k=10$ and $\langle\xi_i\xi_j\rangle=\delta_{ij}$. 
    Parameters are $(s,a)=(5,0)$ for equilibrium (a, c), and $(0,50)$ for nonequilibrium steady state (NESS) (b, d). In 
    (a, b), the force $\bm{F}(\bm{x})$ (black) and current velocity $\bm{v}(\bm{x})$ (yellow) are overlaid on steady-state density $p(\bm{x})$. Crosses indicate sampling locations for the time midpoint $\bm{x}_{m}$. 
    In (c, d), conditional displacements $\Delta\bm{x}|\bm{x}_{m}$ at the marked locations are shown. Circles are centered at sample means with radius $1.96\times\mathrm{SD}$. Dashed lines mark the means, and black solid lines denote the origin. In equilibrium (c), the distributions are indistinguishable. In NESS (d), they are separated by the current velocity $\bm{v}(\bm x_m)\Delta t$, rendering $\bm{x}_{m}$ informative about $\Delta\bm{x}$. In the simulations, $\Delta t=10^{-3}$.
    \textbf{(e, f)} Coincidence of information rates $4\,\mathcal I(\mathrm d\bm{x};\bm{x}_m)$ (dotted) to the total EP rate (solid) for the same linear model with $s=0$ (e), and for the nonlinear model $\dot{x}=\!-x+y+\epsilon y^{3}\!+\!\sqrt 2\xi_{x}$, $\dot{y}=\!-y+\!\sqrt 2\xi_{y}$ (f). Eq.~\eqref{eq:main_identity} remains exact in both linear and nonlinear~models.
    }\label{fig2}
    \vskip -0.2in
\end{figure*}

\textit{Main identity}.---We consider standard overdamped Langevin dynamics with arbitrary nonlinear and time-dependent forces
\vspace{-4pt}
\begin{equation}
    \dot{\bm{x}}_t=\matbf{\mu}_t \bm{F}_t(\bm{x}_t)+\sqrt{2\matbf{D}_t}\,\bm{\xi}_t,
\label{eq:overdamped_Langevin}
\end{equation}
where $\matbf{\mu}_t$ is the mobility, $\bm{F}_t$ the force, and $\bm{\xi}_t$ is Gaussian white noise with
$\langle \bm{\xi}_t\rangle=\bm{0}$ and $\langle \bm{\xi}_t\bm{\xi}_s^{\intercal}\rangle=\mathbf{I}\delta(t-s)$.
We assume position-independent noise $\matbf{D}_t$ in the main text for simplicity; extensions to $\matbf{D}_t(\bm{x})$ are given in the Supplemental Material (SM).
The total EP is standardly defined by the Kullback--Leibler divergence between the forward and time-reversed path measures, quantifying the distinguishability of the process from its time reverse~\cite{Seifert2005Entropy}.
For the dynamics in Eq.~\eqref{eq:overdamped_Langevin}, its instantaneous rate reads $\sigma_t=\langle\bm v_t^\intercal\matbf D_t^{-1}\bm v_t\rangle$, 
where $\langle\cdot\rangle=\int (\cdot)\,p_t(\bm x)\,d\bm x$, $p_t$ is the probability density, and $\bm v_t\coloneq \matbf\mu_t\bm F_t-\mathbf D_t\nabla\ln p_t$ is the current velocity (also called the local mean velocity)~\cite{seifert2012stochastic}.

We show that this path-space distinguishability underlying EP can be reframed as forward within-ensemble dependence between an infinitesimal displacement and its temporal midpoint, both belonging to the same forward trajectory segment, as illustrated in Fig.~\ref{fig2}a--d.
Let $m=t+\frac{\mathrm{d}t}{2}$, and we denote quantities at time $m$ with the subscript $m$. Considering the infinitesimal displacement $\mathrm d\bm{x}_t = \bm{x}_{t+\mathrm{d}t} - \bm{x}_t$ conditioned on the midpoint position $\bm{x}_m$, we obtain the exact identity at the level of rates,
\begin{equation}
    \sigma_t
    = 4\,\mathcal I(\mathrm{d}\bm{x}_t;\bm{x}_m)
    + \langle \bm{v}_t \rangle^\intercal
      \matbf{D}_t^{-1}
      \langle \bm{v}_t \rangle ,
\label{eq:main_identity}
\end{equation}
where $\mathcal I(A;B) := \lim_{\mathrm{d}t\to 0} I(A;B)/\mathrm{d}t$.

To prove Eq.~\eqref{eq:main_identity}, we use the Markov property at the midpoint $m$, where conditioning on $\bm{x}_m$ makes the two half-step increments $(\bm{x}_{t+\mathrm{d}t}-\bm{x}_m)$ and $-(\bm{x}_t-\bm{x}_m)$ statistically independent. 
The forward increment follows the Euler--Maruyama step~\cite{maruyama1955continuous,kloeden1992numerical} with mean $\matbf{\mu}_m\bm{F}_m\mathrm{d}t/2$ and covariance $\matbf{D}_m\mathrm{d}t$. 
The backward increment $\bm{x}_t-\bm{x}_m$ is governed by reverse-time diffusion~\cite{anderson1982reverse, haussmann1986time} with score-corrected drift $\bm{F}^\mathrm{B}=-\bm{F}+2\matbf{\mu}^{-1}\matbf{D}\nabla\ln p$ (and the same diffusion matrix), hence $\langle-(\bm{x}_t-\bm{x}_m)|\bm{x}_m\rangle=-\matbf{\mu}_m\bm{F}^\mathrm{B}_m\mathrm{d}t/2$.
Consequently, in the infinitesimal-time limit $\mathrm{d}t\rightarrow0$, the sum admits a Gaussian core with conditional mean $\langle\mathrm{d}\bm{x}_t|\bm{x}_m\rangle=\bm{v}_m\mathrm{d}t$ and covariance $\mathrm{Cov}(\mathrm{d}\bm{x}_t|\bm{x}_m)=2\matbf{D}_m\mathrm{d}t$. Equivalently, to the order relevant for Eq.~\eqref{eq:main_identity}, $p(\mathrm d\bm x_t|\bm x_m)$ is captured by the effective Gaussian channel
\begin{equation}
    \mathrm{d}\bm{x}_t
    = \bm{v}_m(\bm{x}_m)\mathrm{d}t+\sqrt{2\matbf{D}_m\mathrm{d}t}\,\mathbf{N},
    \qquad \mathbf{N}\sim\mathcal{N}(0,\mathbf{I}).
\label{eq:dx_cond_xm Gaussian channel}
\end{equation}
We emphasize that the midpoint conditioning is essential, as other conditioning does not yield the current velocity. For example, $\langle \mathrm d\bm x_t \mid \bm x_t \rangle=\mathbf\mu_t\bm F_t(\bm x_t)\,\mathrm dt$, whereas $\langle \mathrm d\bm x_t \mid \bm x_m \rangle=\bm v_m(\bm x_m)\,\mathrm dt$.
We also note that Eq.~\eqref{eq:dx_cond_xm Gaussian channel} is not a stochastic differential equation for the original process, but an effective statistical model for $p(\mathrm{d}\bm{x}_t|\bm{x}_m)$.

Since Eq.~\eqref{eq:dx_cond_xm Gaussian channel} operates in the low signal-to-noise ratio (SNR) regime as $\mathrm{d}t\rightarrow0$, a standard small-SNR expansion for Gaussian channels~\cite{guo2005mutual} yields
\begin{equation}
    I(\mathrm{d}\bm{x}_t;\bm{x}_m)
    =\frac{\mathrm{d}t}{4}\langle\delta\bm{v}_m^\intercal
    \matbf{D}_m^{-1}
    \delta\bm{v}_m\rangle
    +o(\mathrm{d}t),
\label{eq:dx_xm_I_MMSE_result}
\end{equation}
with $\delta\bm{v}_m:=\bm{v}_m-\langle\bm{v}_m\rangle$. 
Using $\mathrm{Cov}(\bm{v}_m(\bm{x}_m))=\mathrm{Cov}(\bm{v}_t(\bm{x}_t))+O(\mathrm{d}t)$, dividing by $\mathrm{d}t$ and taking the limit $\mathrm{d}t\to0$ recovers the main identity, Eq.~\eqref{eq:main_identity}.
No assumption of steady state, linearity, or time independence is required beyond standard regularity conditions.
For position-dependent diffusion, the identity retains the same information-theoretic structure in the local diffusion metric; the SM provides a rigorous derivation of the general result, recovering the present additive-noise case, together with a numerical illustration.

The midpoint $\bm x_m$ plays a special role because it is the sole position left invariant by the infinitesimal time-reversal exchange $t\leftrightarrow t+\mathrm dt$. It therefore provides a natural reference for recasting path-space irreversibility in terms of statistics within the forward trajectory ensemble, as conceptualized in Fig.~\ref{fig1}.
%Note that it is simply the state of the same system at the intermediate time: for an equally spaced three-point trajectory window $(\bm x_{n-1},\bm x_n,\bm x_{n+1})$, $\mathrm d\bm x=\bm x_{n+1}-\bm x_{n-1}$ has temporal midpoint $\bm x_m=\bm x_n$.

The mean-flow contribution $\langle\bm v_t\rangle^\intercal\matbf D_t^{-1}\langle\bm v_t\rangle$, which can be written as $\langle\matbf\mu_t\bm F_t\rangle^\intercal\matbf D_t^{-1}\langle\matbf\mu_t\bm F_t\rangle$ under usual boundary conditions, has a simple interpretation. The mutual information $I(\mathrm d\bm x_t;\bm x_m)$ is invariant under a uniform translation, whereas the path-space EP still detects such a global flow. The mean-flow term thus accounts for the mismatch between the forward/backward reference underlying EP and the midpoint-based forward reference underlying $\mathcal I$.
When $\langle\bm v_t\rangle=\bm 0$, the mismatch disappears and Eq.~\eqref{eq:main_identity} reduces to the exact relation $\sigma_t=4\mathcal I(\mathrm d\bm x_t;\bm x_m)$. In this case, path-space distinguishability is fully aligned with distinguishability within the forward trajectory ensemble. This includes, for example, confined steady states with vanishing net mean-flow.

\textit{Self and interaction EP channels}.---Eq.~\eqref{eq:main_identity} provides a new information-theoretic lens for understanding EP.
Consider a bipartite system $\{A,B\}$ with block-diagonal diffusion, and define
$\sigma_A^{\rm mf}:=\langle\bm v_A\rangle^\intercal\matbf D_A^{-1}\langle\bm v_A\rangle$.
Applying the same midpoint approach to subsystem $A$ yields an identity analogous to Eq.~\eqref{eq:main_identity} for the local EP rate of $A$,
\begin{equation}
    \sigma_A=4\,\mathcal I(\mathrm d\bm x_A;\bm x_m)
    +\sigma_A^{\rm mf}.
    \label{eq:local_EP_identity}
\end{equation}

Now, using the chain rule for $\bm{x}_m=(\bm{x}_m^A,\bm{x}_m^B)$ in the natural $A$-first ordering appropriate to $\sigma_A$ gives
\begin{subequations}\label{eq:local_EP_decom}
\renewcommand{\theequation}{\theparentequation\alph{equation}}
\begin{align}
    \sigma_\mathrm{A} 
    &= \underbrace{\sigma_A^{\rm mf} + 4\,\mathcal{I}(\mathrm{d}\bm{x}_A;\bm{x}_m^A)
    }_{\sigma_A^{\rm self}}
    +\underbrace{4\,\mathcal{I}(\mathrm{d}\bm{x}_A;\bm{x}_m^B|\bm{x}_m^A)}_{\sigma_{A|B}^{\rm int}}
    \label{eq:local_EP_decom_A}
    \\
    &=\big\langle\tilde{\bm v}_A^\intercal\matbf D_A^{-1}\tilde{\bm v}_A\big\rangle
    +\big\langle(\bm v_A-\tilde{\bm v}_A)^\intercal\matbf D_A^{-1}(\bm v_A-\tilde{\bm v}_A)\big\rangle,
    \label{eq:local_EP_decom_B}
\end{align}
\end{subequations}
where $\tilde{\bm v}_A(\bm x_A)\coloneq\mathbb E[\bm v_A|\bm x_A]$.
The midpoint identity and the Shannon chain rule thus yield an exact information-theoretic decomposition of the local EP rate into two nonnegative channels, Eq.~\eqref{eq:local_EP_decom_A}.
Equation~\eqref{eq:local_EP_decom_B} is the corresponding current-space representation, obtained by applying the small-SNR expansion~\cite{guo2005mutual} to Eq.~\eqref{eq:local_EP_decom_A}; 
the marginal-current projection underlying this representation and the nonnegative remainder it entails have precedents in earlier treatments of apparent EP and incomplete observation \cite{nakazato2021geometrical,frishman2020learning}.
%The contribution specific to the present construction is that the midpoint identity maps the Shannon chain rule exactly onto this current-space decomposition, and identifying its two contributions with the information rates in Eq.~\eqref{eq:local_EP_decom_A}. See also~\cite{diana2014mutual,lynn2022decomposing} for information-theoretic decompositions based on different path- or transition-level objects.
What Eq.~\eqref{eq:local_EP_decom_A} newly shows is that the midpoint identity maps the Shannon chain rule exactly onto this current-space decomposition, identifying its two terms with distinct midpoint information rates. See also~\cite{diana2014mutual,lynn2022decomposing} for information-theoretic decompositions constructed from different dynamical settings.
%path- or transition-level objects.
%The new content is not the algebraic projection in Eq. (6b) itself, but its exact realization in Eq. (6a) as two independently defined midpoint-information channels generated by the Shannon chain rule.

\begin{figure}[!t]
    \centering
    \includegraphics[width=\linewidth]{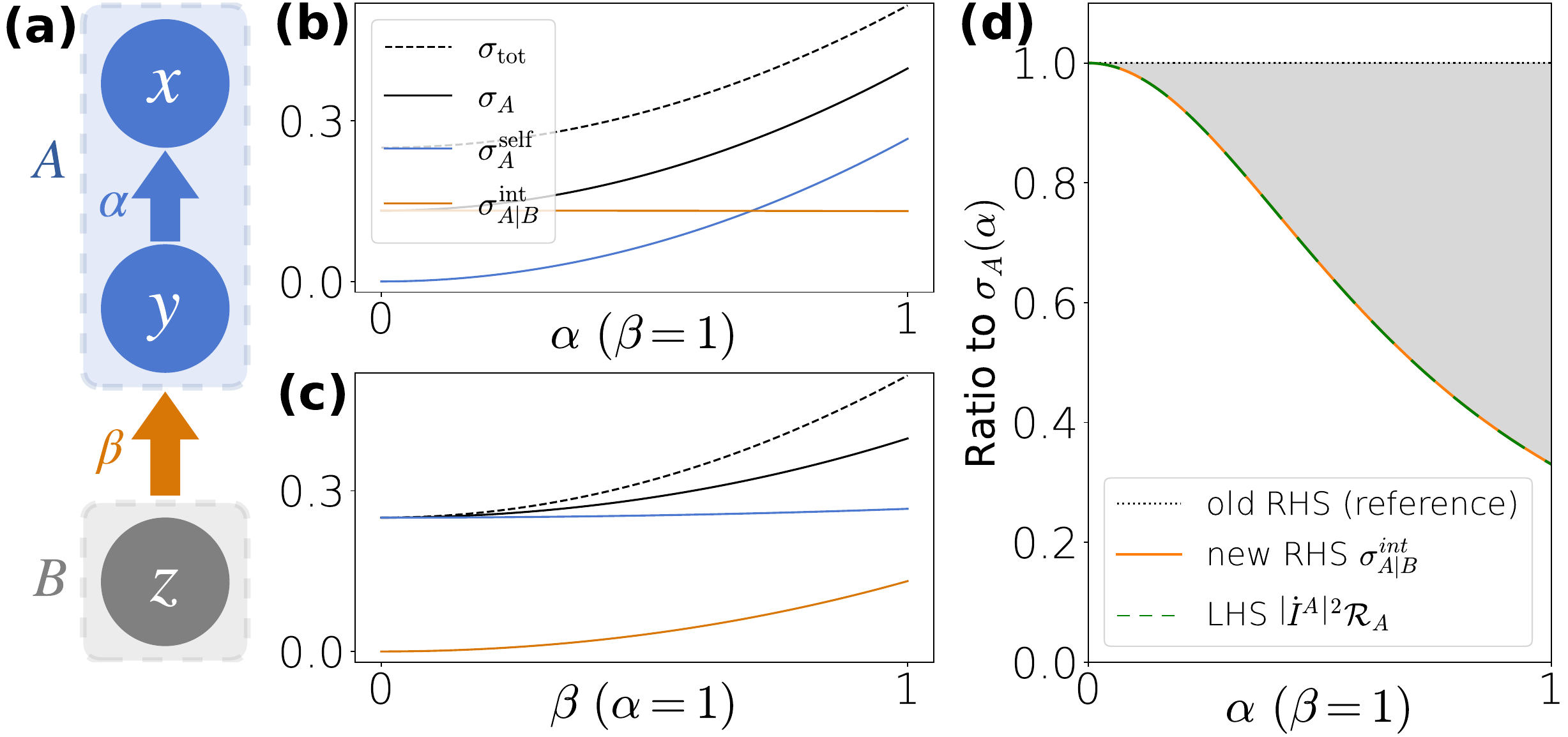}
    \vskip -0.15in
    \caption{
    Demonstration of the self and interaction EP channels and sharpened learning-rate bound.
    \textbf{(a)} 
    Cascade model $\dot{x}=-2 x + \alpha y + \sqrt 2\xi_x$, $\dot{y}=-2 y + \beta z + \sqrt 2\xi_y$, and $\dot{z}=-2 z + \sqrt 2\xi_z$ with $\langle\xi_i\xi_j\rangle=\delta_{ij}$. The system is partitioned into $A=\{x,y\}$ (blue) and $B=\{z\}$ (gray). 
    The coupling constant $\alpha$ controls the internal coupling within $A$, while $\beta$ drives $A$ from $B$.
    \textbf{(b, c)} 
    Local EP rate $\sigma_\mathrm{A}$ (solid black) and its nonnegative channels, self $\sigma_A^{\mathrm{self}}$ (blue) and interaction $\sigma_{A|B}^{\mathrm{int}}$ (orange), as functions of coupling strength. The black dashed line denotes the total EP $\sigma_{\mathrm{tot}}$.
    Varying $\alpha$ (with $\beta=1$) predominantly enhances $\sigma_A^{\mathrm{self}}$, while $\sigma_{A|B}^{\mathrm{int}}$ remains nearly unchanged (b).
    Varying $\beta$ (with $\alpha=1$) selectively enhances $\sigma_{A|B}^{\mathrm{int}}$ with minimal change in $\sigma_A^{\mathrm{self}}$ (c).
    \textbf{(d)}
    Tightening of the learning-rate bound along the same $\alpha$ sweep as in (b), with $\sigma_A^{\rm mf}=0$ in this model: the previous RHS $\sigma_A$ (black dotted), sharpened RHS $\sigma_{A|B}^{\rm int}$ (orange solid), and LHS $|\dot I^A|^2\mathcal R_A$ (green dashed) are shown, all normalized by $\sigma_A$.
    The gray shading marks the gap between the two bounds. 
    All three coincide at $\alpha=0$; the sharpened RHS decreases monotonically to $0.330$ at $\alpha=1$, while the LHS nearly saturates throughout, $|\dot I^A|^2\mathcal R_A/\sigma_{A|B}^{\rm int}>0.999$.
    }\label{fig3}
    \vskip -0.1in
\end{figure}

For a fixed partition, we refer to the two contributions in Eq.~\eqref{eq:local_EP_decom_A} as the \textit{self} and \textit{interaction} channels.
This terminology is resolution-based. The self channel $\sigma_A^{\rm self}$ is the $A$-sector irreversibility resolved from $A$'s own infinitesimal statistics of $(\mathrm d\bm x_A,\bm x_m^A)$
\footnote{The mean-flow contribution $\sigma_A^{\rm mf}$ is included in the self channel because it is likewise determined entirely by $A$'s own infinitesimal statistics: both $\langle\bm v_A\rangle=\lim_{\mathrm dt\to0}\mathbb E[\mathrm d\bm x_A]/\mathrm dt$ and $\matbf D_A=\lim_{\mathrm dt\to0}\operatorname{Cov}(\mathrm d\bm x_A)/(2\,\mathrm dt)$ are fixed by those statistics.}.
It may nevertheless depend on $A$'s coupling to $B$ through those marginal statistics.
Note that when $A$ is the full system, $\sigma_A^{\rm self}$ coincides with the total EP rate of the system. For a proper subsystem, however, $\sigma_A^{\rm self}$ is an instantaneous marginal-current contribution of the full Markovian dynamics and need not equal the complete path-space irreversibility of the generally non-Markovian $A$-only process~\cite{kahlen2018hidden}.
The interaction channel $\sigma_{A|B}^{\rm int}$ is the additional $A$-sector irreversibility revealed by resolving $\bm x_m^B$ in addition to $\bm x_m^A$; it need not imply direct dynamical coupling.
To illustrate these channels, we use a linear cascade model $z\rightarrow y\rightarrow x$, partitioned into $A=\{x,y\}$ and $B=\{z\}$ (Fig.~\ref{fig3}a--c). 
%As shown in Fig.~\ref{fig3}b,c, strengthening the internal coupling within subsystem $A$ primarily elevates the self EP rate, whereas increasing the external drive from $B$ selectively raises the interaction EP rate.
% The self channel is the instantaneous marginal-current contribution of subsystem A, and hence coincides with its apparent EP; we exploit this identification below to resolve a reported coarse-graining reversal.
% For block-diagonal subsystem noise, the self channel coincides with the standard instantaneous apparent EP; this identification will be used below to resolve a reported coarse-graining reversal.

\textit{Interaction EP constrains learning}.---The interaction channel has an operational meaning beyond being the
remainder after subtracting the self channel. To make this statement precise, we begin with the recently derived learning-rate bound~\cite{matsumoto2025learningrate},
\begin{equation}
    \big|\dot I^A(\bm x_A;\bm x_B)\big|^2\mathcal R_A\le
    \sigma_A-\sigma_A^{\rm mf}.
    \label{eq:known_learning_rate_bound}
\end{equation}
Here, $\dot I^A$ is the learning rate and $\mathcal R_A\coloneq\mathrm{Tr}[\matbf D_A\matbf F_A^{|A}]^{-1}$ is the corresponding information-resistance factor, defined from the conditional Fisher information matrix $\matbf F_A^{|A}$ of subsystem $A$.
The channel representation in Eq.~\eqref{eq:local_EP_decom_A} makes explicit that the RHS of Eq.~\eqref{eq:known_learning_rate_bound} decomposes into two midpoint-information channels,
\[
    \sigma_A-\sigma_A^{\rm mf}=4\,\mathcal I(\mathrm d\bm x_A;\bm x_m^A)+4\,\mathcal I(\mathrm d\bm x_A;\bm x_m^B|\bm x_m^A).
\]
Because $\dot I^A$ quantifies the instantaneous change in the statistical dependence between $A$ and $B$ generated by the $A$-dynamics~\cite{horowitz2014thermodynamics,barato2014efficiency}, this decomposition suggests that the corresponding dissipative cost should be controlled not by the $A$-only channel but by the $B|A$ conditional channel $4\,\mathcal I(\mathrm d\bm x_A;\bm x_m^B|\bm x_m^A)=\sigma_{A|B}^{\rm int}$. This structural prediction is rigorously established in current-space (see End Matter), yielding
\begin{equation}
    \big|\dot I^A(\bm x_A;\bm x_B)\big|^2\mathcal R_A\le
    \sigma_{A|B}^{\mathrm{int}}.
    \label{eq:sharpened_learning_rate_bound}
\end{equation}
Thus, any nonzero $A$-learning rate necessarily draws on this channel, providing an operational basis for the term \textit{interaction}.

The improvement over Eq.~\eqref{eq:known_learning_rate_bound} is exactly $4\,\mathcal I(\mathrm d\bm x_A;\bm x_m^A)$. Hence, Eq.~\eqref{eq:sharpened_learning_rate_bound} provides a strictly tighter bound whenever $\mathcal I(\mathrm d\bm x_A;\bm x_m^A)>0$.
Figure~\ref{fig3}d shows that this tightening can be substantial. As $\sigma_A^{\rm self}$ increases from zero, the previous bound becomes increasingly loose, whereas the interaction-channel bound remains nearly saturated. This behavior illustrates that the self channel contributes to the looseness of the previous bound, while the interaction channel captures the dissipation relevant to learning. A similar substantial tightening is observed in a model fitted to red blood cell data~\cite{terlizzi2024variance}; see the SM.

%left upper anchor

\begin{figure}[!t]
    \centering
    \includegraphics[width=\linewidth]{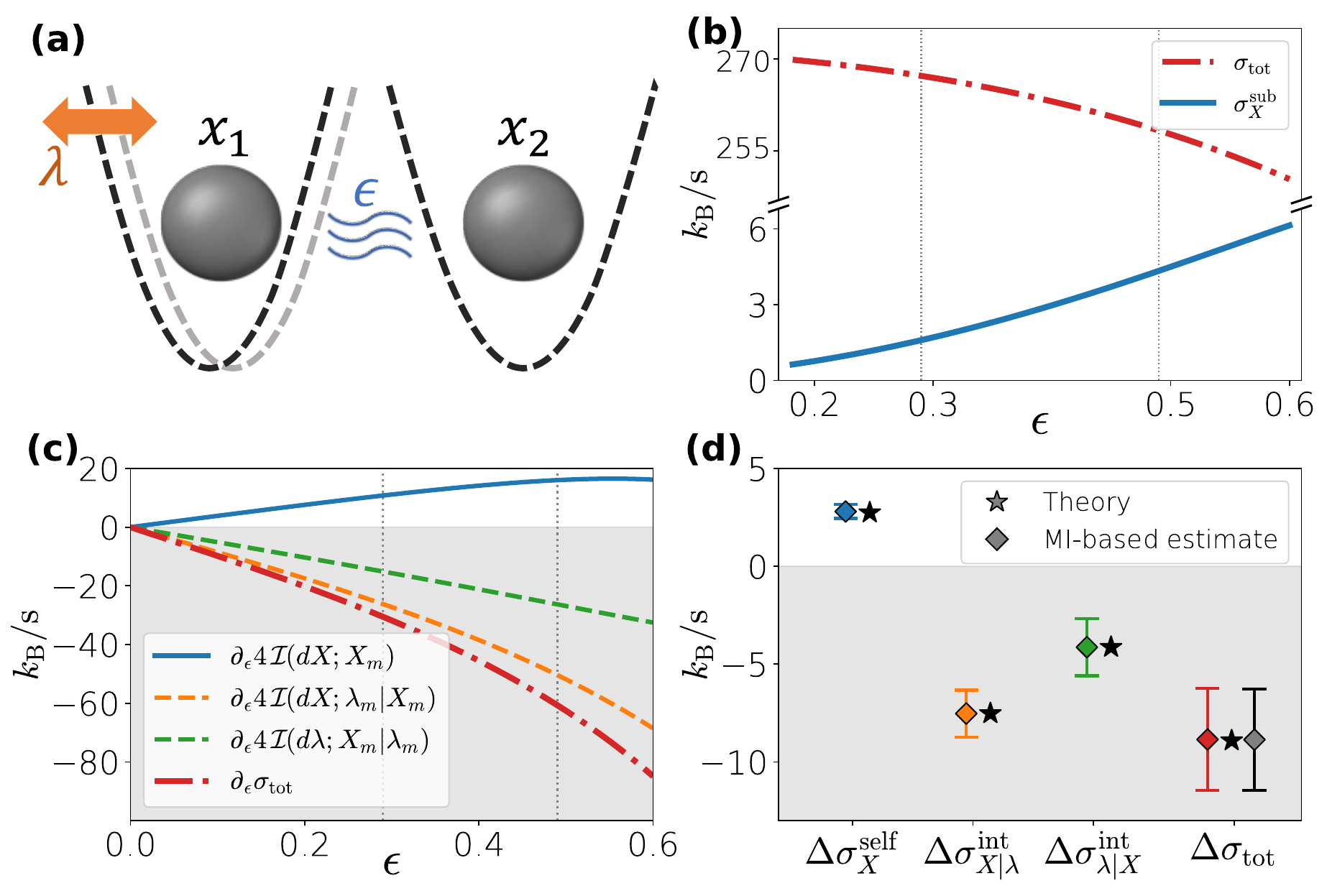}
    \vskip -0.15in
    \caption{
    Midpoint-information resolution of a coarse-graining reversal. Gray shading denotes negative values in panels (c) and (d).
    \textbf{(a)}
    Schematic of two harmonically trapped particles coupled by hydrodynamic interactions of strength \(\epsilon\). The Ornstein--Uhlenbeck drive \(\lambda(t)\) displaces the center of the first trap.
    \textbf{(b)}
    Previously reported coarse-graining reversal. Increasing \(\epsilon\) decreases the full-system EP rate $\sigma_{\rm tot}$ but increases the particle-subspace EP rate \(\sigma_X^{\rm sub}\).
    \textbf{(c)}
    Responses of the three channels in Eq.~\eqref{eq:Das_model_channel_decomposition}, \(\partial_\epsilon\sigma_X^{\mathrm{self}}\) (blue solid), \(\partial_\epsilon\sigma_{X|\lambda}^{\mathrm{int}}\) (orange dashed), and \(\partial_\epsilon\sigma_{\lambda|X}^{\mathrm{int}}\) (green dashed), together with \(\partial_\epsilon\sigma_{\mathrm{tot}}\) (red dash-dotted). Coarse-graining to \(X\) retains only the self channel, which is the only channel with a positive response. Vertical dotted lines mark \(\epsilon=0.29\) and \(0.49\).
    \textbf{(d)}
    Finite-resolution recovery of the channel changes \(\Delta\sigma_\alpha\equiv\sigma_\alpha(0.49)-\sigma_\alpha(0.29)\). Stars denote continuous-time theoretical values. Diamonds show midpoint-MI rate estimates obtained from 40 independent synthetic trajectories at each value of \(\epsilon\), each sampled at \(10\,\mathrm{kHz}\) for \(50\,\mathrm{s}\); error bars are 95\% confidence intervals for the differences of ensemble means. For the total channel, the directly estimated change (red) and the sum of the three channel changes (gray) are shown side by side. Their difference is \(-0.012\,k_{\mathrm B}\,\mathrm{s}^{-1}\), with a 95\% confidence interval of \([-0.146,0.123]\,k_{\mathrm B}\,\mathrm{s}^{-1}\). Panels (b--d) use \(\theta=0.56\).
    }\label{fig4}
    \vskip -0.12in
\end{figure}

\textit{Midpoint-information resolution of a coarse-graining reversal}.---Recently, a counterintuitive coarse-graining reversal was reported in hydrodynamically coupled colloids~\cite{das2025irreversibility}. Two harmonically trapped particles $X=\{x_1,x_2\}$ are coupled with strength $\epsilon$, while an autonomous Ornstein--Uhlenbeck (OU) drive $\lambda(t)$ modulates the center of the first trap. Increasing $\epsilon$ lowers the total EP rate, but raises the particle-subspace EP rate (Fig.~\ref{fig4}a,b). 
A subsequent study traced these trends to redistributed dependencies and information pathways using time-delayed MI and transfer entropy~\cite{das2025uraveling}.
Here we make this connection exact at the EP level by showing how the physical coarse-graining $(X,\lambda)\mapsto X$ acts directly on the midpoint-information budget.
% "this connection"에 information-theoretic을 명시하는 편이 낫지 않나?
%“이미 peer-reviewed literature에서 발견되고, 후속 논문까지 낳았지만 thermodynamic calculation과 information-theoretic mechanism이 분리되어 있던 문제를, 우리의 exact identity가 하나의 quantitative information-theoretic EP budget으로 닫는다.”

The linear dynamics can be written as
\begin{equation}
    \dot{\bm z} = -\matbf F \bm z + \sqrt{2\matbf D}\,\bm\xi,
    \quad \bm z = (x_1,x_2,\lambda)^\intercal,
    \label{eq:Das_model}
\end{equation}
where
\begin{equation*}
    \matbf F\!=\!\!
    \begin{pmatrix}
    1/\tau_1 & \epsilon/{\tau_2} & -1/\tau_1 \\
    \epsilon/\tau_1 & 1/{\tau_2} & -\epsilon/\tau_1 \\
    0 & 0 & 1/\tau_\lambda
    \end{pmatrix}\!\!,
    \;\; 
    \matbf D\!=\!\!
    \begin{pmatrix}
    D_0 & \epsilon D_0 & 0 \\
    \epsilon D_0 & D_0 & 0 \\
    0 & 0 & \theta D_0
    \end{pmatrix}\!.   
    \label{eq:Das_model_details}
\end{equation*}
Here, $\tau_{1,2}$ are the particle relaxation times, $\tau_\lambda$ is the OU correlation time, $D_0$ sets the particle diffusion scale, and $\theta$ controls the OU drive noise strength relative to that scale.
Since $\matbf D_{X\lambda}=0$, the partition $\{X,\lambda\}$ satisfies the block-diagonal condition of Eq.~\eqref{eq:local_EP_identity}.
All mean-flow terms vanish in the centered steady state, so Eq.~\eqref{eq:local_EP_identity} gives
\begin{equation*}
    \sigma_{tot}=4\,\mathcal I(\mathrm dX;X_m,\lambda_m)+4\,\mathcal I(\mathrm d\lambda;X_m,\lambda_m).
    \label{eq:Das_model_local_EP_decom}
\end{equation*}

The key point is that physical coarse-graining acts directly on this information budget, because the variables defining the resolution themselves appear as arguments of the MI terms. Under coarse-graining,
\begin{equation*}
    \{X,\lambda\}\mapsto X
    \implies
    \begin{cases}
        \ \mathcal I(\mathrm dX;X_m,\lambda_m)
        \mapsto
        \mathcal I(\mathrm dX;X_m),
        \\
        \ \mathcal I(\mathrm d\lambda;X_m,\lambda_m)
        \mapsto
        \mathcal I(\emptyset;X_m)=0.
    \end{cases}
    \label{eq:coarse_graining_map}
\end{equation*}
The resulting term is precisely the particle-subspace EP $\sigma_X^{\rm sub}$ considered in the original studies~\cite{das2025irreversibility,das2025uraveling},
\begin{equation}
    \sigma_X^{\rm sub}=\sigma_X^{\rm self}=4\,\mathcal I(\mathrm dX;X_m),
\end{equation}
as verified from its marginal-current definition given in the End Matter.
For the $\lambda$-sector, the symmetric increment $\mathrm d\lambda$ is uncorrelated with $\lambda_m$ in the stationary scalar OU marginal. Joint Gaussianity therefore gives $\mathcal I(\mathrm d\lambda;\lambda_m)=0$.
The Shannon chain rule then identifies two distinct conditional-information channels as exactly what is removed from each sector. 
Hence, the total EP has the exact three-channel budget
\begin{equation}
    \begin{aligned}
        \sigma_{tot}
        =\underbrace{4\,\mathcal I(\mathrm dX;X_m)}_{\sigma_X^{\rm self}=\sigma_X^{\rm sub}}
        +\underbrace{4\,\mathcal I(\mathrm dX;\lambda_m|X_m)}_{\sigma_{X|\lambda}^{\rm int}}
        +\underbrace{4\,\mathcal I(\mathrm d\lambda;X_m|\lambda_m)}_{\sigma_{\lambda|X}^{\rm int}}.
    \end{aligned}
    \label{eq:Das_model_channel_decomposition}
\end{equation}

% right upper anchor

The full-system EP $\sigma_{tot}$ and the particle-subspace EP $\sigma_X^{\rm sub}$ thus belong to the same exact information budget. Hydrodynamic coupling redistributes this budget according to
\begin{equation}
    \partial_\epsilon\sigma_{X|\lambda}^{\rm int}
    ,\ \partial_\epsilon\sigma_{\lambda|X}^{\rm int}<-\partial_\epsilon\sigma_X^{\rm self}<0.
    \label{eq:epsilon_diff_channel}
\end{equation}
This means that increasing $\epsilon$ suppresses both interaction channels more strongly than it enhances the self channel. As a result, $\sigma_{tot}$ falls while $\sigma_X^{\rm self}$ rises.
Coarse-graining to $X$ retains only the growing self channel as $\sigma_X^{\rm sub}$, producing the reported reversal (Fig.~\ref{fig4}c).
%Over the experimentally relevant interval ϵ=0.29→0.49, the decrease in the discarded interaction budget is more than four times the increase in the retained self channel.

%midpoint itself가 곧 EP이다.
These channel responses specify what information is redistributed. Stronger coupling makes $X_m$ more informative about $\mathrm dX$, while $\lambda_m$ supplies less additional information once $X_m$ is known. The decrease of $\mathcal I(\mathrm d\lambda;X_m|\lambda_m)$ has a subtler retrodictive origin; a nested chain rule decomposition resolves how this memory is redistributed between the two particles (see End Matter).
Using three-point trajectory windows and zero-time extrapolation, finite-resolution synthetic records recover the predicted MI and conditional MI channel changes, whose sum agrees with a direct estimate of \(\Delta\sigma_{\rm tot}\) within uncertainty (Fig.~\ref{fig4}d; see SM for estimation details).

\textit{Discussion}.---Equivalent representations of EP need not be equally revealing for a given physical question. For overdamped Langevin dynamics, the path-space representation naturally exposes time reversal and fluctuation relations~\cite{Maes2003time-reversal,Seifert2005Entropy}, whereas the current-quadratic form reveals geometric and projection structures~\cite{dechant2022geometricPRR,frishman2020learning}. Their distinct strengths reflect the native operations of the respective representations. 
The midpoint-information representation adds a third perspective, expressing dissipation itself through ordinary MI and making the Shannon chain rule a native operation on EP. 

In additive bipartite Langevin systems, this algebraic access has two concrete consequences. Coarse-graining onto a subsystem removes a precise interaction channel of irreversibility rather than merely underestimating dissipation, which provides an exact resolution of the reported hydrodynamic coarse-graining reversal. The learning-rate bound, by contrast, couples selectively to the interaction channel and not to the self contribution. Taken together, these results show that, within this channel decomposition, coarse-graining and learning probe complementary parts of the same dissipation decomposition.
The success of the chain rule in exposing this structure suggests that other information-theoretic operations, such as the data-processing inequality and partial information decomposition, could likewise provide new insights into dissipation. Extending the framework to non-bipartite architectures, underdamped dynamics, and other stochastic systems offers natural directions for future work.

% $\langle\ln\mathcal P/\mathcal P^\dagger\rangle$과 $\langle\bm v^\intercal\matbf D^{-1}\bm v\rangle$은 똑같은 EP를 나타내고 똑같은 값을 주지만, equally revealing for a given physical question 하진 않는다. 전자는 fluctuation theorem 등에서 해석력이 더 강하고 후자는 geometric decomposition에서 더 강하다. 이는 각 representation의 native operation이 다르기 때문이다.
% 우리는 Overdamped Langevin dynamics에서, EP가 midpoint information을 통해 새로운 표현 방법을 가짐을 보였으며, 이를 통해 argument단위에서의 Shannon chain rule을 EP의 native operation으로 격상시킨다.
% Coarse-graining therefore removes a precise interaction channel of irreversibility, instead of merely underestimating dissipation.
% Learning rate bound는 흥미롭게도 반대 채널에 선택적으로 결합한다는 통찰도 얻는다.
% 나아가 후속 연구 방향으로서 chain rule 외에도 Data processing inequality나, partial information decomposition 등 MI form이 제공할 수 있는 정보론적 분석이 EP에 대한 새로운 통찰을 제공할 가능성을 제시한다. 또한 extensions to non-bipartite architectures, underdamped dynamics, and other stochastic systems도 흥미로운 연구 방향일 것이다.

\begin{acknowledgments}
\textit{Acknowledgments.---}We thank Jong-Min Park, Takahiro Sagawa, and Biswajit Das for helpful discussions. This work was supported by the Basic Science Research Program through the National Research Foundation of Korea (NRF Grant No. RS-2025-00514776).

\end{acknowledgments}

\bibliography{refs}

% =========================
% Save pre-appendix settings
% =========================
\makeatletter
\let\saved@hangfrom@section\@hangfrom@section
\let\saved@sectioncntformat\@sectioncntformat
\let\saved@thesection\thesection
\let\saved@thesubsection\thesubsection
\let\saved@thesubsubsection\thesubsubsection
\let\saved@cl@section\cl@section
\let\saved@theequation@prefix\theequation@prefix
\@ifundefined{theHequation}{}{%
  \let\saved@theHequation\theHequation
}
\makeatother

% =========================
% End Matter
% =========================
\appendix*
\section*{End Matter}

\section{A. Current-space proof of the sharpened learning-rate bound Eq.~\eqref{eq:sharpened_learning_rate_bound}}

Let $\bm s_A$ denote the conditional score and $\matbf F_A^{|A}$ the corresponding conditional Fisher information matrix for subsystem $A$,
\begin{equation*}
    \bm s_A(\bm x)\coloneq\nabla_{\bm x_A}\ln p(\bm x_B|\bm x_A),
    \quad
    \matbf F_A^{|A}\coloneq\langle\bm s_A\bm s_A^\intercal\rangle.
\end{equation*}
The learning rate satisfies~\cite{horowitz2014thermodynamics}
\begin{equation*}
    \dot I_A(\bm x_A;\bm x_B)=\langle\bm v_A\cdot\bm s_A\rangle.
\end{equation*}
Moreover,
\begin{equation*}
    \mathbb E[\bm s_A|\bm x_A]=\int\mathrm d\bm x_B\,p(\bm x_B|\bm x_A)\nabla_{\bm x_A}\ln p(\bm x_B|\bm x_A)=0,
\end{equation*}
so as anticipated by the midpoint-information representation, only the part of $\bm v_A$ not determined by $\bm x_A$ contributes to the learning rate:
\[
    \dot I_A(\bm x_A;\bm x_B)
    =\big\langle(\bm v_A-\tilde{\bm v}_A)\cdot\bm s_A\big\rangle.
\]
Applying Cauchy--Schwarz inequality in the diffusion metric $\matbf D_A$, we obtain
\begin{equation}
   \left|\dot I_A(\bm x_A;\bm x_B)\right|^2\leq
    \Big\langle(\bm v_A-\tilde{\bm v}_A)^\intercal\matbf D_A^{-1}(\bm v_A-\tilde{\bm v}_A)\Big\rangle
    \Big\langle\bm s_A^\intercal\matbf D_A\bm s_A\Big\rangle.
    \label{eq:learning-rate_CS}
\end{equation}
The equality condition for the Cauchy--Schwarz inequality gives
\begin{equation*}
    \bm v_A-\tilde{\bm v}_A
    \propto \matbf D_A\bm s_A
    %\quad\text{almost surely},
\end{equation*}
with a state-independent proportionality constant.

Equation~\eqref{eq:local_EP_decom_B} identifies the first factor on the RHS of Eq.~\eqref{eq:learning-rate_CS} as the interaction EP channel,
\begin{equation*}
    \Big\langle(\bm v_A-\tilde{\bm v}_A)^\intercal\matbf D_A^{-1}(\bm v_A-\tilde{\bm v}_A)\Big\rangle
    =\sigma_{A|B}^{\rm int}.
\end{equation*}
The second factor is $\big\langle\bm s_A^\intercal\matbf D_A\bm s_A\big\rangle
    =\mathrm{Tr}[\matbf D_A\matbf F_A^{|A}]
    =\mathcal R_A^{-1}$.
Hence we get
\[
    |\dot I_A(\bm x_A;\bm x_B)|^2\mathcal R_A
    \leq \sigma_{A|B}^{\rm int},
\]
which is Eq.~\eqref{eq:sharpened_learning_rate_bound}.

\section{B. Identification of the retained midpoint channel with particle-subspace EP}

Denote by $\sigma_X^{sub}$ the particle-subspace EP considered in~\cite{das2025irreversibility,das2025uraveling}. Its marginal current is
\[
    \bm j_X^{\rm marg}(X)
    \coloneq\int\!\mathrm d\lambda\,\bm j_X(X,\lambda),
\]
with $\tilde{\bm v}_X(X)\coloneq\mathbb E[\bm v_X|X]=\bm j_X^{\rm marg}/p_X(X)$. All mean-flow terms vanish in the centered steady state, giving
\begin{equation*}
\begin{aligned}
    \sigma_X^{\rm sub}
    &=\langle\tilde{\bm v}_X^\intercal\matbf D_X^{-1}\tilde{\bm v}_X\rangle
    =\int dX\frac{(\bm j_X^{\rm marg})^\intercal\matbf D_X^{-1}\bm j_X^{\rm marg}}{p_X}
    \\
    &=\sigma_X^{\rm app}=\sigma_X^{\rm self}
    =4\,\mathcal I(\mathrm dX;X_m).
\end{aligned}
    \label{eq:sub_equal_app_equal_self}
\end{equation*}
The equality $\sigma_X^{\rm sub}=\sigma_X^{\rm app}$ is the known marginal-current projection~\cite{mehl2012role,nakazato2021geometrical}, whereas the final equality is its midpoint-information representation of Eq.~\eqref{eq:local_EP_decom_A}.

Importantly, this identification is not specific to the present linear model. At the level of the instantaneous marginal current, the apparent EP admits the same midpoint-information representation for general nonlinear dynamics and does not rely on the bipartite structure used for the full channel decomposition. More generally, restoring a possible mean-flow gives
\begin{equation*}
    \sigma_A^{\rm app}
    =\sigma_A^{\rm self}
    =4\,\mathcal I(d\bm x_A;\bm x_m^A)+\sigma_A^{\rm mf}.
\end{equation*}
Thus, the midpoint representation provides an information-theoretic interpretation of the conventional apparent EP. As emphasized in the main text, however, this instantaneous marginal-current quantity generally differs from the full path-space irreversibility of the $A$-only trajectory, which is generally non-Markovian. For example, in a stationary confined one-dimensional subsystem, $j_A^{\rm marg}=0$ and hence $\sigma_A^{\rm self}=0$, even though its path-space irreversibility can be positive.

\begin{figure}[h]
    \centering
    \includegraphics[width=0.95\linewidth]{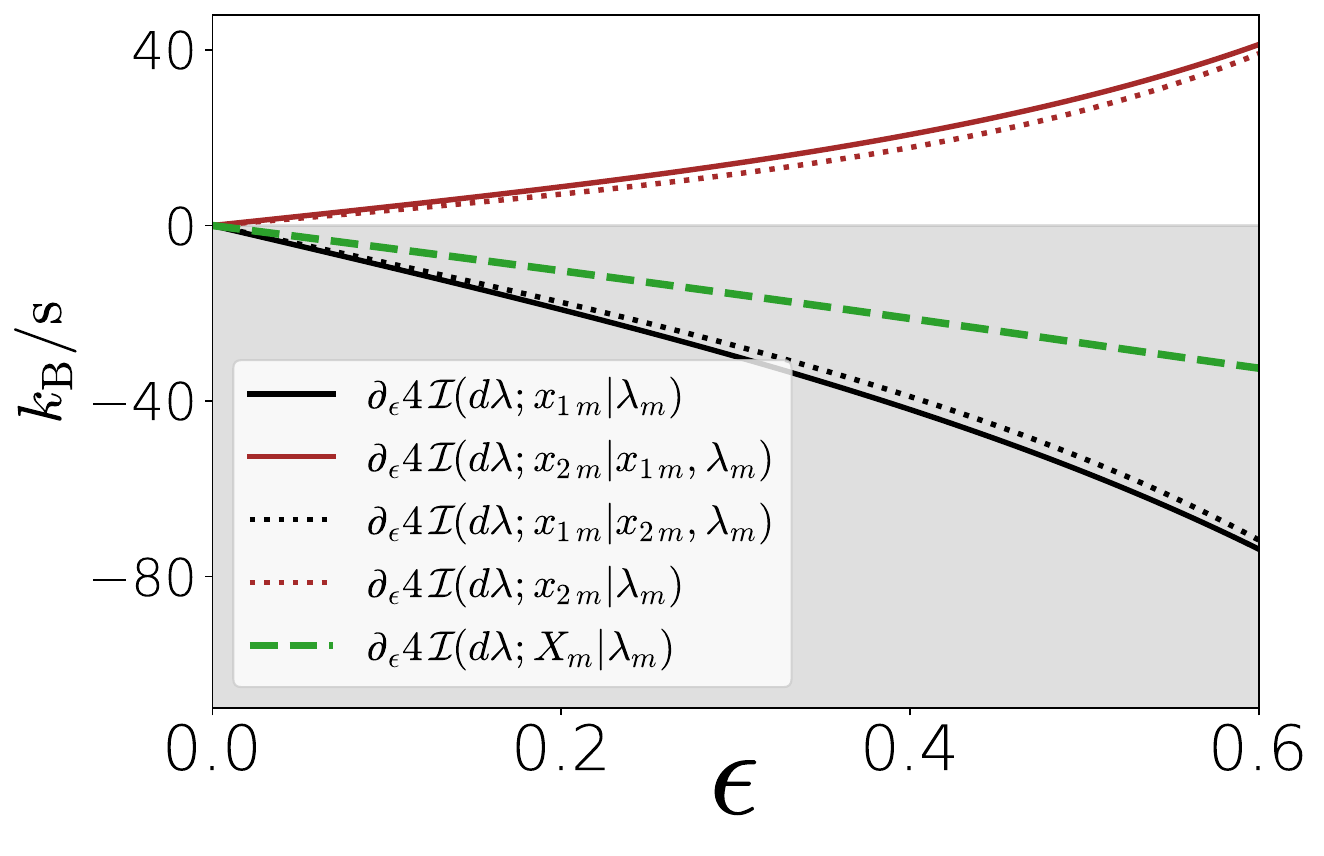}
    \vskip -0.15in
    \caption{
    Nested chain rule decomposition of    $\sigma_{\lambda|X}^{\rm int}    =4\,\mathcal I(\mathrm d\lambda;X_m|\lambda_m)$.
    The curves show the two contributions under the $x_1$-first (solid) and $x_2$-first (dotted) chain rule orderings.
    Regardless of the ordering, the $x_1$ contribution $\partial_\epsilon 4\,\mathcal I(\mathrm d\lambda;{x_1}_m|\cdot)$ (black) is negative, whereas the $x_2$ contribution $\partial_\epsilon 4\,\mathcal I(\mathrm d\lambda;{x_2}_m|\cdot)$ (brown) is positive.
    Their sum is independent of the ordering and equals $\partial_\epsilon 4\,\mathcal I(\mathrm d\lambda;X_m|\lambda_m)$ (green dashed), which remains negative throughout.
    All curves are evaluated at $\theta=0.56$. The horizontal gray line marks zero, and the gray shading denotes negative response. The two vertical gray lines indicate $\epsilon=0.29$ and $0.49$, the coupling strengths analyzed in the experiments.
    }\label{efig1}
    %\vskip 0.9in
\end{figure}

\section{C. Information-theoretic origin of the decrease in $\sigma_{\lambda|X}^{\rm int}=4\,\mathcal I(\mathrm d\lambda;X_m|\lambda_m)$}

Although the causal coupling in the dynamics in Eq.~\eqref{eq:Das_model} runs from $\lambda$ to $X$, $\mathcal I(\mathrm d\lambda;X_m|\lambda_m)$ is nonzero. The reason is that $X_m$ is the actual temporal midpoint. Writing
\begin{equation*}
    \mathrm d\lambda
    = [\lambda(t+\mathrm dt)-\lambda_m]
    + [\lambda_m-\lambda(t)],
\end{equation*}
the earlier half-step $\lambda_m-\lambda(t)$ is generally not independent of $X_m$, even when conditioned on $\lambda_m$, because $X_m$ retains information about the recent history of $\lambda$ through the lagged dynamical response of the particle system. Thus, this channel measures a retrodictive, lagged memory rather than a causal influence from $X$ back to $\lambda$. In this sense, the ``interaction'' captured by $\sigma_{\lambda|X}^{\rm int}$ reflects memory generated by the dynamical coupling.

To resolve how this memory is redistributed between the two particles, consider the $x_1$-first chain rule decomposition
\begin{equation*}
    \mathcal I(\mathrm d\lambda;X_m|\lambda_m)
    =\mathcal I(\mathrm d\lambda;{x_1}_m|\lambda_m)
    +\mathcal I(\mathrm d\lambda;{x_2}_m|{x_1}_m,\lambda_m).
\end{equation*}
As $\epsilon$ increases, the first term decreases whereas the second increases. Stronger hydrodynamic coupling between $x_1$ and $x_2$ effectively loads $x_1$ through its coupling to $x_2$, thereby reducing the memory of $\lambda$ retained in ${x_1}_m$.
At the same time, ${x_2}_m$ acquires part of this memory through the hydrodynamic coupling. This gain, however, is insufficient to compensate for the memory lost by ${x_1}_m$, so the total $\mathcal I(\mathrm d\lambda;X_m|\lambda_m)$ decreases. Reversing the chain rule ordering of $\{{x_1}_m,{x_2}_m\}$ preserves these signs and their qualitative magnitude relation, as shown in Fig.~\ref{efig1}.

% =========================
% Leave appendix mode
% =========================
\makeatletter
\let\@hangfrom@section\saved@hangfrom@section
\let\@sectioncntformat\saved@sectioncntformat
\let\thesection\saved@thesection
\let\thesubsection\saved@thesubsection
\let\thesubsubsection\saved@thesubsubsection
\let\cl@section\saved@cl@section
\let\theequation@prefix\saved@theequation@prefix
\@ifundefined{saved@theHequation}{}{%
  \let\theHequation\saved@theHequation
}
\makeatother

% =========================
% SM
% =========================
\newpage
\begin{widetext}
\setcounter{secnumdepth}{3}

\setcounter{equation}{0}
\setcounter{figure}{0}
\setcounter{table}{0}
\setcounter{page}{1}

\renewcommand{\theequation}{S\arabic{equation}}
\renewcommand{\theHequation}{S\arabic{equation}}
\renewcommand{\thefigure}{S\arabic{figure}}
\renewcommand{\theHfigure}{S\arabic{figure}}

\section*{Supplemental materials}

\section{Derivation of The Main Identity}
\vspace*{-0.3\baselineskip}
The main text focuses on the spatially homogeneous (additive-noise) case for clarity. In this Supplement we establish the general multiplicative-noise result, and recover the additive-noise statement of the main text as a special case.
The derivation proceeds as follows. Secs I.A--I.B set notation and assumptions. Sec. I.C recalls the reverse-time diffusion. Sec. I.D derives the midpoint half-step decomposition and the resulting short-time representation of the midpoint-whitened increment. Sec. I.E proves the cancellation of the $O(\sqrt{\mathrm dt})$ non-Gaussian terms under midpoint conditioning, and Sec. I.F shows that the remaining deviation from the associated Gaussian proxy contributes only an $o(\mathrm dt)$ term and yields the multiplicative-noise identity. Throughout the SM, uppercase $\bm X_t$ denotes the stochastic process, lowercase $\bm x$ a conditioned value or integration variable.

\vspace*{-0.5\baselineskip}
\subsection{Noise convention, model settings, and the midpoint notation}
\vspace*{-0.5\baselineskip}
\noindent\textbf{$\bullet$ Noise convention.} We use a $d$-dimensional Gaussian white noise $\bm{\xi}(t)$ with
\begin{equation}
    \langle\bm{\xi}(t)\bm{\xi}(t')^\intercal\rangle=\mathbf{I}_d\,\delta(t-t').
    \label{seq:noise_def}
\end{equation}
Equivalently, we introduce a $d$-dimensional standard Wiener process $W_t$ satisfying
$\mathbb{E}[\mathrm{d}W_t]=0$ and
$\mathbb{E}[\mathrm{d}W_t\,\mathrm{d}W_t^\intercal]=\mathbf{I}_d\,\mathrm{d}t$.
Formally, one may view $\mathrm{d}W_t$ as the time integral of the white noise,
i.e.\ $\mathrm{d}W_t = \bm{\xi}(t)\,\mathrm{d}t$.

\bigskip
\noindent\textbf{$\bullet$ Dynamics.} We consider the Itô SDE
\begin{equation}
    \mathrm{d}\bm{X}_t=\bm{F}_I(\bm{X}_t,t)\mathrm{d}t+\matbf{B}(\bm{X}_t,t)\mathrm{d}W_t,\quad\bm{X}_t\in\mathbb{R}^d.
    \label{seq:dynamics_def}
\end{equation}
We define the diffusion matrix $\matbf{D}(\bm{x},t)\coloneq\frac{1}{2}\matbf{a}(\bm{x},t)$,
where $\matbf{a}(\bm{x},t)\coloneq\matbf{B}(\bm{x},t)\matbf{B}(\bm{x},t)^\intercal$. Throughout, we fix $\matbf{B}$ to be the symmetric positive-definite square root of $\matbf{a}$ (i.e., $\matbf{B}=\matbf{a}^{1/2}$), which implies $\matbf{B}=\matbf{B}^\intercal$.
(In a physical overdamped Langevin form $\dot{\bm{x}}=\matbf{\mu}(\bm{x},t)\bm{f}(\bm{x},t)+\matbf{B}(\bm{x},t)\bm{\xi}$, one may absorb the mobility $\matbf{\mu}$ into the drift and relabel $\matbf{\mu}\bm{f}\mapsto\bm{F}_I$; our result depends only on the resulting It\^{o} drift $\bm{F}_I$ and diffusion matrix $\matbf{D}$.)

\bigskip
\noindent\textbf{$\bullet$ Current and current velocity.}
Let $p(\bm{x},t)$ denote the one-time density of $\bm{X}_t$. It is well known that $p$ evolves according to the Fokker-Planck equation, which can be written in conservation (continuity) form $\partial_tp=-\nabla\cdot\bm{j}$ with probability current
\begin{equation}
    \bm{j}(\bm{x},t)\coloneq\bm{F}_I(\bm{x},t)p(\bm{x},t)-
    \frac{1}{2}\nabla\cdot(\matbf{a}(\bm{x},t)p(\bm{x},t))=
    \bm{F}_I\,p-\nabla\cdot(\matbf{D}p),
    \label{seq:def_j}
\end{equation}
where $(\nabla\cdot(\matbf{a}p))_i\coloneq\sum_j\partial_{x_j}(\mathsf{a}_{ij}p)$.
This form of the probability current is standard in the Fokker-Planck description~\cite{risken1996fokkerplanck,gardiner2009stochastic}.
Following the standard representation~\cite{nelson1967dynamical, seifert2012stochastic}, we further introduce the current velocity 
\begin{equation}
    \bm{v}(\bm{x},t)\coloneq\frac{\bm{j}(\bm{x},t)}{p(\bm{x},t)}
    =\bm{F}_I-\frac{\nabla\cdot(\matbf{D}p)}{p}.
    \label{seq:def_v}
\end{equation}

\medskip
\noindent\textbf{$\bullet$ Entropy production rate.}
For overdamped diffusion processes, under suitable smoothness and regularity assumptions (to be specified in the next section), 
the instantaneous total entropy production rate $\sigma_t$ can be expressed in terms of the probability current as
\begin{equation}
    \sigma_t
    \;=\;\int \mathrm{d}\bm{x}\;
    \frac{\bm{j}(\bm{x},t)^\intercal \matbf{D}(\bm{x},t)^{-1}\bm{j}(\bm{x},t)}{p(\bm{x},t)}
    \;=\;\big\langle \bm{v}(\bm{X}_t,t)^\intercal \matbf{D}(\bm{X}_t,t)^{-1}\bm{v}(\bm{X}_t,t)\big\rangle,
    \label{seq:ep_rate}
\end{equation}
where we used $\bm{j}=p\bm{v}$. This is a standard result in stochastic thermodynamics~\cite{seifert2012stochastic}.

\bigskip
\noindent\textbf{$\bullet$ Time Midpoint notation.} Fix a small $\mathrm{d}t>0$ and define
\begin{equation}
    \mathrm{d}\bm{x}(t)\coloneq\bm{X}_{t+\mathrm{d}t}-\bm{X}_t,
    \quad
    t_m\coloneq t+\frac{\mathrm{d}t}{2},
    \quad
    \bm{X}_m\coloneq\bm{X}_{t_m}.
    \label{seq:midpoint_notation}
\end{equation}

\noindent\textbf{$\bullet$ Time midpoint whitened increment.} Let $\matbf{B}_m\coloneq\matbf{B}(\bm{X}_m,t_m)$. We define
\begin{equation}
    \bm{\eta}_t\coloneq\frac{1}{\sqrt{\mathrm{d}t}}\matbf{B}_m^{-1}\mathrm{d}\bm{x}(t).
    \label{seq:def_Binv_dx}
\end{equation}

Here, we emphasize that the scaling $\matbf{B}_m^{-1}\mathrm{d}\bm{x}(t)\mapsto\bm{\eta}_t$ is deterministic for fixed $\mathrm{d}t$ and therefore preserves mutual information: $I(\matbf{B}_m^{-1}\mathrm{d}\bm{x}(t);\bm{X}_m)=I(\bm{\eta}_t;\bm{X}_m)$. In contrast, the whitening map $\mathrm{d}\bm{x}\mapsto\matbf{B}_m^{-1}\mathrm{d}\bm{x}$ depends on the random midpoint $\bm{X}_m$ and thus does not preserve $I(\mathrm{d}\bm{x};\bm{X}_m)$ in general. We therefore study $I(\matbf{B}_m^{-1}\mathrm{d}\bm{x};\bm{X}_m)$ as the natural diffusion-metric-normalized quantity, which reduces to $I(\mathrm{d}\bm{x};\bm{X}_m)$ in the additive-noise case (constant $\matbf{B}$).

\subsection{Assumptions ensuring smooth transition densities and well-defined entropy production rates}
\vspace*{-0.8\baselineskip}
We list explicit, checkable sufficient conditions on the coefficients and the law of $\bm{X}_t$ under which all objects defined in Section~I.A are well-posed (in particular the current $\bm{j}$, current velocity $\bm{v}=\bm{j}/p$, and the entropy production rate $\sigma_t$), and under which we may invoke standard results in the diffusion literature (time reversal and short-time density expansions)~\cite{anderson1982reverse,haussmann1986time,li2013maximum,watanabe1987analysis,yang2019anew,kloeden1992numerical,maruyama1955continuous,yoshida1992asymptotic,higa2013estimates,bally2015probabilistic}. Our goal is to establish the main identity rigorously within this broad and physically natural class, rather than to optimize the weakest possible hypotheses.

\medskip
\hypertarget{assumption:A1}{}
\noindent\textbf{(A1) Uniform ellipticity.} There exist $0<\lambda\leq\Lambda<\infty$ such that $\lambda I\leq\matbf{D}(\bm{x},t)\leq\Lambda I$ for all $(\bm{x},t)$.
% \begin{equation*}
%     \lambda I\leq\matbf{D}(\bm{x},t)\leq\Lambda I,
%     \quad
%     \forall(\bm{x},t).
% \end{equation*}

\bigskip
\hypertarget{assumption:A2}{}
\noindent\textbf{(A2) Regularity and smoothness of coefficients.}
For each $t$, $\bm{F}_I(\cdot,t)\in C^\infty(\mathbb{R}^d)$ and
$\matbf{B}(\cdot,t)\in C_b^\infty(\mathbb{R}^d)$ in $\bm{x}$, where $C_b^\infty(\mathbb{R}^d)$ denotes the class of smooth functions
whose spatial derivatives of all orders are globally bounded. Also for each $\bm x$, both $\bm F_I(\bm x,\cdot)$, $\matbf B(\bm x,\cdot)$ and their spatial derivatives are $C^1$ in time on $[t,t+\mathrm dt]$.
We denote by $\partial_{\bm{x}}^a (\cdot)$
and $\partial_{t}^b(\cdot)$ the corresponding spatial and time derivatives of order $a$ and $b$.

\bigskip
\hypertarget{assumption:A3}{}
\noindent\textbf{(A3) Well-posedness and smooth densities.}
The SDE~\eqref{seq:dynamics_def} admits 
a unique strong non-explosive solution. For each $t>0$, the law of $\bm{X}_t$ admits a strictly positive density $p(\cdot,t)$ such that $p(\cdot,t)$ and $\partial_t p(\cdot, t)$ are $C^\infty(\mathbb{R}^d)$ in $\bm x$. Consequently, the Fokker-Planck equation holds in the classical sense and the current defined in~\eqref{seq:def_j} is well-defined.

\bigskip
\hypertarget{assumption:A4}{}
\noindent\textbf{(A4) Polynomial growth bounds (drift, score and their derivatives).} There exist a constant $C>0$ and an integer $r\geq0$ such that, uniformly for $u\in[t,t+\mathrm{d}t]$,
\begin{equation*}
    \|\partial_{\bm{x}}^a \bm{F}_I(\bm{x},u)\|
    +\|\partial_{\bm x}^b \log p(\bm x,u)\|
    \leq
    C(1+\| \bm{x}\|^r),
    \quad
    a,b\in\mathbb N\cup\{0\}
\end{equation*}

\hypertarget{assumption:A5}{}
\noindent\textbf{(A5) Non-explosion and moment control.}
We assume that the process is non-explosive and that sufficiently high moments are finite on the time interval considered. Concretely, we impose either \textbf{(A5-A)} or \textbf{(A5-B)} below.

\medskip
\hypertarget{assumption:A5-A}{}
\noindent\textbullet\ \textbf{(A5-A) Bounded drift and current case.}
The drift $\bm F_I(\bm x,u)$ and current velocity $\bm v(\bm x,u)$ in~\eqref{seq:def_v} are uniformly bounded in $(\bm x,u)\in\mathbb R^d\times[t,t+\mathrm dt]$. Also all moments exist, i.e.
\begin{equation*}
    \sup_{u\in[t,t+\mathrm{d}t]}\mathbb{E}\!\left[\|\bm X_u\|^{m}\right]<\infty,
    \quad\forall m\in\mathbb N.
\end{equation*}

\hypertarget{assumption:A5-B}{}
\noindent\textbullet\ \textbf{(A5-B) Confined system case.}
There exist $\alpha,\theta>0$ such that 
\begin{equation*}
    \sup_{u\in[t,t+\mathrm{d}t]}\mathbb{E}\!
    \left[\exp\!\big(\theta\|\bm X_u\|^{1+\alpha}\big)\right]<\infty,
    \quad u\in[t,t+\mathrm dt]
\end{equation*}
%\langle \bm x,\bm F_I(\bm x,u)\rangle 
    %&\le -c\|\bm x\|^{1+\alpha}+C,
    %\quad\forall \bm x\in\mathbb R^d, u\in[t,t+\mathrm dt],
%\medskip

Assumptions~\hyperlink{assumption:A1}{A1}--\hyperlink{assumption:A5}{A5} should be viewed as an explicit statement of the regularity conditions that are usually left implicit when one says that the dynamics is a smooth overdamped Langevin diffusion. Physically,~\hyperlink{assumption:A1}{A1} is the standard non-degeneracy of the noise,~\hyperlink{assumption:A2}{A2}--\hyperlink{assumption:A3}{(A3)} give the usual smooth well-posed diffusion setting with smooth one-time densities,~\hyperlink{assumption:A4}{A4} restricts the drift/score sector to at most polynomial growth, and~\hyperlink{assumption:A5}{A5} supplies the corresponding non-explosion and moment control. The two alternatives in~\hyperlink{assumption:A5}{A5} cover the most common physical situations:~\hyperlink{assumption:A5-A}{A5-A} describes effectively bounded dynamics, including compact spaces such as motion on a ring, while~\hyperlink{assumption:A5-B}{A5-B} is tailored to unbounded domains with standard Lyapunov-type dissipative drift. Under these assumptions, the time-reversal and short-time expansions used below are standard, and the entropy production rate in~\eqref{seq:ep_rate} is finite.
For later reference,~\hyperlink{assumption:A1}{A1}--\hyperlink{assumption:A3}{A3} justify the time-reversal formula and local short-time expansions,~\hyperlink{assumption:A4}{A4} controls the polynomial-growth coefficients appearing in the remainder bounds, and~\hyperlink{assumption:A5}{A5} is used only for moment/tail control and the localization arguments.

These assumptions are not intended to be minimal. They are a convenient sufficient package without repeated technical bookkeeping. For any fixed truncation order, corresponding finite order hypotheses would suffice. Non-smooth or geometric variants (e.g., periodic identifications, reflecting boundaries, or piecewise-defined drifts) can often be treated by standard localization/truncation and smoothing arguments, but we do not pursue such extensions here.

\subsection{Time reversal diffusion}
\vspace*{-0.5\baselineskip}
We briefly recall the reverse-time diffusion needed for the midpoint construction. Under assumptions~\hyperlink{assumption:A1}{A1}-–\hyperlink{assumption:A3}{A3}, the time reversal of an It\^{o} diffusion is again a diffusion with the same diffusion matrix and an explicit reversed drift~\cite{anderson1982reverse,haussmann1986time}. Specializing that standard formula at the midpoint yields~\eqref{seq:get_v}, the identity used below.

A key ingredient is the drift of the time-reversed diffusion. Consider an It\^{o} diffusion on a finite horizon $\left[0, T\right]$
\begin{equation*}
    \mathrm{d}\bm{X}_t=\bm{b}(\bm{X}_t,t)\mathrm{d}t+\matbf{B}(\bm{X}_t,t)\mathrm{d}W_t\,,
\end{equation*}
which satisfies the assumptions \hyperlink{assumption:A1}{A1}--\hyperlink{assumption:A3}{A3}.
Under these assumptions, let $p(\bm{x}, t)$ denote the one-time density of $\bm{X}_t$. 
We then define the time-reversed process by $\bar{\bm{X}}_t \coloneq \bm{X}_{T-t}$ for $t \in [0, T]$. 
With this definition, $\bar{\bm{X}}_t$ is again a diffusion process with the same diffusion matrix $\matbf{a}\coloneq\matbf{B}\matbf{B}^\intercal=2\matbf{D}$ and reversed drift $\bar{\bm{b}}$ given in component form by

\begin{equation}
    \bar{b}_i(\bm{x}, t)=
    -b_i(\bm{x},T-t)+
    \frac{1}{p(\bm{x},T-t)}
    \sum_{j=1}^d
    \partial_{x_j}(\mathsf{a}_{ij}\left(\bm{x},T-t)\,p(\bm{x},T-t)\right).
    \label{seq:def_reversed_drift}
\end{equation}
Equivalently, in vector notation,
\begin{equation*}
    \bar{\bm{b}}(\bm{x}, t)=
    -\bm{b}(\bm{x},T-t)+
    \frac{
    \nabla\cdot\left(
    \matbf{a}(\bm{x},T-t)\,p(\bm{x},T-t)
    \right)
    }{p(\bm{x},T-t)},
    \label{seq:def_reversed_drift_vec}
\end{equation*}
where $(\nabla\cdot(\matbf{a}p))_i\coloneq\sum_j\partial_{x_j}(\mathsf{a}_{ij}p)$~\cite{haussmann1986time}.

Specializing this general formula to the local reversal around the midpoint time $t_m$ (i.e. set $T=t_m$ and evaluate at reversed time $t=0$), and writing the reverse-time drift at time $t_m$ as $\bm{F}_{rev}(\bm{x},t_m)\coloneq\bar{\bm{b}}(\bm{x},0)$, we obtain the local identity

\begin{equation}
    \bm{F}_{rev}(\bm{x},t_m)=-
    \bm{F}_I(\bm{x},t_m)+
    \frac{\nabla\cdot(\matbf{a}(\bm{x},t_m)\,p(\bm{x},t_m))}{p(\bm{x},t_m)}.
    \label{seq:def_F_rev}
\end{equation}
Combining Eq.~\eqref{seq:def_j},~\eqref{seq:def_v} with~\eqref{seq:def_F_rev} at time $t_m$ yields the exact algebraic relation,
\begin{equation}
    \frac{\bm{F}_I(\bm{x},t_m)-\bm{F}_{rev}(\bm{x},t_m)}{2}=
    \bm{F}_I(\bm{x},t_m)-
    \frac{\nabla\cdot(\matbf{D}\,p)}{p}(\bm{x},t_m)=
    \frac{\bm{j}}{p}(\bm{x},t_m)=\bm{v}(\bm{x},t_m).
    \label{seq:get_v}
\end{equation}

Equation~\eqref{seq:get_v} is the algebraic reason midpoint conditioning is special since it turns the difference between forward and reverse half-step drifts into the current velocity.

\subsection{Time midpoint half-step decomposition}
\vspace*{-0.8\baselineskip}
We now focus on the time-midpoint half-step decomposition, which is the key ingredient of the proof. 
We first introduce the notation and rewrite the normalized time-midpoint whitened increment $\bm\eta_t$ defined in Eq.~(S7) from the perspective of time-midpoint conditioning. 
Hereafter, we use the term ``midpoint'' to refer to the time midpoint. 
For convenience, we define
\[
h := \mathrm{d}t/2, \quad \bm\Delta^+ := \bm{X}_{t_m+h} - \bm{X}_{t_m}, \quad \bm\Delta^- := \bm{X}_{t_m} - \bm{X}_{t_m-h}.
\]
Then, by the definition, the total increment decomposes as $\mathrm{d}\bm{x} = \bm\Delta^+ + \bm\Delta^-$.  
Note that the triplet $(\bm{X}_t, \bm{X}_m, \bm{X}_{t+dt})$ can equivalently be represented as $(\bm{X}_{t_m-h}, \bm{X}_{t_m}, \bm{X}_{t_m+h})$. 
This representation leads to the following key property:

\hypertarget{lem:S1}{}
\textbf{Lemma S1 (conditional independence).} By the Markov property, conditional on $\bm{X}_m\coloneq\bm{X}_{t_m}=\bm{x}$, the past endpoint $\bm{X}_{t_m-h}$ and the future endpoint $\bm{X}_{t_m+h}$ are independent. Consequently, any functionals of the $\bm\Delta^+$ and $\bm\Delta^-$ are conditionally independent given $\bm{X}_m$.

\textit{Proof.} For the joint distribution $p(\bm{X}_{t_m-h}=\bm{x}^-,\bm{X}_m=\bm{x},\bm{X}_{t_m+h}=\bm{x}^+)$, we can write
\begin{equation*}
    p(\bm{X}_{t_m-h}=\bm{x}^-,\bm{X}_m=\bm{x},\bm{X}_{t_m+h}=\bm{x}^+)=p(\bm{X}_{t_m-h}=\bm{x}^-,\bm{X}_{t_m+h}=\bm{x}^+|\bm{X}_m=\bm{x})p(\bm{X}_m=\bm{x}).
\end{equation*}

We also have
%\vspace*{-0.8\baselineskip}
\begin{align*}
    p(\bm{X}_{t_m-h}=\bm{x}^-,\bm{X}_m=\bm{x},\bm{X}_{t_m+h}=\bm{x}^+)
    &=p(\bm{X}_{t_m+h}=\bm{x}^+|\bm{X}_m=\bm{x})p(\bm{X}_{m}=\bm{x},\bm{X}_{t_m-h}=\bm{x}^-)
    \\
    &=p(\bm{X}_{t_m+h}=\bm{x}^+|\bm{X}_m=\bm{x})p(\bm{X}_{t_m-h}=\bm{x}^-|\bm{X}_m=\bm{x})p(\bm{X}_{m}=\bm{x}),
\end{align*}
where the first equality holds because of the Markov property. Hence we get 
\begin{equation*}
    p(\bm{X}_{t_m-h}=\bm{x}^-,\bm{X}_{t_m+h}=\bm{x}^+|\bm{X}_m=\bm{x})=p(\bm{X}_{t_m+h}=\bm{x_+}|\bm{X}_m=\bm{x})p(\bm{X}_{t_m-h}=\bm{x}^-|\bm{X}_m=\bm{x}),
\end{equation*}
which means $\bm{X}_{t_m+h}\perp\!\!\!\perp\bm{X}_{t_m-h}\,|\bm{X}_m$. $\square$

Throughout this subsection, we condition on $\bm{X}_m \coloneq\bm{X}_{t_m}= \bm{x}$.

\bigskip
\noindent\textbf{$\bullet$ Conditional mean and covariance}

\hyperlink{lem:S1}{Lemma S1} makes the conditional mean and covariance of $\mathrm{d}\bm{x}$ easy to compute because $\bm\Delta^+$ and $\bm\Delta^-$ are conditionally independent given $\bm{X}_m=\bm{x}$.
For the forward half-step, by using standard properties of It\^{o} diffusions, we obtain
\begin{equation}
    \mathbb{E}[\bm\Delta^+|\bm{X}_m=\bm{x}]=
    \bm{F}_I(\bm{x},t_m)\,h+o(h).
    \label{seq:E_Delta^+}
\end{equation}

For the backward half-step, define the reversed process $\bar{\bm{X}}_s\coloneq\bm{X}_{t_m-s}$. Then $\bar{\bm{X}}$ is a diffusion whose drift at $s=0$ equals $\bm{F}_{rev}(\bm{x},t_m)$ (Section C). Hence
\begin{equation*}
    \mathbb{E}\left[
    \bm{X}_{t_m-h}-\bm{X}_{t_m}\,|\,\bm{X}_m=\bm{x}
    \right]=
    \mathbb{E}\left[
    \bar{\bm{X}}_h-\bar{\bm{X}}_0\,|\,\bar{\bm{X}}_0=\bm{x}
    \right]=\bm{F}_{rev}(\bm{x},t_m)\,h+o(h),
\end{equation*}
equivalently 
\begin{equation}
    \mathbb{E}\left[\bm\Delta^-|\bm{X}_m=\bm{x}\right]=-\bm{F}_{rev}(\bm{x},t_m)\,h+o(h).
    \label{seq:E_Delta^-}
\end{equation} 
Adding~\eqref{seq:E_Delta^+} and~\eqref{seq:E_Delta^-} with the relation Eq.~\eqref{seq:get_v} yields
\begin{equation}
    \mathbb{E}[\mathrm{d}\bm{x}|\bm{X}_m=\bm{x}]=\bm{v}(\bm{x},t_m)\,\mathrm{d}t+o(\mathrm{d}t).
    \label{seq:cal_dx_xm}
\end{equation}

For the covariance, conditional independence (\hyperlink{lem:S1}{Lemma S1}) gives
\begin{equation}
    \mathrm{Cov}(\mathrm{d}\bm{x}\,|\,\bm{X}_m=\bm{x})=
    \mathrm{Cov}(\bm\Delta^+\,|\bm{X}_m=\bm{x})
    +\mathrm{Cov}(\bm\Delta^-\,|\bm{X}_m=\bm{x}).
    \label{seq:cov_sum}
\end{equation}
By the standard second-moment expansions for Itô diffusions (or Euler-Maruyama with frozen coefficients~\cite{maruyama1955continuous,kloeden1992numerical}),
\begin{equation}
    \mathrm{Cov}(\bm\Delta^+\,|\bm{X}_m=\bm{x})
    =\matbf{a}(\bm{x},t_m)\,h+o(h).
    \label{seq:cov_Delta^+}
\end{equation}
The reversed diffusion has the same diffusion matrix $\matbf{a}(\cdot,t_m)$ at $s=0$, hence likewise
\begin{equation}
    \mathrm{Cov}(\bm\Delta^-\,|\bm{X}_m=\bm{x})
    =\matbf{a}(\bm{x},t_m)\,h+o(h).
    \label{seq:cov_Delta^-}
\end{equation}
Combining~\eqref{seq:cov_sum}-\eqref{seq:cov_Delta^-} gives
\begin{equation}
    \mathrm{Cov}(\mathrm{d}\bm{x}(t)|\bm{X}_m=\bm{x})=
    \matbf{a}(\bm{x},t_m)\,\mathrm{d}t+o(\mathrm{d}t).
    \label{seq:cal_cov_dx_xm}
\end{equation}

% We emphasize that only conditioning on the midpoint $\bm{X}_m$ can yield the current velocity $\bm{v}$ as the conditional mean since~\eqref{seq:get_v}. 
% Other choices, for example conditioning on the endpoints $\bm{X}_t$ and $\bm{X}_{t+\mathrm{d}t}$, 
% give $\bm{F}_I$ and $-\bm{F}_{rev}$ as conditional means respectively.
% Because the total entropy production rate $\sigma_t$ is expressed as the average of a quadratic form in $\bm{v}$
% (see~\eqref{seq:ep_rate}), conditioning on $\bm{X}_m$ is crucial for establishing a connection with $\sigma_t$.

At this stage the midpoint-conditioned increment already has the correct mean and covariance. The only remaining question is whether multiplicative-noise corrections beyond Gaussian order can affect $I(\bm\eta_t;\bm X_m)$ at order $\mathrm dt$.

%\newpage
\bigskip
\noindent\textbf{$\bullet$ Strong It\^{o}-Taylor expansion and normalized sum}

For short-time approximations of an It\^{o} SDE up to order $O(\mathrm{d}t)$, the Euler–Maruyama (EM) scheme is well known.
However, in a strict sense, there exist additional effects of order $O(\mathrm{d}t)$ that influence the distribution beyond the mean drift in multiplicative-noise case.
These include deformations of the local shape of the distribution and, in the presence of multidimensional noise, 
orientation or circulation effects at the pathwise level.
Although these contributions have zero mean and therefore do not appear as drift terms, 
their impact on higher-order statistics beyond the mean and leading-order covariance is of order $O(\mathrm{d}t)$~\cite{kloeden1992numerical,milstein1975strong}.
The mutual information, however, can be sensitive to these effects.
Therefore, to evaluate the mutual information accurately up to order $O(\mathrm{d}t)$, 
it is necessary to go beyond the EM scheme and consider the strong It\^{o}-Taylor expansion, which incorporates all such 
$O(\mathrm{d}t)$ contributions.

We use this framework at the midpoint, where both half-steps are expanded around the same space-time point $(\bm{x},t_m)$ and thus share the same $\matbf{B}(\bm{x},t_m)$ and its spatial derivatives (see subsection C). Because of sufficient smoothness and regularity conditions \hyperlink{assumption:A1}{A1}-\hyperlink{assumption:A4}{A4}, the It\^{o}-Taylor expansion to strong order one is valid locally. Then we can write the normalized and whitened half-steps as
\begin{align}
    \matbf{B}(\bm{x},t_m)^{-1}\frac{\bm\Delta^+}{\sqrt{h}}
        &=\bm{Z}_++\sqrt{h}\,\matbf{B}_m^{-1}\bm{F}_I(\bm{x},t_m)
        +\sqrt{h}\mathcal{T}_1(\bm{x}):\left(\bm{Z}_+\bm{Z}_+^\intercal-\mathbf{I}_d\right)
        +\sqrt{h}\mathcal{T}_2(\bm{x}):\matbf{A}_+
        +O_{L^2}(h),
        \label{seq:Delta^+_ito_taylor}
        \\
        \matbf{B}(\bm{x},t_m)^{-1}\frac{\bm\Delta^-}{\sqrt{h}}
        &=-\bm{Z}_--\sqrt{h}\,\matbf{B}_m^{-1}\bm{F}_{rev}(\bm{x},t_m)
        -\sqrt{h}\mathcal{T}_1(\bm{x}):\left(\bm{Z}_-\bm{Z}_-^\intercal-\mathbf{I}_d\right)
        -\sqrt{h}\mathcal{T}_2(\bm{x}):\matbf{A}_-
        +O_{L^2}(h),
    \label{seq:Delta^-_ito_taylor}
\end{align}
where the symbol ``$:$'' denotes tensor contraction, i.e.
$(\mathcal{T}:\matbf{M})_i\coloneq\sum_{j,k=1}^d\mathcal{T}_{ijk}\matbf{M}_{jk}$.
Here, $\mathcal{T}_1(\bm{x})$ and $\mathcal{T}_2(\bm{x})$ are deterministic tensors depending on
$\mathbf{B}(\bm{x},t_m)$ and its spatial derivatives evaluated at $(\bm{x},t_m)$ (hence they vanish in the additive-noise case).
On the other hand, $\bm{Z}_+,\bm{Z}_-\sim\mathcal{N}(0,\mathbf{I}_d)$ are normalized Gaussian random variables,
and the matrices $\matbf{A}_{\pm}$ represent the normalized  L\'{e}vy-area contribution, defined by
\begin{equation}
    (\matbf{A}_{\pm})_{ij}\coloneq
    \frac{1}{2h}\left(\matbf{J}^{\pm}_{ij}-\matbf{J}^{\pm}_{ji}\right),
    \quad
    \matbf{J}^{\pm}_{ij}\coloneq
    \int_0^h \left(\int_0^s \mathrm{d}W^{\pm,(i)}_u\right)\, \mathrm{d}W^{\pm,(j)}_s,
    \label{seq:def_levy_area_matrix}
\end{equation}
where $\bm{W}^\pm$ denote $d$-dimensional standard Wiener processes used in the auxiliary representation below,
and $d\bm{W}^\pm$ denotes their It\^o increments. Note that $\|\matbf{A}^\pm\|=O_{L^2}(1)$ by definition.
Roughly, while the first two terms on the RHS of~\eqref{seq:Delta^+_ito_taylor}--\eqref{seq:Delta^-_ito_taylor} determine the mean and leading order of covariance of the distribution (and this is exactly what the EM scheme captures), the third term $\mathcal{T}_1:(\bm{Z}\bm{Z}^\intercal-\mathrm{I}_d)$ modulates the ``shape'' of the distribution and the fourth term $\mathcal{T}_2:\matbf{A}$ effectively ``twists'' the individual paths.

We emphasize two key points: First, by \hyperlink{lem:S1}{Lemma S1}, $\bm{\Delta}^+$ and $\bm{\Delta}^-$ are independent conditional on $\bm{X}_m=\bm{x}$.
Therefore, we can represent the two half-step expansions using conditionally independent auxiliary random variables $(\bm{Z}_+,\matbf{A}_+)$ and $(\bm{Z}_-,\matbf{A}_-)$, which are conditionally independent given $\bm{X}_m=\bm{x}$, such that~\eqref{seq:Delta^+_ito_taylor}--\eqref{seq:Delta^-_ito_taylor} hold.
Second, the signs in~\eqref{seq:Delta^-_ito_taylor} flip because $\bm\Delta^-=\bm{x}-\bm{X}_{t_m-h}$ is the negative of a forward increment of the reversed diffusion, while the diffusion matrix (hence the coefficients of the third and fourth terms) remain unchanged.

Now, one can rewrite the normalized whitened-increment $\bm\eta_t$ in \eqref{seq:def_Binv_dx} as a normalized sum at the midpoint,
\begin{equation}
    \bm\eta_t\coloneq
    \frac{\matbf{B}_m^{-1}\mathrm{d}\bm{x}}{\sqrt{\mathrm{d}t}}
    =\frac{\matbf{B}_m^{-1}(\bm\Delta^++\bm\Delta^-)}{\sqrt{2h}}
    =\frac{1}{\sqrt{2}}\left(\matbf{B}_m^{-1}\frac{\bm\Delta^+}{\sqrt{h}}
    +\matbf{B}_m^{-1}\frac{\bm\Delta^-}{\sqrt{h}}\right),
    \label{seq:def_midpoint_normalized_sum}
\end{equation}
and adding~\eqref{seq:Delta^+_ito_taylor}-\eqref{seq:Delta^-_ito_taylor} yields
\begin{equation}
    \bm\eta_t
    =\frac{\bm{Z}_+-\bm{Z}_-}{\sqrt{2}}
    +\sqrt{\mathrm{d}t}\,\matbf{B}_m^{-1}\bm{v}(\bm{x},t_m)
    +\sqrt{\mathrm{d}t}\,\mathcal{T}_1(\bm{x})\!:\!
    \frac{\bm{Z}_+\bm{Z}_+^\intercal-\bm{Z}_-\bm{Z}_-^\intercal}{2}
    +\sqrt{\mathrm{d}t}\,\mathcal{T}_2(\bm{x})\!:\!\frac{
    \matbf{A}_+-\matbf{A}_-}{2}
    +O_{L^2}(\mathrm{d}t).
    \label{seq:eta_ito_taylor}
\end{equation}
Here, the drift term $\matbf{B}_m^{-1}\bm{v}(\bm{x},t_m)$ comes from~\eqref{seq:get_v}. Note that \eqref{seq:eta_ito_taylor} is a strict extension of Eq.~(3) in the main text for the multiplicative-noise case.
This equation isolates the entire multiplicative-noise difficulty in two potentially dangerous $O(\sqrt{\mathrm dt})$ terms. We will show that such terms are harmless at the order relevant for mutual information in the next section.

The $O_{L^2}$ notation in~\eqref{seq:Delta^+_ito_taylor}--\eqref{seq:Delta^-_ito_taylor},~\eqref{seq:eta_ito_taylor} means there exist $C>0$, $r\in\mathbb N$ for all $\mathrm dt<\mathrm dt_0$ (or $h<h_0$) such that
\begin{equation}
    \|\bm R_{\bm x}\|_{L^2}\leq
    C\big(1+\|\bm x\|^r\big)\,\mathrm dt
    \qquad
    \forall\bm x\in \mathbb R^d,
\end{equation}
and this is ensured by assumption \hyperlink{assumption:A1}{A1}--\hyperlink{assumption:A4}{A4}. Especially, the polynomial growth bounds assumption \hyperlink{assumption:A4}{A4} gives the polynomial term $(1+\|\bm x\|^r)$.

\subsection{Cancellation of the $\sqrt{\mathrm{d}t}$-order correction term in small-time density expansions}
\vspace*{-0.6\baselineskip}
Small-time density expansions of multivariate diffusions are naturally organized in half-integer powers of $\mathrm dt$~\cite{watanabe1987analysis,yoshida1992asymptotic,li2013maximum,yang2019anew}. For the present problem, this means that an $O(\sqrt{\mathrm dt})$ correction in the conditional distribution of $\bm\eta_t$ could in principle contribute to mutual information at order $O(\mathrm dt)$. The purpose of this section is to show that midpoint conditioning removes exactly this contribution. Using~\eqref{seq:eta_ito_taylor}, we therefore isolate the two potentially dangerous $O(\sqrt{\mathrm dt})$ terms and prove that both are conditionally centered.

% In general multivariate diffusions, 
% in the small-time limit $\mathrm{d}t \to 0$ with the normalized increment
% $\mathrm{d}\bm{X}_t/\sqrt{\mathrm{d}t}$ held fixed,
% small-time density expansions are naturally organized in half-integer powers
% of the time step (powers of $\sqrt{\mathrm{d}t}$, where $\mathrm{d}t$ is the time step).
% This is not just a heuristic: Li (2013) explicitly constructs a density expansion whose terms scale like $\mathrm{d}t^{k/2}$ and proves uniform convergence on compacts using Watanabe/Yoshida theory~\cite{li2013maximum,watanabe1987analysis,yoshida1992asymptotic}. Independently, Yang-Chen-Wan (2019) derive a delta expansion for multivariate diffusions via It\^{o}-Taylor, obtaining Hermite-polynomial-based terms with explicit half-power indexing, and prove convergence with remainder orders showing $(t'-t)^{k/2}$-type behavior~\cite{yang2019anew}. Therefore, a priori, each half-step density can contain $O(\sqrt{\mathrm{d}t})$ corrections beyond the normalized Gaussian, as discussed in the previous subsection. Those corrections can make non-zero $O(\mathrm{d}t)$ contribution in mutual information (MI) terms, so we should check that its effect really disappears for our setting.

Now, for convenience, we denote the dangerous terms of~\eqref{seq:eta_ito_taylor} as follows:
\begin{equation}
    \bm{G}_{\bm{x}}(\bm{Z}_+,\bm{Z}_-)\coloneq
    \mathcal{T}_1(\bm{x})\!:\!
    \frac{\bm{Z}_+\bm{Z}_+^\intercal-\bm{Z}_-\bm{Z}_-^\intercal}{2},
    \quad
    \bm{H}_{\bm{x}}(\matbf{A}_+,\matbf{A}_-)\coloneq
    \mathcal{T}_2(\bm{x})\!:\!\frac{
    \matbf{A}_+-\matbf{A}_-}{2}.
    \label{seq:def_G_and_H}
\end{equation}
We now analyze these $O(\sqrt{\mathrm{d}t})$ non-Gaussian terms $\bm{G}_{\bm{x}}$ and $\bm{H}_{\bm{x}}$ in the normalized and whitened increment. Although these terms are present, we show that their conditional expectations vanish under midpoint conditioning, implying that the $O(\sqrt{\mathrm{d}t})$ coefficient in the characteristic function is exactly zero. Consequently, the short-time density expansion contains no non-Gaussian corrections at $O(\sqrt{\mathrm{d}t})$.

\bigskip
\noindent\textbf{\textbullet\ Step 1. Conditional Expectation of} $\bm{G}_{\bm{x}}$

Under conditioning $\bm X_m=\bm x$, we first reparameterize the independent and normalized Gaussian variables $\bm{Z}_+,\bm{Z}_-$ in~\eqref{seq:eta_ito_taylor} using
\begin{equation}
    \bm{Z}\coloneq\frac{\bm{Z}_+-\bm{Z}_-}{\sqrt{2}},
    \quad
    \bm{U}\coloneq\frac{\bm{Z}_++\bm{Z}_-}{\sqrt{2}},
    \label{seq:def_ZU}
\end{equation}
so $\bm{Z}, \bm{U}$ are independent $\mathcal{N}(0,\mathbf{I}_d)$. A direct algebra gives
\begin{equation*}
    \bm{Z}_+\bm{Z}_+^\intercal-\bm{Z}_-\bm{Z}_-^\intercal
    =\bm{Z}\bm{U}^\intercal+\bm{U}\bm{Z}^\intercal.
\end{equation*}
Hence the $O(\sqrt{\mathrm{d}t})$ non-Gaussian term in~\eqref{seq:eta_ito_taylor} becomes a bilinear form in $(\bm{Z},\bm{U})$:
\begin{equation*}
    \sqrt{\mathrm{d}t}\,\mathcal{T}_1(\bm{x}):
    \frac{1}{\sqrt 2}\frac{\bm{Z}_+\bm{Z}_+^\intercal-\bm{Z}_-\bm{Z}_-^\intercal}{\sqrt{2}}
    =\sqrt{\mathrm{d}t}\,\bm{G}_{\bm{x}}(\bm{Z},\bm{U}),
\end{equation*}
with $\bm{G}_{\bm{x}}(\bm{Z},\bm{U})$ linear in $\bm{U}$ for fixed $\bm{Z}$. In particular, because $\bm{U}\perp\!\!\!\perp\bm{Z}$ and $\mathbb{E}[\bm{U}]=0$, we have
\begin{equation}
    \mathbb{E}[\bm{G}_{\bm{x}}(\bm{Z},\bm{U})\,|\bm{X}_m=\bm{x},\bm{Z}\,]
    =\bm 0.
    \label{seq:G_cond_mean_zero}
\end{equation}

This cancellation relies crucially on midpoint conditioning. Under other conditionings, such as conditioning on $\bm{X}_t=\bm{x}$, the corresponding $O(\sqrt{\mathrm{d}t})$ term involves the quadratic structure $\bm{Z}_+\bm{Z}_+^\intercal-\mathbf{I}_d$, which cannot be rewritten into a bilinear form in two independent Gaussians, so the above conditional-centering argument does not apply.

\bigskip
\noindent\textbf{\textbullet\ Step 2. Conditional Expectation of} $\bm{H}_{\bm{x}}$

We work under the auxiliary representation introduced in Section D: conditional on $\bm{X}_m=\bm x$, the two half-steps can be realized using independent Brownian motions, i.e. the Wiener process $\bm{W}^\pm$. Under conditioning $\bm X_m=\bm x$, fix one half-step and write $\Delta\bm W^\pm:=\bm W^\pm_h$ and $\bm Z_\pm:=\Delta\bm W^\pm/\sqrt h$. Now consider the time-reversal map on paths on $[0,h]$,
\[
\widetilde{\bm W}^\pm_s := \Delta\bm W^\pm - \bm W^\pm_{h-s},\qquad s\in[0,h].
\]
Then $\widetilde{\bm W}^\pm_0=0$ and $\widetilde{\bm W}^\pm_h=\Delta\bm W^\pm$.
Moreover, conditional on the endpoint $\Delta\bm W^\pm$, the Brownian bridge law is invariant under this map:
$\bm W^\pm \mid \Delta\bm W^\pm \stackrel{d}{=} \widetilde{\bm W}^\pm\mid \Delta\bm W^\pm$.
(For completeness, this follows since the Brownian bridge is a Gaussian process and the map preserves its conditional mean and covariance.)
Hence for any integrable functional $f$,
\begin{equation}
\mathbb E\!\left[f(\bm W^\pm)\mid \Delta\bm W^\pm\right]=\mathbb E\!\left[f(\widetilde{\bm W}^\pm)\mid \Delta\bm W^\pm\right].
\label{seq:bridge_invariance}
\end{equation}

Also, under time-reversal map and by definition of matrix $\matbf{A}$~\eqref{seq:def_levy_area_matrix}, we get the sign-flip property because
\begin{equation}
    \begin{aligned}
        \left(\matbf{A}_\pm\left(\widetilde{\bm{W}}^\pm\right)\right)_{ij}
    &=\frac{1}{2h}\int_0^h\left(
    \int_0^s \mathrm{d}\widetilde{W}^{\pm,(i)}_u
    \right)\mathrm{d}\widetilde{W}^{\pm,(j)}_s
    -(i\leftrightarrow j)
    \\
    &=\frac{1}{2h}\int_0^h\left(
    W_h^{\pm,(i)}-W_{h-s}^{\pm,(i)}
    \right)\left(
    -\mathrm{d}W_{h-s}^{\pm,(j)}\right)
    -(i\leftrightarrow j)
    \\
    &=\frac{1}{2h}\int_0^h\left(
    W_h^{\pm,(i)}-W_\tau^{\pm,(i)}
    \right)
    \mathrm{d}W_\tau^{\pm,(j)}
    -(i\leftrightarrow j)
    \\
    &=\frac{1}{2h}\left(W_h^{\pm,(i)}W_h^{\pm,(j)}-\int_0^h
    W_\tau^{\pm,(i)}\mathrm{d}W_\tau^{\pm,(j)}\right)
    -(i\leftrightarrow j)
    \\
    &=\frac{1}{2h}\left(-
    \int_0^h\left(\int_0^\tau
    \mathrm{d}W_u^{\pm,(i)}
    \right)\mathrm{d}W_\tau^{\pm,(j)}
    \right)-(i\leftrightarrow j)
    \\
    &=-\left(\matbf{A}_\pm\left(\bm{W}^\pm\right)\right)_{ij} \,\,\,\forall i,j.
    \end{aligned}
    \label{seq:Levy_matrix_sign_flip}
\end{equation}
In the second equality, we use the definition of the time-reversed Wiener
process \(\widetilde W_s = W_h - W_{h-s}\), which formally yields
\(\mathrm{d}\widetilde W_s = -\,\mathrm{d}W_{h-s}\).
This computation is understood in a formal sense and can be rigorously
justified by approximating the It\^{o} integrals by Riemann sums and applying the change of variables.
In the third equality, we use the change of variables
\(s \mapsto \tau \coloneq h-s\). The $W_h^{\pm,(i)}W_h^{\pm,(j)}$ term cancels out in the fourth line, since $W_h^{\pm,(i)}W_h^{\pm,(j)}-(i\leftrightarrow j)=0$.

Combining \eqref{seq:bridge_invariance} with \eqref{seq:Levy_matrix_sign_flip} yields
\[
\mathbb E[\matbf A_\pm(\bm W^\pm)\mid \Delta\bm W^\pm]
=\mathbb E[\matbf A_\pm(\widetilde{\bm W}^\pm)\mid \Delta\bm W^\pm]
=\mathbb E[-\matbf A_\pm(\bm W^\pm)\mid \Delta\bm W^\pm]
=-\mathbb E[\matbf A_\pm(\bm W^\pm)\mid \Delta\bm W^\pm],
\]
hence $\mathbb E[\matbf A_\pm\mid \Delta\bm W^\pm]=\bm 0$, equivalently $\mathbb E[\matbf A_\pm\mid \bm Z_\pm]=\bm 0.$ 
Therefore,
\begin{equation*}
    \mathbb{E}\!\left[
    \bm{H}_{\bm{x}}\,\middle|\,\bm{Z}_+,\bm{Z}_-\right]
    = \mathcal{T}_2(\bm x):
    \frac{\mathbb E[\matbf A_+\mid \bm Z_+]-\mathbb E[\matbf A_-\mid \bm Z_-]}{2}=\bm 0,
\end{equation*}
and since $\mathcal T_2(\bm x)$ is deterministic conditional on $\bm X_m=\bm x$, the same conclusion holds under midpoint conditioning:
\begin{equation}
    \mathbb E\!\left[\bm H_{\bm x}\,\middle|\,\bm X_m=\bm x,\,\bm Z_+,\,\bm Z_-\right]=\bm 0.
    \label{seq:H_cond_mean_zero}
\end{equation}

Unlike $\bm{G}_{\bm x}$, the centering of the L\'evy-area term $\bm{H}_{\bm x}$ does not rely on midpoint conditioning: it follows from the Brownian-bridge property that the L\'evy area has zero conditional mean given the endpoint increment, i.e. $\mathbb{E}[\matbf{A}\,\vert\Delta\bm{W}]=0$. Consequently, vanishing $\bm{H}_{\bm{x}}$ holds regardless of the particular time-point conditioning; it only relies on the endpoint-centered Brownian-bridge property.
Intuitively, $\bm{H}_{\bm{x}}$ influences how individual trajectories are ``twisted'' along the path; however, once the endpoints are fixed, these twisting effects average out and do not contribute to one-time marginal distributions.

\bigskip
\noindent\textbf{\textbullet\ Step 3. Cancellation of the $\sqrt{\mathrm{d}t}$-order non-Gaussian corrections in characteristic function} 

Now we show that the coefficient of the $O(\sqrt{\mathrm{d}t})$ term in the non-Gaussian correction to the conditional characteristic function, 
\begin{equation*}
    \varphi_{\bm\eta_t|\bm{x}}(\bm{u})\coloneq\mathbb{E}[e^{i\langle\bm{u},\bm\eta_t\rangle}|\bm{X}_m=\bm{x}].
\end{equation*}
Applying the tower property under conditioning on $\bm{Z}$ in~\eqref{seq:def_ZU}, we have
\begin{equation}
    \varphi_{\bm\eta_t|\bm{x}}(\bm{u})
    =\mathbb{E}[e^{i\langle\bm{u},\bm\eta_t\rangle}|\bm{X}_m=\bm{x}]
    =\mathbb{E}\big[
    \mathbb{E}[e^{i\langle\bm{u},\bm\eta_t\rangle}|\bm{X}_m=\bm{x},\bm{Z}\,]
    \mid\bm{X}_m=\bm{x}\big].
    \label{seq:def_phi_cond_Z}
\end{equation}
Also we can write $\bm\eta_t$ by using~\eqref{seq:eta_ito_taylor}--\eqref{seq:def_ZU},
\begin{equation*}
    \bm\eta_t=\left(\bm{Z}+\sqrt{\mathrm{d}t}\,\bm{s}(\bm{x},t_m)\right)
    +\sqrt{\mathrm{d}t}\,\bm{G}_{\bm{x}}(\bm{Z},\bm{U})
    +\sqrt{\mathrm{d}t}\,\bm{H}_{\bm{x}}(\matbf{A}_+,\matbf{A}_-)
    +\bm{R}_{\bm{x}}
    ,
    \quad
    \|\bm{R}_{\bm{x}}\|_{L^2}=O(\mathrm{d}t),
\end{equation*}
for $\bm x$ in compact where $\bm{s}(\bm x,t_m)$ denotes $\matbf{B}_m^{-1}\bm v(\bm x, t_m)$.
Expanding~\eqref{seq:def_phi_cond_Z} for non-Gaussian term to the second order in $\sqrt{\mathrm{d}t}$, we get the first-order correction terms
\begin{equation}
    \begin{aligned}
    &\sqrt{\mathrm{d}t}\,i\,\mathbb{E}\left[
    e^{i\langle \bm{u},\bm{Z}+\sqrt{\mathrm{d}t}\,\bm{s} \rangle}\,
    \mathbb{E}\big[
    \langle\bm{u},\bm{G}_{\bm{x}}(\bm{Z},\bm{U})
    +\bm{H}_{\bm{x}}(\matbf{A}_+,\matbf{A}_-)
    \rangle
    \vert\bm{X}_m=\bm{x},\bm{Z}\,\big]\middle|\bm{X}_m=\bm{x}
    \right]
    \\
    &=\sqrt{\mathrm{d}t}\,i\mathbb{E}\left[e^{i\langle \bm{u},\bm{Z}+\sqrt{\mathrm{d}t}\,\bm{s}\rangle}
    \left(\langle \bm{u},\mathbb{E}[\bm{G}_{\bm{x}}(\bm{Z},\bm{U})\,|\,\bm{X}_m=\bm{x},\bm{Z}\,]\rangle
    +\langle \bm{u},\mathbb{E}[\bm{H}_{\bm{x}}(\matbf{A}_+,\matbf{A}_-)\,|\,\bm{X}_m=\bm{x},\bm{Z}\,]\rangle
    \right)\,\middle|\bm{X}_m=\bm{x}\right].
    \end{aligned}
    \label{seq:phi_first_order}
\end{equation}
Applying~\eqref{seq:G_cond_mean_zero} and~\eqref{seq:H_cond_mean_zero} on RHS of~\eqref{seq:phi_first_order}, we have
%\vspace*{-0.6\baselineskip}
\begin{equation*}
    \begin{aligned}
    \mathbb{E}[
    \bm{G}_{\bm{x}}(\bm{Z},\bm{U})\,|\bm{X}_m=\bm{x},\bm{Z}\,]
    &=0,
    \\
    \mathbb{E}\big[
    \bm{H}_{\bm{x}}(\matbf{A}_+,\matbf{A}_-)|\bm{X}_m=\bm{x},\bm{Z}\,\big]
    &=\mathbb{E}\big[\mathbb{E}[
    \bm{H}_{\bm{x}}(\matbf{A}_+,\matbf{A}_-)|\bm{X}_m=\bm{x},\bm{Z}_+,\bm{Z}_-\,]
    \,|\,\bm{X}_m=\bm{x},\bm{Z}\,\big]
    =0.
    \end{aligned}
\end{equation*}
This means the non-Gaussian corrections of order $\sqrt{\mathrm{d}t}$ in the conditional characteristic function $\varphi_{\bm\eta_t|\bm{x}}(\bm{u})$ is perfectly vanishing under midpoint conditioning even though each half-step may have an $O(\sqrt{\mathrm dt})$ non-Gaussian correction. 
Therefore, the conditional characteristic function has the expansion
\begin{equation}
    \varphi_{\bm\eta_t|\bm{x}}(\bm{u})
    =\underbrace{\exp\left(-\frac{\|\bm{u}\|^2}{2} \right)
    \exp\left(i\sqrt{\mathrm{d}t}\,\langle\bm{u},\bm{s}(\bm{x},t_m) \rangle\right)}_{\text{(normalized) Gaussian part}}
    \,\,\,\underbrace{\left(1+O(\mathrm{d}t)(1+\|\bm{u}\|^m)\right)}_{\text{non-Gaussian corrections}},
    \label{seq:varphi_sqrt_dt_cancel}
\end{equation}
uniformly for $\bm{x}$ in compacts. Note that the non-Gaussian corrections start with the $\mathrm dt$-order, which means the conditional characteristic function also has no non-Gaussian $O(\sqrt{\mathrm dt})$ term.

The $O(\mathrm dt)$ notation in~\eqref{seq:varphi_sqrt_dt_cancel} can be expressed in a more rigorous form by using~\eqref{seq:eta_ito_taylor} as follows:
There exist constants $\mathrm{d}t_0>0$, $C>0$, and $m,l\in\mathbb N$
and a function $r_{\mathrm dt}(\bm u;\bm x)$ such that,
\begin{equation}
    | r_{\mathrm dt}(\bm u;\bm x) |
    \le C \big(1+\|\bm x\|^l\big)\big(1+\|\bm u\|^m\big)
    \qquad
    \forall\,\bm x,\bm u\in\mathbb R^d,\,
    0<\mathrm dt<\mathrm dt_0,
    \label{seq:r_dt_bound}
\end{equation}
and
\begin{equation}
    \varphi_{\bm\eta_t|\bm x}(\bm u)
    =\exp\left(-\frac{\|\bm u\|^2}{2} \right)
    \exp\left(
    i\sqrt{\mathrm dt}\,\langle \bm u, \bm s(\bm x,t_m)\rangle
    \right)
    \big(1+r_{\mathrm dt}(\bm u;\bm x)\,\mathrm dt\,\big).
    \label{seq:varphi_expansion_remainder}
\end{equation}

Indeed, by~\eqref{seq:eta_ito_taylor} we have $\bm\eta_t=\bm Z+\sqrt{\mathrm dt}\,\bm s+\sqrt{\mathrm dt}\,\bm Q_{\bm x}+\bm R_{\bm x}$ with $\|\bm R_{\bm x}\|_{L^2(\cdot|\bm X_m=\bm x)}=O(\mathrm dt)\big(1+\|\bm x\|^r\big)$ where $\bm Q_{\bm x}=\bm G_{\bm x}+\bm H_{\bm x}$. 
The $\sqrt{\mathrm dt}$-order non-Gaussian correction vanishes, as just shown. 
The remainder term is controlled by $\|\bm R_{\bm x}\|_{L^2(\cdot|\bm X_m=\bm x)}=O(\mathrm dt)\big(1+\|\bm x\|^r\big)$ together with finite-moment bounds for finite Wiener chaoses (and the moment assumption~\hyperlink{assumption:A5}{A5}): using $|e^{i\langle\bm u,\bm \delta\rangle}-1|\leq\min\{2,|\langle\bm u,\bm\delta\rangle|\}\leq\min\{2,\|\bm u\|\,\|\bm\delta\|\}$ for $\bm\delta=\bm R_{\bm x}$, one obtains $\mathbb{E}[\|\bm R_{\bm x}\|]\leq\mathbb{E}\left[\|\bm R_{\bm x}\|^2\right]^{\frac{1}{2}}=O(\mathrm dt)\big(1+\|\bm x\|^r\big)$ 
therefore $\left|r_{\mathrm dt}(\bm u;\bm x)\right|\leq C\big(1+\|\bm x\|^l\big)(1+\|\bm u\|^m)$ uniformly for $\bm x$. Note that for compact $K\subset\mathbb R^d$, we can take the sufficiently large constant $C_K<\infty$ such that $C_K>C\big(1+\|\bm x\|^l\big)$ for all $\bm x\in K$.

We also remark that the polynomial factor $(1+\|\bm u\|^m)$ in~\eqref{seq:varphi_sqrt_dt_cancel} and~\eqref{seq:r_dt_bound} comes from bounding the Taylor remainder of
$e^{i\langle \bm u,\bm\eta_t\rangle}$ after conditioning on $(\bm X_m=\bm x,\bm Z)$.
Since the $\sqrt{\mathrm{d}t}$-order term vanishes by \eqref{seq:G_cond_mean_zero}--\eqref{seq:H_cond_mean_zero},
the leading non-Gaussian contribution starts at order $\mathrm{d}t$ and is controlled by moments of
$|\langle \bm u,\bm Q_{\bm x}\rangle|^j\le \|\bm u\|^j\|\bm Q_{\bm x}\|^j$ for some finite $j$.
Under (\hyperlink{assumption:A1}{A1}) the whitening matrix $\mathbf B^{-1}$ is uniformly bounded, while (\hyperlink{assumption:A2}{A2})--(\hyperlink{assumption:A4}{A4}) imply that the strong Taylor
coefficients $\mathcal T_1(\bm x),\mathcal T_2(\bm x)$ in~\eqref{seq:eta_ito_taylor} have at most polynomial growth in $\|\bm x\|$.
Together with the moment bound (\hyperlink{assumption:A5}{A5}), this yields finite conditional moments of $\bm Q$ up to order $m$
(and uniformity for $\bm x$ on compact sets), leading to a remainder bound of the form
$O(\mathrm dt)(1+\|\bm u\|^m)$.

\bigskip
\noindent\textbf{\textbullet\ Step 4. Cancellation of the $\sqrt{\mathrm{d}t}$-order non-Gaussian corrections in probability density function} 

\noindent$-$\ \textbf{4.A Bounded drift case} 

As stated at the beginning of this subsection, Yang et al. (2019) established that, under Assumptions~\hyperlink{assumption:A1}{A1}--\hyperlink{assumption:A4}{A4}, the transition probability density of the It\^{o} SDE admits an asymptotic expansion in half-integer powers of the time increment $\mathrm dt$~\cite{yang2019anew}.
Since Yang et al.'s result is applicable only to the bounded-drift case, we proceed in two steps. We first follow their argument in the case (Assumption~\hyperlink{assumption:A5-A}{A5-A}), where the drift is uniformly bounded. We then extend the conclusion to the unbounded-drift regime (Assumption~\hyperlink{assumption:A5-B}{A5-B}) by means of a standard truncation-localization argument: after truncating the drift outside a ball $\|\bm x\|\leq R$, we apply the result on that compact subset.

For convenience in carrying out this extension, we follow the structure of Yang et al.'s proof but adopt a different representation of the remainder term, based on Kohatsu-Higa et al. (2013) and Bally et al. (2015)~\cite{higa2013estimates,bally2015probabilistic}, in order to make the drift-dependent prefactor explicit because that form is better suited to the localization step.
Applying the resulting expansion to each half-steps $\bm Y_\pm \coloneqq \matbf{B}_m^{-1}\bm\Delta^\pm/\sqrt h$ defined in~\eqref{seq:Delta^+_ito_taylor}--\eqref{seq:Delta^-_ito_taylor} for bounded-drift case (Assumption~\hyperlink{assumption:A5-A}{A5-A}), we obtain the expansion up to order $J\in\mathbb N$ for $0<h\le h_0$ (Note that $h=\mathrm dt/2$):

\begin{equation}
    f_{\pm}(\bm Y_\pm=\bm z_{\pm}|\bm X_m=\bm x)=
    \phi\!\left(\bm z_\pm^{(c)}\right)
    \!\!\left(1+\sum_{j=1}^{J}h^{j/2}\,
    c_{\pm,j}\!\left(\bm z_{\pm}^{(c)}\,;\bm x\right)\right)
    +\mathrm{Rem}_\pm\!\!\left(\bm z_{\pm}^{(c)},\bm x;J\right),
    \quad
    \bm z_{\pm}^{(c)}\!\coloneq\bm z_{\pm}-\sqrt h\,\bm\mu_{\pm}(\bm x)
    \label{seq:f_pm_yang_expansion_original}
\end{equation}

Here, $\phi(\cdot)$ denotes the probability density function of the $d$-dimensional standard normal distribution. The vectors $\bm\mu_+$ and $\bm\mu_-$ denote $\matbf{B}_m^{-1}\bm F_I$ and $-\matbf{B}_m^{-1}\bm F_{rev}$, respectively, and each $c_{\pm,j}(\cdot;\bm x)$ is a finite linear combination of shifted Hermite polynomials (i.e. Hermite polynomials of $\bm z_\pm-\sqrt h\,\bm\mu_\pm$) with coefficients depending on $\bm x$.
More precisely, under assumptions~\hyperlink{assumption:A1}{A1}--\hyperlink{assumption:A5-A}{A5-A}, there exist $C_\pm,c_\pm,\lambda_\pm,k^\pm_1,k^\pm_2,h_\pm>0$ such that
\begin{equation}
    \left|\mathrm{Rem}_\pm\!\!\left(\bm z_{\pm}^{(c)},\bm x;J\right)\right|
    \leq
    C_\pm\,h^{\frac{J+1}{2}}
    e^{c_\pm\|\bm\mu_\pm\|_\infty^2 h}
    \!\left(1+\|\bm x\|^{k^\pm_1}\right)
    \!\left(1+\big\|\bm z_{\pm}^{(c)}\big\|^{k^\pm_2}\right)
    \exp\!\left(-\lambda_\pm\!\left\|\bm z_{\pm}^{(c)}\right\|^2\right),
    \label{seq:f_remainder_bound}
\end{equation}
for all $\bm x,\bm z\in\mathbb R^d$ and $h<h_\pm$, where $\|\bm\mu_\pm\|_\infty$ denotes the supremum of $\|\bm\mu_\pm\|$.
Note that $C_\pm,c_\pm>0$ are constant. Under~\hyperlink{assumption:A1}{A1}--\hyperlink{assumption:A4}{A4}, the remainder admits a shifted Gaussian-tail bound with polynomial prefactors with some integers $k_1^\pm$, while the drift enters only through the exponential prefactor with $\|\bm\mu_\pm\|_\infty$~\cite{higa2013estimates,bally2015probabilistic}. 
Although the bounded-drift assumption~\hyperlink{assumption:A5-A}{A5-A} allows these drift-dependent prefactors to be absorbed into a single constant $C_{\mathrm{drift}}$, we keep them explicit to streamline the extension to the unbounded-drift regime.
An integer $k^\pm_2$ and constant $\lambda_\pm>0$ follow from repeated derivatives of Gaussian kernel structures, as shown in~\cite{yang2019anew,bally2015probabilistic}.
Accordingly, the remainder can be bounded by a polynomial in the centered increment $\bm z_\pm^{(c)}\coloneq\bm z_\pm-\sqrt h\,\bm\mu_\pm$ multiplied by a Gaussian tail.
Here, $\lambda_\pm$ governs the tail decay and depends only on the diffusion coefficient $\matbf B$ through the uniform ellipticity bounds (\hyperlink{assumption:A1}{A1}) and finitely many bounds on spatial derivatives of $\matbf B$ (\hyperlink{assumption:A2}{A2}), and is independent of the drift magnitude.

By~\hyperlink{lem:S1}{Lemma S1}, $\bm Y_+$ and $\bm Y_-$ are conditionally independent given $\bm{X}_m=\bm{x}$. We consider the normalized total increment $\bm\eta_t=(\bm Y_+ + \bm Y_-)/\sqrt{2}$.
Its density function can be established via the scaled convolution of $f_+$ and $f_-$: 
\begin{equation*}
    p_{\mathrm dt}(\bm\eta_t=\bm z \mid \bm X_m=\bm x)
    = \sqrt 2^d\!\!\int f_+(\bm s \mid \bm x)\, 
    f_-\!\!\left(\sqrt{2}\,\bm z - \bm s\,\middle|\, \bm x\right)\mathrm d\bm s.
\end{equation*}

Since the class of functions defined by the product of a Gaussian and polynomials (with Gaussian-bounded errors) is closed under convolution, the density of $\bm\eta_t$ retains the same asymptotic structure. More precisely, expanding $f_+$ and $f_-$ up to order $J$, convolving term-by-term and collecting powers of $\sqrt{\mathrm dt}$ (using standard Gaussian/Hermite identities) and re-express polynomials by combinations of shifted Hermite, we obtain the small-time expansion for $\mathrm dt<\mathrm dt_0$:
\begin{equation}
    p_{\mathrm dt}(\bm\eta_t=\bm z \mid \bm X_m=\bm x)
    =\phi\!\left(\bm z_c\right)
    \left(1+\sum_{j=1}^{J}\mathrm dt^{\frac{j}{2}}\,
    c_j\!\left(\bm z_c;\bm x\right)\right)
    +\mathrm{Rem}\!\left(\bm z_c,\bm x;J\right),
    \quad
    \bm z_c\coloneqq\bm z-\sqrt{\mathrm dt}\,\bm s(\bm x)
    \label{seq:p_dt_sqrt_dt_expansion}
\end{equation}
where $\bm s=(\bm\mu_++\bm\mu_-)/2=\matbf{B}_m^{-1}\bm v$, and each $c_j$ denotes a finite linear combination of shifted Hermite polynomials with coefficients depending on $\bm x$. The remainder term $\mathrm{Rem}(\bm z,\bm x;J)$ satisfies a global bound analogous to~\eqref{seq:f_remainder_bound}:
\begin{equation}
    \left\vert\mathrm{Rem}\!\left(\bm z_c,\bm x;J\right)\right\vert
    \leq
    C\,\mathrm dt^{\frac{J+1}{2}}
    e^{c\,\left(\|\bm\mu_+\|_\infty^2+\|\bm\mu_-\|_\infty^2\right)\mathrm dt}
    \!\left(1+\|\bm x\|^{k_1}\right)
    \big(1+\|\bm z_c\|^{k_2}\big)
    \exp\!\left(-\lambda\left\|\bm z_c\right\|^2\right),
    \label{seq:p_remainder_bound}
\end{equation}
for all $\bm x,\bm z\in\mathbb R^d$ and $\mathrm dt<\mathrm dt_0$,
where $C,c,k_1, k_2$, and $\lambda$ are defined similarly to their counterparts in~\eqref{seq:f_remainder_bound}. Note that the remainder term $\mathrm{Rem}(\bm z,\bm x;J)$ is still bounded under the Gaussian tails.

Now applying Fourier transformation $\mathcal F$ on~\eqref{seq:p_dt_sqrt_dt_expansion}, we obtain the conditional characteristic function of $\bm\eta_t$ conditioning $\bm X_m=\bm x$ as
\begin{equation}
    \begin{aligned}
        \varphi_{\bm\eta_t|\bm x}(\bm u)
        &=\int e^{i\langle\bm u,\bm z\rangle}
        p_{\mathrm dt}(\bm z|\bm x)\mathrm d\bm z
        \eqqcolon\mathcal F[p_{\mathrm dt}](\bm u)        
        \\
        &=\underbrace{\mathcal F\big[\phi(\bm z-\sqrt{\mathrm dt}\,\bm s)\big](\bm u)}_{\eqqcolon\,\widetilde{\phi}(\bm u,\bm s)}
        +\sum_{j=1}^J\sqrt{\mathrm dt}^{\,j}
        \mathcal F\left[
        c_j\!\left(\bm z-\sqrt{\mathrm dt}\,\bm s;\bm x\right)\,
        \phi\!\left(\bm z-\sqrt{\mathrm dt}\,\bm s\right)\right](\bm u)
        +\underbrace{\mathcal F\left[\mathrm{Rem}(\bm z,\bm x;J)\right](\bm u)}_{\eqqcolon\,\widetilde{\mathrm{Rem}}(\bm u,\bm x;J)}
        \\
        &=\widetilde{\phi}(\bm u,\bm s)
        +\sum_{j=1}^J\sqrt{\mathrm dt}^{\,j}
        \mathcal F\left[\left(\sum_{k=0}^{k_j} a_k^{(j)}(\bm x)H_k\big(\bm z-\sqrt{\mathrm dt}\,\bm s\big)\right)\phi\!\left(\bm z-\sqrt{\mathrm dt}\,\bm s\right)\right](\bm u)
        +\widetilde{\mathrm{Rem}}(\bm u,\bm x;J)
        \\
        &=\widetilde{\phi}(\bm u,\bm s)
        +\sum_{j=1}^J\sqrt{\mathrm dt}^{\,j}
        \underbrace{\left(\sum_{k=0}^{k_j}a_k^{(j)}(\bm x)\,(i\bm u)^k\right)}_{\eqqcolon\,\widetilde{c_j}(\bm u;\bm x)}
        \widetilde{\phi}(\bm u,\bm s)
        +\widetilde{\mathrm{Rem}}(\bm u,\bm x;J)
        \\
        &=\underbrace{
        \exp\left(-\frac{\|\bm u\|^2}{2} \right)   \exp\left(i\sqrt{\mathrm dt}\,
        \langle \bm u, \bm s \rangle
        \right)}_{=\,\widetilde{\phi}(\bm u,\bm s)}
        \left(1+\sum_{j=1}^J
        \sqrt{\mathrm dt}^{\,j}\widetilde{c_j}(\bm u,\bm x)
        \right)
        +\widetilde{\mathrm{Rem}}(\bm u,\bm x;J).
    \end{aligned}
    \label{seq:p_Fourier_to_varphi}
\end{equation}
In the fourth equality, we use $\mathcal F[H_k(\bm z-\bm\mu)\phi(\bm z-\bm\mu)](\bm u)=\mathcal F[(-1)^k \frac{\mathrm d^k}{\mathrm d\bm z^k}\phi(\bm z-\bm\mu)](\bm u)=(i\bm u)^k\mathcal F[\phi(\bm z-\bm\mu)](\bm u)$ where $H_k$ is Hermite polynomial of order $k$. We emphasize that the remainder $\widetilde{\mathrm{Rem}}(\bm u,\bm x;J)$ cannot contribute to the terms of order $\mathrm dt^{(j/2)}$ when $j\leq J$ because
\begin{equation*}
\begin{aligned}
    \left|\widetilde{\mathrm{Rem}}(\bm u,\bm x;J)\right|
    &\leq\int\left|e^{i\langle\bm u,\bm z\rangle}\right|
    \left|\mathrm{Rem}\!\left(\bm z_c,\bm x;J\right)\right|\mathrm d\bm z
    \\
    &=\int C\,\mathrm dt^{\frac{J+1}{2}}
    e^{c\left(\|\bm\mu_+\|_\infty+\|\bm\mu_-\|_\infty\right)\mathbf dt}
    \!\left(1+\|\bm x\|^{k_1}\right)
    \big(1+\|\bm z_c\|^{k_2}\big)
    \exp\left(-\lambda\|\bm z_c\|^2\right)\mathrm d\bm z
    \quad\text{by~\eqref{seq:p_remainder_bound}}
    \\
    &=C\,\mathrm dt^{\frac{J+1}{2}}
    e^{c\left(\|\bm\mu_+\|_\infty+\|\bm\mu_-\|_\infty\right)\mathbf dt}
    \!\left(1+\|\bm x\|^{k_1}\right)
    \int\big(1+\|\bm z_c\|^{k_2}\big)\exp\left(-\lambda\|\bm z_c\|^2\right)\mathrm d\bm z_c
    \\
    &\leq C_Z\,\mathrm dt^{\frac{J+1}{2}}
    e^{c\left(\|\bm\mu_+\|_\infty+\|\bm\mu_-\|_\infty\right)\mathbf dt}
    \!\left(1+\|\bm x\|^{k_1}\right).
\end{aligned}
\end{equation*}
In third line, the integral variable is changed: $\bm z\mapsto\bm z_c$.
The last inequality holds because all moments of the Gaussian distribution are finite.

This means that since there is no non-Gaussian $\sqrt{\mathrm dt}$-order term in $\varphi_{\bm\eta_t|\bm x}$ by~\eqref{seq:varphi_sqrt_dt_cancel}, $\widetilde{c_1}(\bm u;\bm x)$ in~\eqref{seq:p_Fourier_to_varphi} must vanish. The injectivity of Fourier transform ensures that $c_1\big(\bm z-\sqrt{\mathrm dt}\,\bm s\big)$ in~\eqref{seq:p_dt_sqrt_dt_expansion} must also be zero. Moreover, $\mathrm dt^{(j/2)}c_j$ cannot contribute to terms of order $\mathrm dt^{k/2}$ when $k<j$ because $c_j$ is polynomial in $\bm z-\sqrt{\mathrm dt}\,\bm s$. Hence, one can conclude that there is no non-Gaussian $\sqrt{\mathrm dt}$-order term in the conditional density function $p_{\mathrm dt}$. 

\medskip
\noindent$-$\ \textbf{4.B Truncation and localization for unbounded drift case} 

We emphasize that the arguments from~\eqref{seq:f_pm_yang_expansion_original} to~\eqref{seq:p_remainder_bound} hold rigorously only for the bounded-drift case. Therefore, while they apply seamlessly under Assumption~\hyperlink{assumption:A5-A}{A5-A}, accommodating the unbounded drift allowed by Assumption~\hyperlink{assumption:A5-B}{A5-B} requires an additional truncation and localization procedure. 
Specifically, let $B(0,R)$ denote a ball of fixed radius $R$ centered at the origin. We introduce an auxiliary process with a variable $\hat{\bm X}_m$, whose drift and diffusion coefficients perfectly match the original system within $B(0,2R)$, but transition smoothly to bounded boundary values outside $B(0,2R+\delta)$ with $\delta>0$. If we restrict our focus to the dynamics within $B(0,R)$, this auxiliary process is nearly identical to the original one; the discrepancy is bounded exactly by the probability that the process exits $B(0,2R)$ during the short-time interval $\mathrm dt$. 

Quantitatively, let $\hat p_{dt}(\cdot|\bm x)$ denote the conditional density of the localized auxiliary system with bounded coefficients
and $p_{dt}(\cdot|\bm x)$ that of the original system. Since the two SDEs have identical coefficients on $B(0,2R)$,
the two conditional laws coincide on the event that the trajectory does not exit $B(0,2R)$ during the time window of length $\mathrm dt$.
Therefore, for $\|\bm x\|\le R$,
\begin{equation}
\begin{aligned}
    |p_{dt}(\bm z|\bm x)-\hat p_{dt}(\bm z|\bm x)|
    &\le \mathbb{P}_{\bm x}\!\big(\tau_{2R}\le\mathrm dt\big)
    \sup_{\bm y\in\mathbb R^d}\hat p_{dt}(\bm z|\bm y)\\
    &\le \mathbb{P}_{\bm x}\!\big(\tau_{2R}\le\mathrm dt\big)
    \Big(C\,e^{\bar c\big(1+R^{k_1}\big)\mathrm dt} (1+R^{k_2})(1+\|\bm z\|^m)e^{-\gamma_{\mathrm{tr}}\|\bm z\|^2}\Big),
    \label{seq:p_trun_error_bound}
\end{aligned}
\end{equation}
for certain constants $C,\bar c, k_1, k_2, m, \gamma_{\mathrm tr}>0$ where $\tau_{2R}$ is the first exit time from $B(0,2R)$. 
Intuitively, this is because the deviation in probability density is bounded by the escape probability weighted by the maximal density of the auxiliary process within $\bm x\in B(0,R)$.
That maximal density is given by bounded-drift expansion~\eqref{seq:p_dt_sqrt_dt_expansion}--\eqref{seq:p_remainder_bound} with an envelope whose $R$-dependence is explicit in the prefactors (with numerical constants independent of $R$), hence one can get the second inequality of~\eqref{seq:p_trun_error_bound}.
Moreover, since midpoint is the starting point of the forward and backward half-step, $\mathbb{P}_{\bm x}(\tau_{2R}\le\mathrm dt)\le C_{\mathrm{tr}} \exp\big(\!\!-\!c_{\mathrm{tr}}(R-C_F(1+R^m)\mathrm dt)^2_+/\mathrm dt\big)$ for $\|\bm x\|\le R$ by exponential martingale and polynomial growth bounds of drift (Assumption \hyperlink{assumption:A4}{A4}), where $(\cdot)_+\coloneq\max\{\cdot,0\}$. Note that the $R$ and $C_F(1+R^m)\mathrm dt$ term are coming from the minimum distance to the truncation boundary and the maximum magnitude of the drift, respectively. Combining these yields the truncation error term and we finally obtain
\begin{equation}
    p_{\mathrm dt}(\bm\eta_t=\bm z \mid \bm X_m=\bm x)
    =\phi\!\left(\bm z_c\right)
    \!\!\left(1+\mathrm dt\,c_{\mathrm dt}\big(\bm z_c;\bm x\big)\right)
    +\mathrm{Rem}_{\mathrm dt}\!\left(\bm z_c,\bm x;R\right),
    \quad\forall\|\bm x\|\leq R,\,\bm z\in\mathbb R^d,
    \label{seq:p_dt_expansion_cancel_sqrt_dt}
\end{equation}
where $\bm z_c=\bm z-\sqrt{\mathrm dt}\,\bm s(\bm x)$ and $c_{\mathrm dt}(\bm z_c;\bm x)$ is a finite linear combination of shifted Hermite polynomials in $\bm z_c$ whose coefficients are deterministic functions of $\bm x$ of at most polynomial growth. 
Equation~\eqref{seq:p_dt_expansion_cancel_sqrt_dt} is the key output of this subsection since on the localization region $\|\bm x\|\leq R(\mathrm dt)$, the conditional distribution is a shifted Gaussian plus $O(\mathrm dt)$ corrections, with no $O(\sqrt{\mathrm dt})$ contribution.

The remainder $\mathrm{Rem}_{\mathrm dt}$ satisfies the bounds
\begin{subequations}\label{seq:rem_dt}
\renewcommand{\theequation}{\theparentequation-\Alph{equation}}
\begin{align}
  \mathrm{Rem}_{\mathrm dt}&
  \leq C_A\,\mathrm dt^{3/2}
  \big(1+\|\bm x\|^{a_1}\big)
  \big(1+\|\bm z_c\|^{a_2}\big)
  e^{-\gamma_A\|\bm z_c\|^2}
  \quad\text{under Assumption~\hyperlink{assumption:A5-A}{A5-A}}
  \label{seq:rem_dt_A},
  \\
  \mathrm{Rem}_{\mathrm dt}
  &\leq
  \begin{aligned}
    C_B\!\Bigg(\!
    \mathrm dt^{3/2}
    &+\underbrace{e^{-c_{B}^{(1)}\frac{\big(R-C_F(1+R^{b_1})\mathrm dt\big)_+^2}{\mathrm dt}}
    \,(1+R^{b_2})
    }_{\text{truncation error}}
    \Bigg)
    \\
    &\times
    e^{c_{B}^{(2)}(1+R^{b_3})\mathrm dt}\,
    (1+\|\bm x\|^{b_4})
    (1+\|\bm z_c\|^{b_5})\,e^{-\gamma_B\|\bm z_c\|^2}
  \end{aligned}
  \quad\text{under Assumption~\hyperlink{assumption:A5-B}{A5-B}},
  \label{seq:rem_dt_B}
\end{align}
\end{subequations}
for certain constants $C_A,C_B,C_F,c_B^{(1)},c_B^{(2)},a_1,a_2,b_1,b_2,b_3,b_4,b_5,\gamma_A,\gamma_B>0$, which are independent of $R$.
We emphasize that there are no $\sqrt{\mathrm dt}$-order non-Gaussian correction terms even for the unbounded-drift case.

Note that in~\eqref{seq:rem_dt}, all numerical constants are independent of $R$, and the $R$-dependence appears explicitly only in the prefactors. Hence the bound~\eqref{seq:rem_dt} remains valid for any $R$. Nevertheless, sending $R\rightarrow\infty$ allows the prefactors (e.g., $(1+\|\bm x\|^b)$) to become arbitrarily large. 
This implies that the exact finite expansion formula~\eqref{seq:p_dt_expansion_cancel_sqrt_dt} may cease to be asymptotically informative in the joint regime $\mathrm dt\rightarrow0$ and $R\rightarrow\infty$—specifically, the bound does not guarantee $\mathrm{Rem}_{\mathrm dt}=o(\mathrm dt)$ uniformly over $\|\bm x\|\leq R$, even though the expansion identity~\eqref{seq:p_dt_expansion_cancel_sqrt_dt} and the estimate~\eqref{seq:rem_dt} remain formally valid. 
To prevent the estimate from becoming vacuous, one must choose an appropriate $R=R(\mathrm dt)$ to ensure~\eqref{seq:p_dt_expansion_cancel_sqrt_dt} is meaningful. Therefore, we set $R(\mathrm dt)$ hereafter as
\begin{subequations}\label{seq:R_dt}
\renewcommand{\theequation}{\theparentequation-\Alph{equation}}
\begin{align}
    R(\mathrm dt)&=
    \mathrm dt^{-\kappa},
    \quad0<\kappa<\frac{\epsilon}{a_1}\text{ with }
    \epsilon>0
    \text{ and $a_1$ in~\eqref{seq:rem_dt_A} under Assumption~\hyperlink{assumption:A5-A}{A5-A}},
    \label{seq:R_dt_A}
    \\
    R(\mathrm dt)&=
    \left(\ln\frac{1}{\mathrm dt}\right)^\beta,
    \quad\beta>\frac{1}{1+\alpha}
    \text{ with $\alpha$ in~\hyperlink{assumption:A5-B}{A5-B} under Assumption~\hyperlink{assumption:A5-B}{A5-B}}.
    \label{seq:R_dt_B}
\end{align}
\end{subequations}

This choice allows us to control the truncation error term in~\eqref{seq:rem_dt_B} as $o(\mathrm dt^{3/2})$ for sufficiently small $\mathrm dt$. Indeed, since $C_F(1+R^{b_1})\mathrm dt=O(\mathrm dt\,|\ln\mathrm dt|^{\beta b_1})\rightarrow0$ as $\mathrm dt\rightarrow 0$, we have
\begin{equation*}
    \exp\!\!\left(-c_B^{(1)}\frac{\big(|\ln \mathrm dt|^\beta
    -C_F(1+|\ln \mathrm dt|^{\beta b_1})\mathrm dt
    \big)_+^2}{\mathrm dt}\right)
    \leq \exp\left(-\bar c\frac{|\ln \mathrm dt|^{2\beta}}{\mathrm dt}\right)
    = o(\mathrm dt^m)
    \quad\forall m>0,
\end{equation*}
and $\mathrm dt^m(1+R^k)=O(\mathrm dt^m\mathrm |\ln\mathrm dt|^{\beta k})=o(\mathrm dt^{m-\epsilon})$ for all $\epsilon>0$.
Similarly, we can rewrite~\eqref{seq:rem_dt} in a much simpler form under~\eqref{seq:R_dt}: There exist $C_{\beta,\kappa,\epsilon}>0$ and $\mathrm dt_0>0$ for every $\epsilon>0$, $\beta>1/(1+\alpha)$, and $0<\kappa<\epsilon/a_1$ such that
\begin{equation}
  \mathrm{Rem}_{\mathrm dt}
  \leq C_{\beta,\kappa,\epsilon}\,
  \mathrm dt^{\frac{3}{2}-\epsilon}
  \big(1+\|\bm z_c\|^m\big)
  e^{-\gamma\|\bm z_c\|^2}
  \qquad\forall
  \mathrm dt<\mathrm dt_0,
  \|\bm x\|\leq R(\mathrm dt)\text{ in~\eqref{seq:R_dt}},
  \label{seq:rem_dt_Rdt_choosen}
\end{equation}
for certain constants $m,\gamma>0$; since for~\eqref{seq:rem_dt_A}, $\mathrm dt^{3/2}\sup_{\|\bm x\|\in B(0,R)}(1+\|\bm x\|^{a_1})=\mathrm dt^{3/2}(1+R^{a_1})=o(\mathrm dt^{\frac{3}{2}-\epsilon})$;
for~\eqref{seq:rem_dt_B}, $e^{c_{B}^{(2)}(1+R^{b_3})\mathrm dt}=e^{o(1)}$ and $\mathrm dt^{3/2}\sup_{\|\bm x\|\in B(0,R)}(1+\|\bm x\|^{b_4})=\mathrm dt^{3/2}(1+R^{b_4})=o(\mathrm dt^{\frac{3}{2}-\epsilon})$ for all $\epsilon>0$. This result is valid for both~\hyperlink{assumption:A5-A}{A5-A} and~\hyperlink{assumption:A5-B}{A5-B}. 

We also remark that even when the exact decomposition~\eqref{seq:p_dt_expansion_cancel_sqrt_dt} is not asymptotically informative for large $R$, the representation and the bound~\eqref{seq:rem_dt} remain useful; for example, they provide a non-asymptotic control of $\|p(\cdot|\bm x)\|_\infty$ on prescribed radii $R$ (e.g., dyadic scales $R=2^k R_0$).

\subsection{Local Gaussian channel proxy and discrepancy of mutual information}
\vspace*{-0.6\baselineskip}
From this point on the proof is bookkeeping. We compare the exact midpoint-conditioned channel with its Gaussian proxy, decompose the mutual information difference and show that each resulting term is $o(\mathrm dt)$.
To this end, we first introduce an auxiliary random variable $\bm\eta_t^G$ as a surrogate for the whitened increment $\bm\eta_t$.
We define $\bm\eta_t^G$ as the output of an idealized Gaussian channel driven by the local signal $\sqrt{\mathrm dt}\,\bm s(\bm X_m,t_m)$,
\begin{equation}
    \bm\eta_t^G=\sqrt{\mathrm dt}\,\bm s(\bm X_m,t_m)+\bm N,
    \quad
    \bm N\sim\mathcal N(0,\mathbf I_d),
    \quad\bm N\perp\!\!\!\perp\bm X_m,
    \label{seq:Gaussian_channel_proxy}
\end{equation}
where $\bm s=\matbf B_m^{-1}\bm v$ represents the diffusion-metric normalized current velocity.
For brevity, hereafter we write $p_{\mathrm dt}(\bm z|\bm x)=p(\bm\eta_t=\bm z|\bm X_m=\bm x)$, $p_{\mathrm dt}(\bm z,\bm x)=p(\bm\eta_t=\bm z,\bm X_m=\bm x)$, and $p_{\mathrm dt}(\bm z)=\int p_{\mathrm dt}(\bm z,\bm x)\mathrm d\bm x$ and similarly for $q_{\mathrm dt}$. The arguments will always indicate which object is meant.

Note that the conditional density of $\bm\eta_t^G$ is exactly given by the shifted Gaussian kernel appearing in the leading order of our small-time expansion~\eqref{seq:p_dt_expansion_cancel_sqrt_dt}. Explicitly,
\begin{equation}
    q_{\mathrm dt}(\bm z|\bm x)\coloneq
    q_{\mathrm dt}\big(\bm\eta_t^G=\bm z|\bm X_m=\bm x\big) 
    = \phi\!\left(\bm z - \sqrt{\mathrm dt}\bm s(\bm x)\right) 
    = \phi\big(\bm z_c\big),
    \quad
    \bm z_c\coloneq\bm z-\sqrt{\mathrm dt}\,\bm s(\bm x),
    \label{seq:q_dt_equal_phi}
\end{equation}
where $\phi(\cdot)$ is the probability density function of the $d$-dimensional standard normal distribution. This identifies $\bm\eta_t^G$ as the ``Gaussian core'' of the dynamics, stripped of the non-Gaussian corrections and the remainder terms $\mathrm{Rem}_{\mathrm dt}$ established in~\eqref{seq:p_dt_expansion_cancel_sqrt_dt}. In the subsequent steps, $\bm\eta_t^G$ will serve as the reference for quantifying the mutual information discrepancy induced by the non-Gaussianity of the actual dynamics.

We now demonstrate that the discrepancy $\Delta I\coloneq I\big(\bm\eta_t;\bm X_m) - I\big(\bm\eta_t^G;\bm X_m)$ is of order $o(\mathrm dt)$ under assumptions \hyperlink{assumption:A1}{A1}--\hyperlink{assumption:A5}{A5} by the following steps.

\bigskip
\noindent\textbf{\textbullet\ Step 1. Decomposing upper bound of the} $|\Delta I|$

As identified in Eq.~\eqref{seq:rem_dt}, the polynomial growth of the coefficients (Assumption \hyperlink{assumption:A4}{A4}) implies that the remainder terms scale with $\|\bm X_m\|$, making a global uniform bound difficult to establish. 
To circumvent this, we consider a compact ball $B(0, R)$ of radius $R=R(\mathrm dt)$ defined in~\eqref{seq:R_dt} and centered at the origin, then decompose the discrepancy $\Delta I$ into contributions from the interior and exterior of this region. We explicitly handle this by introducing an indicator variable $S$ defined as $S=1$ if $\bm X_m \in B(0, R)$ and $S=0$ otherwise.

Applying the information chain rule $I(\bm \eta_t; \bm X_m) = I(\bm \eta_t; (\bm X_m,S)) = I(\bm \eta_t; S) + I(\bm \eta_t; \bm X_m | S)$ to both the actual dynamics and the proxy, we can decompose the magnitude of the discrepancy as:
\begin{equation}
    |\Delta I| \le 
    \big| I(\bm \eta_t; S) - I(\bm \eta_t^G; S) \big| 
    + \mathbb P(S=1)|\Delta I_1| + \mathbb P(S=0)|\Delta I_0|,
    \label{seq:deltaI_first_decom}
\end{equation}
where $\Delta I_k \coloneq I(\bm \eta_t; \bm X_m | S=k) - I(\bm \eta_t^G; \bm X_m | S=k)$ denotes the conditional information difference for each region.

The first term represents the information about the region index carried by the increments. Since $S$ is a discrete binary variable, this term is bounded by twice the Shannon entropy of $S$, i.e., $|I(\bm \eta_t; S) - I(\bm \eta_t^G; S)| \le 2H(S)$.

The core of the analysis thus reduces to the conditional terms. For each region ($S=0,1$), let $p_k$ and $q_k$ denote the conditional joint densities of the actual and proxy processes, respectively (i.e., $p_k(\cdot)=p_{\mathrm dt}(\cdot|S=k)$ and $q_k(\cdot)=q_{\mathrm dt}(\cdot|S=k)$). We utilize the following exact decomposition identity for the mutual information difference:
\begin{equation*}
    \Delta I_1 = 
    D_{\mathrm{KL}}\big(p_1(\bm z,\bm x)\|q_1(\bm z,\bm x)\big)
    - D_{\mathrm{KL}}\big(p_1(\bm z) \| q_1(\bm z)\big) 
    + \int_{B(0,R)}\!\!\!\!\mathrm d\bm x\!
    \int_{\mathbb R^d}\!\!\mathrm d\bm z\,\big(p_1(\bm z,\bm x)-q_1(\bm z, \bm x)\big)
    \ln\frac{q_1(\bm z|\bm x)}{q_1(\bm z)},
\end{equation*}
where $D_{\mathrm{KL}}(p||q)\coloneq\int p\ln(p/q)$ is the relative entropy.
Note that since $p_1(\bm z)$ and $q_1(\bm z)$ are marginals, the data processing inequality yields $D_{\mathrm{KL}}\big(p_1(\bm z)\|q_1(\bm z)\big)\leq D_{\mathrm{KL}}\big(p_1(\bm z,\bm x)\|q_1(\bm z,\bm x)\big)$. 
Also because  $p_1(\bm x)=q_1(\bm x)$, the equality $D_{\mathrm{KL}}\big(p_1(\bm z,\bm x)\|q_1(\bm z,\bm x)\big)=\mathbb E_{\bm x\in B(0,R)}\!\left[D_{\mathrm{KL}}\big(p_1(\bm z|\bm x)\|q_1(\bm z|\bm x)\big)\right]$ holds. Therefore we obtain
\begin{equation}
    |\Delta I_1| \leq 
    \underbrace{
    \mathbb E_{\bm x\in B(0,R)}\!\left[D_{\mathrm{KL}}\big(p_1(\bm z|\bm x)\|q_1(\bm z|\bm x)\big)\right] }_{A}
    + \underbrace{
    \left|\mathbb E_{\bm x\in B(0,R)}\!\!\left[
    \int_{\mathbb R^d}\!\mathrm d\bm z\,\big(p_1(\bm z|\bm x)-q_1(\bm z|\bm x)\big)
    \ln\frac{q_1(\bm z|\bm x)}{q_1(\bm z)}\right]\right|}
    _{B}.
    \label{seq:MI_decomposition_bound}
\end{equation}

Now for $|\Delta I_0|$, we divide the ``outside'' $B(0,R)^c$ into infinitely many shells by introducing $R_k\coloneq 2^k R(\mathrm dt)$. More precisely, let $A_k$ denote the shell $2^{k-1}R<\|\bm X_m\|\leq2^kR$ for $k\in\mathbb N$. Then $B(0,R)^c=\cup_{k=1}^\infty A_k$. This division is useful to obtain the upper bound of $I(\bm\eta_t;\bm X_m|S=0)$ because
\begin{equation}
\begin{aligned}
    I(\bm\eta_t;\bm X_m|S=0)
    &= h(\bm\eta_t|S=0)-\mathbb E_{\bm x|S=0}
    \left[h(\bm\eta_t|\bm X_m=\bm x,S=0)\right]
    \\
    &= h(\bm\eta_t|S=0)-
    \sum_{k=1}^\infty\frac{\mathbb P(\bm X_m\in A_k)}{\mathbb P(S=0)}\,
    \mathbb E_{\bm x\in A_k}\!
    \left[h(\bm\eta_t|\bm X_m=\bm x,S=0)\right]
    \\
    &\leq
    \underbrace{
    \frac{1}{2}\ln\left((2\pi e)^d \det\mathrm{Cov}(\bm\eta_t|S=0)\right)}
    _{G_0}
    +\frac{1}{\mathbb P(S=0)}\sum_{k=1}^\infty\,\underbrace{
    \mathbb P(\bm X_m\in A_k)\sup_{\bm x\in A_k}\ln\!\|p_0(\cdot|\bm x)\|_\infty}
    _{S_k},
    \label{seq:outside_real_I_bound}
\end{aligned}
\end{equation}
where $h(\cdot)$ is differential entropy and $\mathrm{Cov}(\bm\eta_t|S=0)$  denotes the conditional covariance matrix of $\bm\eta_t$ given $S=0$. We use the fact that Gaussian distribution maximizes the differential entropy for fixed covariance and $-h(\cdot)=\int p\ln p\leq\int p\ln\!\| p\|_\infty=\ln\!\|p\|_\infty$ for the last inequality in~\eqref{seq:outside_real_I_bound}.

Since $|\Delta I_0|\leq I(\bm\eta_t;\bm X_m|S=0)+I(\bm\eta_t^G;\bm X_m|S=0)$, we obtain
\begin{equation}
    \mathbb P(S=0)|\Delta I_0|\ \leq\ 
    \underbrace{
    \mathbb P(S=0)\!\left(I\big(\bm\eta_t^G;\bm X_m|S=0\big)+G_0\right)}
    _{K}
    +\sum_{k=1}^\infty S_k,
    \label{seq:delta_I_outside_bound}
\end{equation}
and therefore
\begin{align}
    |\Delta I|\leq
    2H(S)+A+B+K+\sum_{k=1}^\infty S_k.
    \label{seq:delta_I_upper_bound_decomposition}
\end{align}

In the remaining steps, we show that $H(S),A,B,K$ and $\sum_{k=1}^\infty S_k$ are $o(\mathrm dt)$. Here $H(S)$ controls the region label, $A$ and $B$ the interior relative-entropy and information-density errors, $K$ the exterior Gaussian contribution, and $\sum_k S_k$ the far-tail shells.

\bigskip
\noindent\textbf{\textbullet\ Step 2. Bounding the Indicator Entropy:} $H(S)=o(\mathrm dt)$

Let $P_0$ denote the tail probability $\mathbb{P}(S=0) = \mathbb{P}(\|\bm{X}_m\| > R)$. The entropy of the indicator $S$ is given by $H(S) = -P_0\ln P_0 - (1-P_0)\ln(1-P_0)$. Note that if $P_0=0$ or $P_0=1$, then $H(S)=0$, so the condition $H(S)=o(\mathrm{dt})$ holds trivially. For $P_0 \in (0,1)$, we utilize the inequality
\begin{equation}
    H(S) \leq -P_0\ln P_0 + P_0.
    \label{seq:H(S)_bound}
\end{equation}

Now, applying Markov's inequality $\mathbb P(X>c)\leq\mathbb E[X]/c$ with $R(\mathrm dt)$ defined in~\eqref{seq:R_dt} yields
\begin{subequations}\label{seq:P0_bound}
\renewcommand{\theequation}{\theparentequation-\Alph{equation}}
\begin{align}
    P_0&=\mathbb P(\|\bm X_m\|^n>R^n)
    \leq\frac{\mathbb E[\|\bm X_m\|^n]}{R^n}
    =\mathbb E[\|\bm X_m\|^n]\mathrm dt^{\kappa n}
    \quad\forall n\in\mathbb N
    \text{ under Assumption~\hyperlink{assumption:A5-A}{A5-A}},
    \label{seq:P0_bound_A}
    \\
    P_0&=\mathbb P(e^{\theta\|\bm X_m\|^{1+\alpha}}\!>e^{\theta R^{1+\alpha}})
    \leq\frac{\mathbb E[e^{\theta\|\bm X_m\|^{1+\alpha}}]}{e^{\theta R^{1+\alpha}}}
    =\mathbb E\!\left[e^{\theta\|\bm X_m\|^{1+\alpha}}\right]
    e^{-\theta(\ln(1/\mathrm dt))^{\beta(1+\alpha)}}
    \text{ under Assumption~\hyperlink{assumption:A5-B}{A5-B}}.
    \label{seq:P0_bound_B}
\end{align}
\end{subequations}
Assumption~\hyperlink{assumption:A5}{A5} ensures that each RHS of~\eqref{seq:P0_bound} remains finite.

In~\eqref{seq:P0_bound_A}, for any target order $k\in\mathbb N$, we can choose a sufficiently large $n$ such that $\kappa n > k$, which implies $P_0=o(\mathrm dt^k)$. % Revised: Improved logical phrasing
Similarly, in~\eqref{seq:P0_bound_B}, since $\beta(1+\alpha)>1$ implies that the exponential factor dominates any polynomial power $\mathrm dt^k=e^{-k|\ln\mathrm dt|}$, we also have $P_0=o(\mathrm dt^k)$ for any $k\in\mathbb N$.
Therefore, we obtain the uniform bound
\begin{equation}
    P_0=\mathbb P(\|\bm X_m\|>R)=o(\mathrm dt^k)
    \quad\forall k\in\mathbb N,
    \label{seq:P0_bound_uniform}
\end{equation}
which is valid for both~\hyperlink{assumption:A5-A}{A5-A} and~\hyperlink{assumption A5-B}{A5-B}.

Using~\eqref{seq:P0_bound_uniform} with $k=2$, we write $P_0=\mathrm dt^2\epsilon(\mathrm dt)$, where $\epsilon(\mathrm dt)\rightarrow0$ as $\mathrm dt\rightarrow 0$.
This yields $P_0\ln P_0=o(\mathrm dt)$ since
\begin{equation}
\begin{aligned}
    \lim_{\mathrm dt\rightarrow 0}
    \frac{P_0\ln P_0}{\mathrm dt}
    &=\lim_{\mathrm dt\rightarrow 0}
    \mathrm dt\,\epsilon(\mathrm dt)
    \left[2\ln\mathrm dt+\ln\epsilon(\mathrm dt)\right]
    \\
    &=\lim_{\mathrm dt\rightarrow 0}
    2\epsilon(\mathrm dt)\big(\mathrm dt\ln\mathrm dt)
    +\lim_{\mathrm dt\rightarrow 0}
    \big(\epsilon(\mathrm dt)\ln\epsilon(\mathrm dt)\big)\mathrm dt
    \\
    &=0+0=0.
    \label{seq:P0lnP0_odt}
\end{aligned}
\end{equation}

Consequently, we verify that
\begin{equation}
    H(S)\leq -P_0\ln P_0+P_0=o(\mathrm dt).
    \label{seq:H(S)_odt}
\end{equation}

\bigskip
\noindent\textbf{\textbullet\ Step 3. Bounding the Interior Relative Entropy:} $A=o(\mathrm dt)$

To establish $A\coloneq\mathbb E_{\bm x\in B(0,R)}\!\left[D_{\mathrm{KL}}\big(p_1(\bm z|\bm x)\|q_1(\bm z|\bm x)\big)\right]=o(\mathrm dt)$, it suffices to show that $D_{\mathrm{KL}}\big(p_1(\bm z|\bm x)\|q_1(\bm z|\bm x)\big)=o(\mathrm dt)$ holds uniformly for $\bm x\in B(0,R)$. Within the region $\|\bm X_m\|\in B(0,R)$, we have $p_1=p_{\mathrm dt}$ and $q_1=q_{\mathrm dt}$; hence, we may directly invoke the expansion~\eqref{seq:p_dt_expansion_cancel_sqrt_dt} and the bound~\eqref{seq:rem_dt_Rdt_choosen}. However, a direct application of the Chi-square approximation is technically subtle since the tail behavior of $\mathrm{Rem}_{\mathrm dt}$ relative to the Gaussian $q_1$ does not strictly guarantee the integrability of the ratio $p_1/q_1$. Therefore, we explicitly compute $D_{KL}(p_1\|q_1)$ by decomposing the domain to demonstrate the $o(\mathrm dt)$ bound.

To this end, we introduce a cutoff radius $L(\mathrm dt)\coloneq \sqrt{(5/2)\ln(1/\mathrm dt)}$ for the centered increment variable $\bm z_c$, distinct from the spatial truncation $B(0,R)$ for $\bm X_m$. Thus, there are two cutoffs in the proof: $R(\mathrm dt)$ truncates the position variable $\bm x$, whereas $L(\mathrm dt)$ truncates the centered increment $\bm z_c$. 
We further assume that $\epsilon$ in~\eqref{seq:R_dt} is chosen sufficiently small ($<0.1$). For notational brevity, let $B_R$ and $B_L$ denote $B(0,R)$ and $B(0,L)$, respectively. The KL divergence can then be decomposed as:
\begin{equation}
\begin{aligned}
    D_{\mathrm{KL}}\big(p_1(\bm z|\bm x)\|q_1(\bm z|\bm x)\big)
    &=\int_{\mathbb R^d}p_1(\bm z|\bm x)\ln\frac{p_1(\bm z|\bm x)}{q_1(\bm z|\bm x)}\mathrm d\bm z
    =\int_{\mathbb R^d}p_1(\bm z|\bm x)\ln\Bigg(1+
    \underbrace{
    c_{\mathrm dt}(\bm z_c)\,\mathrm dt+\frac{\mathrm{Rem}_{\mathrm dt}(\bm z_c)}{q_1(\bm z|\bm x)}
    }_{u(\bm z_c;\bm x)}
    \Bigg)\mathrm d\bm z_c
    \\
    &=
    \underbrace{
    \int_{B_L}\!\!p_1(\bm z|\bm x)\ln(1+u)\mathrm d\bm z_c
    }_{A_L}
    +
    \underbrace{
    \int_{(B_L)^c}\!\!p_1(\bm z|\bm x)\ln(1+u)\mathrm d\bm z_c
    }_{A_{L^c}}.
    \label{seq:A_D_KL_integral}
\end{aligned}
\end{equation}
In the first line, the change of variables $\bm z\mapsto\bm z_c=\bm z-\sqrt{\mathrm dt}\,\bm s(\bm x)$ was employed.

We first examine the term $A_L\coloneq\int_{B_L}q_1(\bm z|\bm x)(1+u)\ln(1+u)\mathrm d\bm z_c$. In the region $\bm z_c\in B_L$, since $c_{\mathrm dt}(\cdot;\bm x)$ is a polynomial in $\bm z_c$ 
with coefficients uniformly bounded for $\bm x\in B_R$, it follows that $|c_{\mathrm dt}\,\mathrm dt|\rightarrow0$ as $\mathrm dt\rightarrow0$ uniformly for $(\bm z_c,\bm x)\in B_L\times B_R$. Furthermore, using the bound~\eqref{seq:rem_dt_Rdt_choosen} and the definition $\phi(\bm z_c)=(2\pi)^{-d/2}e^{-\|\bm z_c\|^2/2}$, we obtain
\begin{equation}
\begin{aligned}
    \sup_{(\bm z_c,\bm x)\in B_L\times B_R}
    \left|\frac{\mathrm{Rem}_{\mathrm dt}(\bm z_c,\bm x;R)}{\phi(\bm z_c)}\right|
    &\leq
    \sup_{(\bm z_c,\bm x)\in B_L\times B_R}
    C\,\mathrm dt^{\frac{3}{2}-\epsilon}(1+\|\bm z_c\|^m)
    \exp\!\left(\!\left(\frac{1}{2}-\gamma\right)\|\bm z_c\|^2\right)
    \\
    &\leq
    C\,\mathrm dt^{\frac{3}{2}-\epsilon}\,
    \text{polylog}(1/\mathrm dt)
    \exp\!\left(-\frac{5}{2}\left(\frac{1}{2}-\gamma\right)\!\!_+ \ln\mathrm dt\right)
    \\
    &\leq C\,\mathrm dt^{\frac{1}{4}-\epsilon}\,
    \text{polylog}(1/\mathrm dt)
    \rightarrow0
    \text{ as }\mathrm dt\rightarrow 0.
\end{aligned}
\label{seq:rem_over_q_bound_compact_RL}
\end{equation}
This implies that $u(\bm z_c)\rightarrow0$ uniformly in $A_L$ as $\mathrm dt\rightarrow0$. For notational clarity and to distinguish the conditional density from joint or marginal forms, we hereafter denote $q_1(\bm z|\bm x)$ as $q_{1|\bm x}$. By applying the Taylor expansion of $\ln(1+u)$ with respect to $u$, we obtain the following bound for sufficiently small $\mathrm dt$:
\begin{equation}
    |A_L|
    =
    \left|\int_{B_L}q_{1|\bm x}(1+u)\ln(1+u)\mathrm d\bm z_c\right|
    \leq
    \left|\int_{B_L}q_{1|\bm x}u\,\mathrm d\bm z_c\right|
    +O\!\left(
    \int_{B_L}q_{1|\bm x}u^2\mathrm d\bm z_c\right)
    \label{seq:A_L_expansion}.
\end{equation}

Since the probability densities integrate to unity, $\int_{\mathbb R^d} p_1(\cdot|\bm x)=\int_{\mathbb R^d} q_1(\cdot|\bm x)=1$, and noting that $p_1(\cdot|\bm x)=q_{1|\bm x}+q_{1|\bm x}u$, we have the identity $\int_{\mathbb R^d} q_{1|\bm x}u=0$. Consequently, the first term on the RHS of~\eqref{seq:A_L_expansion} can be rewritten as an integral over the complement set:
\begin{equation}
    \left|\int_{B_L}q_{1|\bm x}u\,\mathrm d\bm z_c\right|
    =\left|\int_{(B_L)^c}q_{1|\bm x}u\,\mathrm d\bm z_c\right|
    \leq\mathrm dt\int_{(B_L)^c}q_{1|\bm x}|c_{\mathrm dt}|\mathrm d\bm z_c
    +\int_{(B_L)^c}|\mathrm{Rem}_{\mathrm dt}|\mathrm d\bm z_c.
    \label{seq:q1_u_integral_bound}
\end{equation}

For the first term of~\eqref{seq:q1_u_integral_bound}, since $c_{\mathrm dt}$ is a polynomial in $\bm z_c$ with bounded coefficients (for $\bm x\in B_R$), the Gaussian tail $\exp(-\|\bm z_c\|^2/2)$ of $q_1(\bm z|\bm x)=\phi(\bm z_c)$ in the region $\bm z_c\in (B_L)^c$ (i.e., $\|\bm z_c\|>\sqrt{2.5\ln(1/\mathrm dt)}$) ensures that the integral is of order $O(\mathrm dt^{1+(5/4)})=o(\mathrm dt)$. Similarly, for the second term, incorporating the factor $\mathrm dt^{3/2-\epsilon}$ from~\eqref{seq:rem_dt_Rdt_choosen} and the Gaussian tail decay $\exp(-\gamma\|\bm z_c\|^2)$ of $\mathrm{Rem}_{\mathrm dt}$, we find it is of order $o(\mathrm dt^{3/2-\epsilon})=o(\mathrm dt)$ (given our choice of $\epsilon<0.1$). Thus, the first term of~\eqref{seq:A_L_expansion} is $o(\mathrm dt)$.

Next, we analyze the second term of~\eqref{seq:A_L_expansion}. Using~\eqref{seq:rem_dt_Rdt_choosen} and the inequality $(a+b)^2\leq2a^2+2b^2$, we have
\begin{equation*}
    \int_{B_L}q_{1|\bm x}\,u^2\mathrm d\bm z_c
    \leq2\mathrm dt^2\!\int_{B_L}q_{1|\bm x}\,c_{\mathrm dt}^2\,\mathrm d\bm z_c
    +2\int_{B_L}\frac{\mathrm{Rem}_{\mathrm dt}^2}{q_{1|\bm x}}\,\mathrm d\bm z_c.
\end{equation*}

The first integral on the RHS is $O(\mathrm dt^2)=o(\mathrm dt)$, as moments of polynomials under a Gaussian measure are finite. For the second integral, proceeding analogously to~\eqref{seq:rem_over_q_bound_compact_RL} by utilizing the bound $\|\bm z_c\|\leq L(\mathrm dt)$, we estimate its order as $o(\mathrm dt^{3-2\epsilon-5/4})$, which is $o(\mathrm dt)$. Thus, all terms on the RHS of~\eqref{seq:A_L_expansion} vanish faster than $\mathrm dt$, leading to the conclusion $A_L=o(\mathrm dt)$.

Next, we consider the contribution from the outer region, $A_{L^c}\coloneq\int_{(B_L)^c}p_1(\cdot|\bm x)\ln(p_1(\cdot|\bm x)/q_1(\cdot|\bm x))$. Recalling that $\ln p \leq p$ for all $p > 0$, we have $p_1(\cdot|\bm x)\ln p_1(\cdot|\bm x)\leq p_1(\cdot|\bm x)^2$. Furthermore, given that $-\ln q_{1|\bm x}=\text{const}+\|\bm z_c\|^2/2$, we can combine these with the expansion~\eqref{seq:p_dt_expansion_cancel_sqrt_dt} to obtain
\begin{equation}
\begin{aligned}
    A_{L^c}&\leq
    \int_{(B_L)^c}p_1(\bm z|\bm x)^2\,\mathrm d\bm z_c
    +\int_{(B_L)^c}p_1(\bm z|\bm x)\left(
    C+\frac{\|\bm z_c\|^2}{2}\right)\mathrm d\bm z_c
    \\
    &\leq
    \begin{aligned}
    &2\underbrace{
    \int_{\|\bm z_c\|>L}q_{1|\bm x}^2(1+c_{\mathrm dt}\,\mathrm dt)^2\mathrm d\bm z_c
    }_{T_1}
    +2\underbrace{
    \int_{\|\bm z_c\|>L}\mathrm{Rem}_{\mathrm dt}^2\mathrm d\bm z_c
    }_{T_2}
    \\
    &\qquad\qquad\qquad\qquad\quad
    +C\underbrace{
    \int_{\|\bm z_c\|>L}q_{1|\bm x}(1+|c_{\mathrm dt}|\mathrm dt)
    (1+\|\bm z_c\|^2)\mathrm d\bm z_c
    }_{T_3}
    +C\underbrace{
    \int_{\|\bm z_c\|>L}|\mathrm{Rem}_{\mathrm dt}|
    (1+\|\bm z_c\|^2)\,\mathrm d\bm z_c
    }_{T_4},
    \end{aligned}
\end{aligned}
    \label{seq:A_Lc_bound}
\end{equation}
where the inequality $(a+b)^2\leq2a^2+2b^2$ is used for the first two terms.

Analyzing $T_1$, since $c_{\mathrm dt}(\bm z_c)$ is a polynomial in $\bm z_c$, the integrand takes the form $\text{poly}(\bm z_c)\exp(-\|\bm z_c\|^2)$. Integrating over the region $\|\bm z_c\|>L$, the integral scales as $\text{poly}(L)\exp(-L^2)$, which is $O(\text{polylog}(1/\mathrm dt)\,\mathrm dt^{5/2})=o(\mathrm dt)$. Similarly, $T_3$ is characterized by the form $\text{poly}(L)\exp(-L^2/2)$, yielding $O(\text{polylog}(1/\mathrm dt)\,\mathrm dt^{5/4})=o(\mathrm dt)$.

For the remainder terms, we utilize~\eqref{seq:rem_dt_Rdt_choosen}, which gives $\mathrm{Rem}_{\mathrm dt}^2\leq C \mathrm dt^{3-2\epsilon}\text{poly}(\bm z_c)\exp(-2\gamma\|\bm z_c\|^2)$. Integrating this bound leads to $T_2\leq O(\text{polylog}(1/\mathrm dt)\,\mathrm dt^{3-2\epsilon+5\gamma})=o(\mathrm dt)$. By a similar logic, we obtain $T_4\leq O(\text{polylog}(1/\mathrm dt)\,\mathrm dt^{3/2-\epsilon+5\gamma/2})=o(\mathrm dt)$ (recalling our choice of $\epsilon < 0.1$). As all terms on the RHS of~\eqref{seq:A_Lc_bound} are $o(\mathrm dt)$, it follows that $A_{L^c}=o(\mathrm dt)$. Consequently, we arrive at the final conclusion for Step 3,
\begin{equation}
    A=\mathbb E_{\bm x\in B(0,R)}[A_L+A_{L^c}]
    =\mathbb E_{\bm x\in B(0,R)}[o(\mathrm dt)]
    =o(\mathrm dt).
    \label{seq:A_odt}
\end{equation}

\noindent\textbf{\textbullet\ Step 4. Bounding the Interior Information Density Difference:} $B=o(\mathrm dt)$

Now, let us examine the term $B\coloneq|\mathbb E_{\bm x\in B_R}[\int (p_{1|\bm x}(\bm z)-q_{1|\bm x}(\bm z))\ln(q_{1|\bm x}(\bm z)/q_1(\bm z))\,\mathrm d\bm z]|$. We begin by analyzing the information density, which is given by logarithmic ratio $\ln(q_{1|\bm x}/q_1)$. Recalling that $q_{1|\bm x}(\bm z)=\phi(\bm z_c)=\phi(\bm z-\sqrt{\mathrm dt}\,\bm s(\bm x))$, the marginal density $q_1(\bm z)$ is given by the mixture:
\begin{equation}
    q_1(\bm z)=\phi(\bm z)\,
    \mathbb E_{\bm x\in B_R}\!\!\left[
    \exp\left(\bm z\cdot\big(\sqrt{\mathrm dt}\,\bm s\big)
    -\frac{\|\bm s\|^2\mathrm dt}{2}
    \right)\right].
    \label{seq:q_1_expression_by_q_1x}
\end{equation}

Here, $\bm s(\bm x)=\matbf B_m^{-1}\bm v(\bm x)$. The inverse diffusion matrix $\matbf B_m^{-1}$ is globally bounded due to the uniform ellipticity condition in Assumption~\hyperlink{assumption:A1}{A1}. If we assume bounded $\bm v$ via~\hyperlink{assumption:A5-A}{A5-A}, then $\bm s$ is globally bounded. Alternatively, under Assumption~\hyperlink{assumption:A5-B}{A5-B}, the relation $\bm v=(\bm F_I-\bm F_{rev})/2$ from~\eqref{seq:get_v}, combined with the structural form of $\bm F_{rev}$ in~\eqref{seq:def_F_rev} (involving linear combinations of $\bm F_I$, $\nabla\ln p$, $\matbf B_m$, and their derivatives), implies that $\bm s(\bm x)$ exhibits at most polynomial growth in $\bm x$. This follows because $\matbf B_m$ and its derivatives are globally bounded by~\hyperlink{assumption:A2}{A2}, while the drift and score satisfy polynomial growth bounds by~\hyperlink{assumption:A4}{A4}. Thus, strictly speaking, we have the global bounds
\begin{subequations}\label{seq:s_bound}
\renewcommand{\theequation}{\theparentequation-\Alph{equation}}
\begin{align}
    \|\bm s(\bm x)\|&\leq C_{s,A}
    \text{ under Assumption~\hyperlink{assumption:A5-A}{A5-A}},
    \label{seq:s_bound_A}
    \\
    \|\bm s(\bm x)\|&\leq C_{s,B}
    (1+\|\bm x\|^m)   
    \text{ under Assumption~\hyperlink{assumption:A5-B}{A5-B}}.
    \label{seq:s_bound_B}
\end{align}
\end{subequations}

Under~\hyperlink{assumption:A5-A}{A5-A}, $\bm s(\bm x)$ is globally bounded. Under~\hyperlink{assumption:A5-B}{A5-B}, by choosing $R(\mathrm dt)$ as in~\eqref{seq:R_dt_B}, for $\bm x\in B_R$ (i.e., $\|\bm x\|\leq (\ln\mathrm (1/dt))^\beta$), the term $(1+\|\bm x\|^m)$ scales as $\text{polylog}(1/\mathrm dt)$. Consequently, regardless of the specific assumption (\hyperlink{assumption:A5-A}{A5-A} or~\hyperlink{assumption:A5-B}{A5-B}), we can uniformly bound $\|\bm s(\bm x)\|\leq C_s\,\text{polylog}(1/\mathrm dt)$ within the region $B_R$. This implies $\|\sqrt{\mathrm dt}\,\bm s(\bm x)\|\leq C_s\sqrt{\mathrm dt}\ \text{polylog}(1/\mathrm dt)$. Therefore, for any arbitrary $\epsilon>0$, there exists $C_{s,\epsilon},\mathrm dt_0(\epsilon)>0$ such that
\begin{equation}
    \|\sqrt{\mathrm dt}\,\bm s\|\leq \mathrm C_{s,\epsilon}\,dt^{1/2-\epsilon}
    \quad\forall\mathrm dt<\mathrm dt_0(\epsilon).
    \label{seq:sqrt_dt_s_bound}
\end{equation}

In the remainder of Step 4, we assume $\mathrm dt$ is sufficiently small to satisfy~\eqref{seq:sqrt_dt_s_bound}. Using the inequalities $\big|\bm z\cdot(\sqrt{\mathrm dt}\,\bm s)\big|\leq\|\bm z\|\|\sqrt{\mathrm dt}\bm s\|\leq C_{s,\epsilon}\|\bm z\|\mathrm dt^{1/2-\epsilon}$ and $\|\bm s\|^2\mathrm dt\leq C_{s,\epsilon}^2\,\mathrm dt^{1-2\epsilon}$, we can derive upper and lower bounds for the expectation term in~\eqref{seq:q_1_expression_by_q_1x}:
\begin{equation}
    \phi(\bm z)\exp\big(\!-C_{s,\epsilon}\|\bm z\|dt^{1/2-\epsilon}
    -C_{s,\epsilon}^2\,dt^{1-2\epsilon}/2\big)
    \leq q_1(\bm z)
    \leq
    \phi(\bm z)\exp\big(C_{s,\epsilon}\|\bm z\|dt^{1/2-\epsilon}\big),
    \label{seq:p1_sandwitch_bound}
\end{equation}
which leads to the following bound on the log-ratio $\ln(q_1(\bm z)/\phi(\bm z))$,
\begin{equation}
    \left|\ln\frac{q_1(\bm z)}{\phi(\bm z)}\right|
    \leq
    C_{s,\epsilon}\|\bm z\|dt^{1/2-\epsilon}+
    \frac{C_{s,\epsilon}^2}{2}dt^{1-2\epsilon}.
    \label{seq:p1_z_bound}
\end{equation}

Similarly, applying the same logic to $\phi(\bm z_c)=\phi(\bm z)\exp\!\big(\bm z\cdot\bm s\sqrt{\mathrm dt}-\|\bm s\|^2\mathrm dt/2\big)$ yields an analogous bound
\begin{equation}
    \left|\ln\frac{\phi(\bm z_c)}{\phi(\bm z)}\right|
    \leq
    C_{s,\epsilon}\|\bm z\|dt^{1/2-\epsilon}+
    \frac{C_{s,\epsilon}^2}{2}dt^{1-2\epsilon}.
    \label{seq:phi_z_c_bound}
\end{equation}

Combining~\eqref{seq:p1_z_bound} and~\eqref{seq:phi_z_c_bound}, we obtain
\begin{equation}
    \left|\ln\frac{q_1(\bm z|\bm x)}{q_1(\bm z)}\right|
    \leq
    \left|\ln\frac{\phi(\bm z_c)}{\phi(\bm z)}\right|
    +\left|\ln\frac{q_1(\bm z)}{\phi(\bm z)}\right|
    \leq
    2C_{s,\epsilon}\|\bm z\|\mathrm dt^{\frac{1}{2}-\epsilon}
    +C_{s,\epsilon}^2\mathrm dt^{1-2\epsilon}
    \leq2C_{s,\epsilon}\|\bm z_c\|\mathrm dt^{\frac{1}{2}-\epsilon}
    +3C_{s,\epsilon}^2\mathrm dt^{1-2\epsilon},
    \label{seq:ln_q1x_q1_bound}
\end{equation}
where the last inequality utilizes $\bm z = \bm z_c + \sqrt{\mathrm dt}\,\bm s$ and the bound~\eqref{seq:sqrt_dt_s_bound}. 

Finally, combining this estimate with the density expansion~\eqref{seq:p_dt_expansion_cancel_sqrt_dt} and the remainder bound~\eqref{seq:rem_dt_Rdt_choosen}, we obtain the bound of the integral for $B$:
\begin{equation}
    \begin{aligned}
    \int|p_1(\bm z|\bm x)-q_1(\bm z|\bm x)|
    \left|\ln\frac{q_1(\bm z|\bm x)}{q_1(\bm z)}\right|
    \mathrm d\bm z_c
    \leq
    \int\left(\mathrm dt\,|c_{\mathrm dt}|\,\phi(\bm z_c)
    +C\,\mathrm dt^{\frac{3}{2}-\epsilon}(1+\|\bm z_c\|^m)e^{-\gamma\|\bm z_c\|^2}\right)
    \left|\ln\frac{q_1(\bm z|\bm x)}{q_1(\bm z)}\right|
    \\
    \leq
    \begin{aligned}
        &\quad C_1\,\mathrm dt^{\frac{3}{2}-\epsilon}
        \underbrace{
        \int |c_{\mathrm dt}|\|\bm z_c\|\phi(\bm z_c)\mathrm d\bm z_c}_{B_1}
        +
        C_2\,\mathrm dt^{2-2\epsilon}
        \underbrace{
        \int |c_{\mathrm dt}|\,\phi(\bm z_c)\mathrm d\bm z_c}_{B_2}
        \\
        &\qquad 
        +C_3\,\mathrm dt^{2-2\epsilon}\underbrace{
        \int(1+\|\bm z_c\|^{m+1})e^{-\gamma\|\bm z_c\|^2}\mathrm d\bm z_c}_{B_3}
        +C_4\,\mathrm dt^{\frac{5}{2}-3\epsilon}
        \underbrace{\int(1+\|\bm z_c\|^m)e^{-\gamma\|\bm z_c\|^2}\mathrm d\bm z_c}_{B_4}.
    \end{aligned}
    \end{aligned}
    \label{seq:b_bound}
\end{equation}
Since $c_{\mathrm dt}$ is a polynomial in $\bm z_c$ with coefficients bounded for $\bm x\in B_R$, the integrals $B_1$ and $B_2$ are finite due to the finiteness of Gaussian moments. Similarly, $B_3$ and $B_4$ are finite as they involve integrals of polynomials against Gaussian tails $\exp(-\gamma\|\bm z_c\|^2)$. Consequently, the RHS of~\eqref{seq:b_bound} is dominated by the leading order term $O(\mathrm dt^{3/2-\epsilon})$, which is $o(\mathrm dt)$ when taking small $\epsilon$. This yields the conclusion for $B$,
\begin{equation}
    B=\big|\mathbb E_{\bm x\in B_R}[o(\mathrm dt)]\big|=o(\mathrm dt).
    \label{seq:B_odt}
\end{equation}

\noindent\textbf{\textbullet\ Step 5. Bounding the Exterior Information:} $K=o(\mathrm dt)$

We now turn our attention to the term $K\coloneq \mathbb P(S=0)\big[I(\bm\eta_t^G;\bm X_m|S=0)+\frac{1}{2}\ln\!\big((2\pi e)^d\det\mathrm{Cov}(\bm\eta_t|S=0)\big)\big]$, where $\mathrm{Cov}(\bm\eta_t|S=0)$ denotes the conditional covariance matrix of $\bm\eta_t$ given $S=0$. Conditioned on $S=0$, $\bm\eta_t^G$ represents the output of a Gaussian channel with input $\sqrt{\mathrm dt}\,\bm s(\bm X_m)|S=0$ (as defined in~\eqref{seq:Gaussian_channel_proxy}). Therefore, invoking the standard capacity bound for Gaussian channels yields:
\begin{equation}
    I(\bm\eta_t^G;\bm X_m|S=0)
    \leq
    \frac{1}{2}\ln\det\!\big(\mathbf I_d
    +\mathrm dt\,\mathrm{Cov}(\bm s(\bm X_m)|S=0)\big)
    \leq
    \frac{\mathrm dt}{2}\mathbb E\!\left[
    \|\bm s(\bm X_m)\|^2|S=0\right].
    \label{seq:I_gauss_out_bound}
\end{equation}
Here, the final inequality follows from the relation $\ln\det(\mathbf I_d+\matbf A)\leq\mathrm{Tr}(\matbf A)$ for any positive semi-definite matrix $\matbf A$.

Multiplying both sides of~\eqref{seq:I_gauss_out_bound} by $\mathbb P(S=0)$, we obtain
\begin{equation}
    \mathbb P(S=0)I(\bm\eta_t^G;\bm X_m|S=0)
    \leq
    \frac{\mathrm dt}{2}\mathbb P(S=0)\mathbb E[\|\bm s(\bm X_m)\|^2|S=0]
    =\frac{\mathrm dt}{2}\mathbb E[\|\bm s(\bm X_m)\|^2\bm 1_{S=0}],
    \label{seq:P(S=0)I_G_bound}
\end{equation}
where $\bm 1_{S=0}$ denotes the indicator function. Applying the Cauchy-Schwarz inequality to the RHS of~\eqref{seq:P(S=0)I_G_bound} gives
\begin{equation}
    \mathbb E[\|\bm s\|^2\bm 1_{S=0}]\leq\mathbb E[\|\bm s\|^4]^{\frac{1}{2}}\mathbb E[\bm 1_{S=0}]^{\frac{1}{2}}
    =\mathbb E[\|\bm s\|^4]^{\frac{1}{2}}\mathbb P(S=0)^{\frac{1}{2}}.
    \label{seq:E_s^2_1_CS_bound}
\end{equation}

As discussed in Step 4, $\bm s(\bm x)$ exhibits at most polynomial growth in $\bm x$; combined with the moment assumption~\hyperlink{assumption:A5}{A5}, this ensures $\mathbb E[\|\bm s\|^4]<\infty$. Recalling from~\eqref{seq:P0_bound_uniform} that $\mathbb P(S=0)^{1/2}=o(\mathrm dt)$, the estimates~\eqref{seq:P(S=0)I_G_bound}--\eqref{seq:E_s^2_1_CS_bound} imply $\mathbb P(S=0)I(\bm\eta_t^G;\bm X_m|S=0)=o(\mathrm dt)$.

Next, to bound $G_0\coloneq\frac{1}{2}\ln\!\big((2\pi e)^d\det\mathrm{Cov}(\bm\eta_t|S=0)\big)$, we invoke the inequality $\det\matbf A\leq(\mathrm{Tr}\matbf A/d)^d$ for positive semi-definite matrices. This yields:
\begin{equation}
    G_0\leq
    C+\frac{d}{2}\ln\mathrm{Tr}[\mathrm{Cov}(\bm\eta_t|S=0)]
    \leq
    C+\frac{d}{2}\ln\mathbb E[\|\bm\eta_t\|^2|S=0]
    \leq
    C+\frac{d}{2}\ln\frac{\mathbb E[\|\bm\eta_t\|^2\bm1_{S=0}]}{\mathbb P(S=0)}
    \leq
    C_1+C_2\ln\frac{1}{\mathbb P(S=0)}.
    \label{seq:G_0_bound}
\end{equation}

The last inequality utilizes $\mathbb E[\|\bm\eta_t\|^2\bm1_{S=0}]\leq\mathbb E[\|\bm\eta_t\|^2]$ and the finiteness of the second moment $\mathbb E[\|\bm\eta_t\|^2]<\infty$, which follows from the Itô-Taylor expansion~\eqref{seq:eta_ito_taylor}. Multiplying~\eqref{seq:G_0_bound} by $\mathbb P(S=0)$ and applying the decay rates established in~\eqref{seq:P0_bound_uniform} and~\eqref{seq:P0lnP0_odt}, we find $\mathbb P(S=0)G_0 \leq C_1\mathbb P(S=0)+C_2\mathbb P(S=0)\ln(1/\mathbb P(S=0))=o(\mathrm dt)$. Consequently, we arrive at the desired conclusion:
\begin{equation}
    K=\mathbb P(S=0)(I(\bm\eta_t^G;\bm X_m|S=0)+G_0)=o(\mathrm dt).
    \label{seq:K_odt}
\end{equation}

\bigskip
\noindent\textbf{\textbullet\ Step 6. Bounding the Infinite Shell sum:} $\sum_{k=1}^\infty S_k=o(\mathrm dt)$ \textbf{and Concluding} $\Delta I=o(\mathrm dt)$

To analyze the total sum $\sum_k S_k$, we first consider the individual terms defined by $S_k\coloneq\mathbb P(\bm X_m\in A_k)\times\sup_{\bm x\in A_k}\!\ln\|p_0(\cdot|\bm x)\|_\infty$. The event $\bm X_m\in A_k$ corresponds to the shell $2^{k-1}R<\|\bm X_m\|\leq 2^k R$; hence, we have the bound $\mathbb P(\bm X_m\in A_k)\leq P_k\coloneq\mathbb P(\|\bm X_m\|>2^{k-1}R)$. Applying Markov's inequality analogously to~\eqref{seq:P0_bound} in Step 2, we obtain the following upper bounds:
\begin{subequations}\label{seq:Pk_bound}
\renewcommand{\theequation}{\theparentequation-\Alph{equation}}
\begin{align}
    P_k&=\mathbb P(\|\bm X_m\|^n>2^{n(k-1)}R^n)
    \leq\frac{\mathbb E[\|\bm X_m\|^n]}{2^{n(k-1)}R^n}
    =\mathbb E[\|\bm X_m\|^n]\,2^{-n(k-1)}\mathrm R^{-n}
    \quad\forall n\in\mathbb N
    \text{ under Assumption~\hyperlink{assumption:A5-A}{A5-A}},
    \label{seq:Pk_bound_A}
    \\
    P_k&=\mathbb P(e^{\theta\|\bm X_m\|^{1+\alpha}}\!>e^{\theta (2^{k-1}R)^{1+\alpha}})
    \leq\frac{\mathbb E[e^{\theta\|\bm X_m\|^{1+\alpha}}]}{e^{\theta (2^{k-1}R)^{1+\alpha}}}
    =\mathbb E\!\left[e^{\theta\|\bm X_m\|^{1+\alpha}}\right]
    \!e^{-\theta (2^{k-1}R)^{1+\alpha}}
    \text{ under Assumption~\hyperlink{assumption:A5-B}{A5-B}}.
    \label{seq:Pk_bound_B}
\end{align}
\end{subequations}
Assumption~\hyperlink{assumption:A5}{A5} ensures that each RHS of~\eqref{seq:Pk_bound} remains finite.

Next, we examine the term $\sup_{\bm x\in A_k}\ln\|p_0(\cdot|\bm x)\|_\infty$. First, utilizing the expansion~\eqref{seq:p_dt_expansion_cancel_sqrt_dt}, for $\bm X_m\in A_k$, we can express the density as $p_0(\bm z|\bm x)=q_0(\bm z|\bm x)(1+c_{\mathrm dt}(\bm z_c;\bm x)\,\mathrm dt)+\mathrm{Rem}_{\mathrm dt}(\bm z_c;\bm x,R_k)$. Noting that $q_0(\bm z|\bm x)=\phi(\bm z_c)$, we obtain
\begin{equation}
    \sup_{\bm x\in A_k}
    \|p_0(\cdot|\bm x)\|_\infty
    \leq
    \sup_{\bm x\in A_k}
    \|\phi(\cdot)\|_\infty
    +\sup_{\bm x\in A_k}
    \|\phi(\cdot)\,c_{\mathrm dt}(\cdot;\bm x)\|_\infty\mathrm dt
    +\sup_{\bm x\in A_k}
    \|\mathrm{Rem}_{\mathrm dt}(\cdot;\bm x,R_k)\|_\infty.
    \label{seq:p0_sup_in_Ak}
\end{equation}

Here, the first term on the RHS of~\eqref{seq:p0_sup_in_Ak} is the supremum of the Gaussian density $\phi(\bm z_c)$, which is bounded by 1. The second term involves $c_{\mathrm dt}(\bm z_c;\bm x)$, which is a polynomial in $\bm z_c$ with coefficients growing at most polynomially in $\bm x$. Since this polynomial is multiplied by the Gaussian decay, the term $\sup_{\bm z_c}\|\phi(\cdot)\,c_{\mathrm dt}(\cdot;\bm x)\|_\infty$ can be bounded by $C(1+\|\bm x\|^r)$ for some finite global constant $C$. Given that $\sup_{\bm x\in A_k}\!\|\bm x\|=2^k R$, the second term is bounded by $C(1+(2^kR)^r)\mathrm dt$. Thus, \eqref{seq:p0_sup_in_Ak} can be rewritten as
\begin{equation}
    \sup_{\bm x\in A_k}
    \|p_0(\cdot|\bm x)\|_\infty
    \leq
    1
    +C\big(1+(2^kR)^r\big)\mathrm dt
    +\sup_{\bm x\in A_k}
    \|\mathrm{Rem}_{\mathrm dt}(\cdot;\bm x,R_k)\|_\infty.
    \label{seq:p0_sup_in_Ak_second}
\end{equation}

The remaining term $\sup_{A_k}\|\mathrm{Rem}_{\mathrm dt}\|_\infty$ is bounded using~\eqref{seq:rem_dt}. Using the fact that $\|\text{poly}(\bm z_c)\exp(-\gamma\|\bm z_c\|^2)\|_\infty<\infty$ for $\gamma>0$ and noting that the truncation radius applied here is $R_k=2^kR$ (note that we cannot use~\eqref{seq:rem_dt_Rdt_choosen} uniformly in $k$ since $R_k=2^kR(\mathrm dt)$ can be arbitrary large for fixed $\mathrm dt$), there exists a finite uniform constant $C_{\mathrm{Rem}}$ such that
\begin{subequations}\label{seq:rem_dt_bound_Ak}
\renewcommand{\theequation}{\theparentequation-\Alph{equation}}
\begin{align}
  \|\mathrm{Rem}_{\mathrm dt}\|_\infty&
  \leq C_{\mathrm{Rem}}\,\mathrm dt^{3/2}
  \big(1+(2^kR)^{a_1}\big)
  \quad\text{under Assumption~\hyperlink{assumption:A5-A}{A5-A}}
  \label{seq:rem_dt_A_Ak},
  \\
  \|\mathrm{Rem}_{\mathrm dt}\|_\infty
  &\leq C_{\mathrm{Rem}}
  \Big(\mathrm dt^{3/2}+\big(1+(2^kR)^{b_2}\big)\Big)
  \exp\!\Big(c_B^{(2)}\big(1+(2^kR)^{b_3}\big)\mathrm dt\Big)
  \big(1+(2^kR)^{b_4}\big)
  \quad\text{under~\hyperlink{assumption:A5-B}{A5-B}},
  \label{seq:rem_dt_B_Ak}
\end{align}
\end{subequations}
where the truncation term of the form $\exp(-(\cdot)_+^2/\mathrm dt)\leq1$ is omitted from the product for simplicity.

For convenience, we choose $\mathrm dt<1$ sufficiently small such that $R(\mathrm dt)>1$ hereafter. Combining the results above into~\eqref{seq:p0_sup_in_Ak_second}, there exist a uniform constant $C_\infty$ and exponents $r_A, r_B > 0$ such that
\begin{subequations}\label{seq:p0_sup_in_Ak_simple}
\renewcommand{\theequation}{\theparentequation-\Alph{equation}}
\begin{align}
    \sup_{\bm x\in A_k}
    \|p_0(\cdot|\bm x)\|_\infty
    &\leq
    C_\infty\,2^{r_A k}R^{r_A}
    \quad\text{under Assumption~\hyperlink{assumption:A5-A}{A5-A}}
    \label{seq:p0_sup_in_Ak_simple_A},
    \\
    \sup_{\bm x\in A_k}
    \|p_0(\cdot|\bm x)\|_\infty
    &\leq
    C_\infty\exp\!\big(2^{r_Bk}R^{r_B}\big)
    \quad\text{under Assumption~\hyperlink{assumption:A5-B}{A5-B}},
  \label{seq:p0_sup_in_Ak_simple_B}
\end{align}
\end{subequations}
where we replace $\mathrm dt$ by 1 using $\mathrm dt\leq 1$ to simplify the upper bounds.

~\eqref{seq:p0_sup_in_Ak_simple} yield the bound of each $S_k$ with uniform constant $C_A,C_B>0$,
\begin{subequations}\label{seq:Sk_bound}
\renewcommand{\theequation}{\theparentequation-\Alph{equation}}
\begin{align}
    S_k=\mathbb P(\bm X_m\in A_k)\sup_{\bm x\in A_k}\ln\|p_0(\cdot|\bm x)\|_\infty
    &\leq
    C_A\,2^{-n(k-1)}\mathrm R^{-n}
    (r_Ak\ln2+r_A\ln R)
    \quad\text{under Assumption~\hyperlink{assumption:A5-A}{A5-A}}
    \label{seq:Sk_bound_A},
    \\
    S_k=\mathbb P(\bm X_m\in A_k)\sup_{\bm x\in A_k}\ln\|p_0(\cdot|\bm x)\|_\infty
    &\leq
    C_B\,e^{-\theta (2^{k-1}R)^{1+\alpha}}
    2^{r_Bk}R^{r_B}
    \quad\text{under Assumption~\hyperlink{assumption:A5-B}{A5-B}}.
  \label{seq:Sk_bound_B}
\end{align}
\end{subequations}

We now proceed to evaluate the infinite sum of these terms for a fixed $R(\mathrm dt)$. First, in the case of~\eqref{seq:Sk_bound_A}, the exponential decay of $P_k$ with respect to $k$ completely dominates the polynomial growth of log-supnorm term $\sup_{A_k}\|\ln p_0(\cdot|\bm x)\|_\infty$.
Indeed, letting $r\coloneq2^{-n}<1$ and applying the standard geometric series formulas $\sum_{j=1}^\infty r^{j-1}=\frac{1}{1-r}$ and $\sum_{j=1}^\infty j\,r^{j-1}=\frac{1}{(1-r)^2}$,
we can simplify the upper bound of the infinite series. Substituting the choice $R(\mathrm dt)=\mathrm dt^{-\kappa}$ from~\eqref{seq:R_dt_A} yields
\begin{equation}
    \sum_{k=1}^\infty S_k
    \leq
    C_Ar_A R^{-n}\left(
    \frac{\ln R}{1-2^{-n}}
    +\frac{\ln 2}{(1-2^{-n})^2}\right)
    \leq
    4C_Ar_AR^{-n}\ln(2R)
    =4C_Ar_A\mathrm dt^{\kappa n}
    \ln\frac{2}{\mathrm dt^\kappa}
    \quad\text{under~\hyperlink{assumption:A5-A}{A5-A}}.
    \label{seq:sum_Sk_bound_A}
\end{equation}

By choosing an integer $n\in\mathbb N$ such that $\kappa n>1$ (this is always possible since all moments are finite under~\hyperlink{assumption:A5-A}{A5-A}), we ensure that $\sum_{k=1}^\infty S_k=o(\mathrm dt)$.

For the case of~\eqref{seq:Sk_bound_B}, although the log-supnorm term exhibits exponential growth with respect to $k$, this growth is dominated by the double-exponential decay of $P_k$. To demonstrate this rigorously, let $T_k$ denote the upper bound of $S_k$ given in the RHS of~\eqref{seq:Sk_bound_B}. Clearly, the inequality $\sum_k S_k\leq\sum_k T_k$ holds. To bound the sum $\sum_k T_k$, we evaluate the ratio of consecutive terms $\rho(k)\coloneq T_{k+1}/T_k$:
\begin{equation}
    \rho(k)=
    \frac{T_{k+1}}{T_k}
    =\frac{C_B\,e^{-\theta (2^{k}R)^{1+\alpha}}
    2^{r_B(k+1)}R^{r_B}}
    {C_B\,e^{-\theta (2^{k-1}R)^{1+\alpha}}
    2^{r_Bk}R^{r_B}}
    =2^{r_B}\exp\!\left(
    -\theta(2^{1+\alpha}-1)R^{1+\alpha}2^{(k-1)(1+\alpha)}
    \right).
    \label{seq:rho_k}
\end{equation}

Since $\theta, a, R > 0$, the ratio $\rho(k)$ is strictly decreasing in $k$, implying $\rho(k)\leq\rho(1)$ for all $k\ge 1$. Furthermore, observing that $\rho(1)\rightarrow0$ as $R\rightarrow\infty$ (which corresponds to $\mathrm dt\rightarrow0$), there exists a global threshold $\mathrm dt_0>0$ such that $\rho(k)\leq\rho(1)\leq1/2$ holds for all $k$ whenever $\mathrm dt<\mathrm dt_0$. Under this condition, the sum is bounded by a geometric series:
\begin{equation*}
    \sum_{k=1}^\infty T_k
    =T_1+\sum_{k=2}^\infty\left(\prod_{j=1}^{k-1} \rho(j)\right)T_1
    \leq T_1+\sum_{k=1}^\infty\rho(1)^kT_1
    \leq T_1+\sum_{k=1}^\infty\left(\frac{1}{2}\right)^k T_1
    =2T_1.
\end{equation*}

Substituting $R(\mathrm dt)=(\ln(1/\mathrm dt))^\beta$ with $\beta(1+\alpha)>1$ from~\eqref{seq:R_dt_B} into the expression for $T_1$, we obtain
\begin{equation}
    \sum_{k=1}^\infty S_k
    \leq\sum_{k=1}^\infty T_k\leq2T_1
    =2^{r_B+1}C_B\left(\ln\frac{1}{\mathrm dt}\right)^{\beta r_B}
    \exp\!\left(-\theta\left(\ln\frac{1}{\mathrm dt}\right)^{\beta(1+\alpha)}\right)
    =o(\mathrm dt)
    \quad\forall\mathrm dt<\mathrm dt_0
    \text{ under~\hyperlink{assumption:A5-B}{A5-B}},
    \label{seq:sum_Sk_bound_B}
\end{equation}
where the limit holds because the ratio $2T_1/\mathrm dt = 2T_1 \exp(\ln(1/\mathrm dt))$ vanishes as $\mathrm dt\rightarrow0$. Specifically, considering the exponent in terms of $x=\ln(1/\mathrm dt)$, we have $x+\beta r_B\ln x-\theta x^{\beta(1+\alpha)}\rightarrow-\infty$ as $x\rightarrow\infty$, provided that $\beta(1+\alpha)>1$.
Consequently, combining~\eqref{seq:sum_Sk_bound_A} and~\eqref{seq:sum_Sk_bound_B}, we establish the target conclusion for all cases,
\begin{equation}
    \sum_{k=1}^\infty S_k=o(\mathrm dt).
    \label{seq:sum_Sk_odt}
\end{equation}

Having established via~\eqref{seq:H(S)_odt},~\eqref{seq:A_odt},~\eqref{seq:B_odt}, and~\eqref{seq:K_odt}, together with the result~\eqref{seq:sum_Sk_odt} just derived. Thus every term on the RHS of the decomposition~\eqref{seq:delta_I_upper_bound_decomposition} is $o(\mathrm dt)$, we have successfully proven the main objective of this subsection:
\begin{equation}
    I(\bm\eta_t;\bm X_m)-I(\bm\eta_t^G;\bm X_m)=o(\mathrm dt)
    \quad\text{as }\mathrm dt\rightarrow0.
    \label{seq:deltaI_odt}
\end{equation}
Therefore the Gaussian proxy is asymptotically exact at order $\mathrm dt$ for mutual information.

\subsection{The main identity}
\vspace*{-0.6\baselineskip}

We now determine the exact leading-order term of $I(\bm\eta_t;\bm X_m)$ in powers of $\mathrm dt$. Here, the previously derived estimate~\eqref{seq:deltaI_odt} plays a pivotal role. By Eq.~\eqref{seq:Gaussian_channel_proxy}, the proxy variable $\bm\eta_t^G$ can be viewed as the output of an additive Gaussian channel with a small signal-to-noise ratio (SNR) of order $\mathrm{d}t$. In this regime, the exact leading term of the mutual information can be calculated using the I-MMSE relation established by Guo et al.~\cite{guo2005mutual}.

Specifically, Guo et al. showed that for a Gaussian channel with input $\bm S$ and output $\bm Y \coloneq \sqrt{\mathrm{snr}}\bm S + \bm N$, the derivative of the mutual information with respect to $\mathrm{snr}$ equals half of the minimum mean-squared error (MMSE)~\cite{guo2005mutual}. Since $\mathrm{MMSE}(0) = \mathrm{Tr}\,\mathrm{Cov}(\bm S)$ at $\mathrm{snr}=0$, integrating this over small $\mathrm{snr}$ yields $I(\bm Y; \bm S) = \frac{\mathrm{snr}}{2}\mathrm{Tr}\,\mathrm{Cov}(\bm S) + o(\mathrm{snr})$. Applying this to our model~\eqref{seq:Gaussian_channel_proxy}, we obtain the exact leading term:
\begin{equation}
    I(\bm{\eta}_t^G;\bm{X}_m)
    =\frac{\mathrm{d}t}{2}\mathrm{Tr}\,\mathrm{Cov}\left[
    \bm{s}(\bm{X}_m,t_m)
    \right]+o(\mathrm{d}t).
    \label{seq:I_G_I_MMSE}
\end{equation}
By the bound~\eqref{seq:deltaI_odt}, the actual information $I(\bm\eta_t;\bm X_m)$ shares this identical leading term. Here, $\bm s = \matbf B_m^{-1}\bm v$. Using the continuity assumptions for $\bm F_I$ and $\matbf B$ (\hyperlink{assumption:A2}{A2}) and for the density (\hyperlink{assumption:A3}{A3}), we have $\bm v_m = \bm v(\bm X_t, t) + o(1)$ and $\matbf B_m^{-1} = \matbf B(\bm X_t, t)^{-1} + o(1)$. Furthermore, utilizing the definition of the diffusion tensor $\matbf{D} = \frac{1}{2}\matbf{B}\matbf{B}^\intercal$, we expand the covariance term as follows:

\begin{equation}
\begin{aligned}
    I(\bm\eta_t;\bm{X}_m)
    &=\frac{\mathrm dt}{2}\mathrm{Tr}\,\mathrm{Cov}\left[
    \matbf B(\bm X_t,t)^{-1}\bm v(\bm X_t,t)\right]
    +o(\mathrm dt)
    \\
    &=\frac{\mathrm dt}{2}\left(
    \frac{1}{2}\mathbb E\!\left[
    \bm v_t^\intercal\matbf D_t^{-1}\bm v_t
    \right]
    -\frac{1}{2}\left\|\mathbb E[\matbf D_t^{-1/2}\bm v_t]\right\|^2
    \right)+o(\mathrm dt)
    \\
    &=\frac{\mathrm dt}{4}\sigma_t
    -\frac{\mathrm dt}{4}\left\|\mathbb E[\matbf D_t^{-1/2}\bm v_t]\right\|^2
    +o(\mathrm dt).
\end{aligned}
    \label{seq:I_EP_relation}
\end{equation}
where $(\cdot)_t := (\cdot)(\bm X_t, t)$. The second equality uses the identity $\mathrm{Tr}\,\mathrm{Cov}(\bm w) = \mathbb E[\|\bm w\|^2] - \|\mathbb E[\bm w]\|^2$ for a random vector $\bm w$ and the relation $\|\matbf B^{-1}\bm v\|^2 = \bm v^\intercal (\matbf B \matbf B^\intercal)^{-1} \bm v = \frac{1}{2}\bm v^\intercal \matbf D^{-1} \bm v$. The final equality substitutes the definition of the entropy production rate~\eqref{seq:ep_rate}.

Dividing by $\mathrm dt$ and taking the limit $\mathrm dt \to 0$, we establish the generalized main identity for multiplicative noise, extending Eq.~(2) of the main text:
\begin{equation}
    \sigma_t
    =4\lim_{\mathrm dt\rightarrow0}
    \frac{I(\bm\eta_t;\bm X_m)}{\mathrm dt}
    +\|\mathbb E[\matbf D_t^{-1/2}\bm v_t]\|^2.
    \label{seq:main_identity_multi}
\end{equation}

If $\matbf D_t$ is spatially constant (i.e., additive noise), then $\mathbb E[\matbf D_t^{-1/2}\bm v_t] = \matbf D_t^{-1/2}\mathbb E[\bm v_t]$, and also $I(\bm\eta_t;\bm X_m)=I(\mathrm d\bm x;\bm X_m)$ since $\bm\eta_t=\matbf B_m^{-1}\mathrm d\bm x/\sqrt{\mathrm dt}$ becomes linear transformation with coordinate-independent matrix $\frac{1}{\sqrt{\mathrm dt}}\matbf B(t_m)^{-1}$. Hence Eq.~\eqref{seq:main_identity_multi} reduces exactly to the main identity in the main text, Eq.~(2).

The rest of the SM only propagates this result to projections, subsystem decompositions, and the RBC applications.

\section{Proofs of the remaining results}
\vspace*{-0.6\baselineskip}

In this section, we prove the remaining main-text results---Eqs.~(5) and (6b)---by first establishing how the identity behaves under linear transformations and then applying that result to bipartite subsystem projections.

\subsection{Extension of the Main identity to Linear Transformations}
\vspace*{-0.6\baselineskip}

While~\eqref{seq:I_EP_relation} provides a powerful result for the full system, it is necessary to extend this framework to accommodate subsystems or their linear combinations. To this end, let $T_{\bm\eta}$ and $T_{\bm X}$ be deterministic linear transformations acting independently on $\bm\eta_t$ and $\bm X_m$, respectively. These transformations need not be identical nor invertible; in particular, they can represent projection maps onto subspaces. In this setting, we first aim to establish the following asymptotic equivalence:
\begin{equation}
    \Delta I_T\coloneq
    I\big(T_{\bm\eta}(\bm\eta_t);T_{\bm X}(\bm X_m)\big)
    -I\big(T_{\bm\eta}(\bm\eta_t^G);T_{\bm X}(\bm X_m)\big)
    =o(\mathrm dt).
    \label{seq:deltaI_T_odt}
\end{equation}

This generalizes~\eqref{seq:deltaI_odt}. The proof follows a structure almost identical to that presented in Section I.F. As in Section I.F, we introduce the same compact ball $B(0,R)$ of radius $R=R(\mathrm dt)$ defined in~\eqref{seq:R_dt}, along with the same indicator variable $S$ such that $S=1$ if $\bm X_m\in B(0,R)$ and $S=0$ otherwise. 
Now we introduce the shorthand variables
\[
    \bm U\coloneq T_{\bm\eta}(\bm\eta_t),
    \quad
    \bm U^G\coloneq T_{\bm\eta}(\bm\eta_t^G),
    \quad
    \bm W\coloneq T_{\bm X}(\bm X_m),
\]
for simplicity. Using the standard information-theoretic identity $I(A;\bm W)=I(A;S)-I(A;S|\bm W)+I(A;\bm W|S)$ for $A=\bm U, \bm U^G$, we can rearrange $\Delta I_T$ to obtain a decomposition analogous to~\eqref{seq:deltaI_first_decom},
\begin{equation}
    |\Delta I_T| \le
    \big| I(\bm U; S) - I(\bm U^G; S) 
    - I(\bm U; S|\bm W) + I(\bm U^G; S|\bm W)\big|
    + \mathbb P(S=1)|\Delta {I_T}_1| + \mathbb P(S=0)|\Delta {I_T}_0|,
    \label{seq:deltaI_first_decom_trans}
\end{equation}
where $\Delta {I_T}_k\coloneq I(\bm U;\bm W|S=k)-I(\bm U^G;\bm W|S=k)$. Furthermore, since $S$ is a binary random variable, the universal bound $I(X;S)\leq H(S)$ and $I(X;S|Y)\leq H(S|Y)\leq H(S)$ hold for any arbitrary random variables $X$ and $Y$. Consequently, the magnitude of the terms in the first four terms on RHS of~\eqref{seq:deltaI_first_decom_trans} is strictly bounded by $4H(S)$, hence $o(\mathrm dt)$ because of~\eqref{seq:H(S)_odt} in step 2 of Section I.F.

Moreover, we have $|\Delta {I_T}_0|\le I\big(\bm U;\bm W|S=0\big) + I\big(\bm U^G;\bm W|S=0\big)$. By the data processing inequality (DPI), this sum is bounded above by $I(\bm\eta_t;\bm X_m|S=0) + I(\bm\eta_t^G;\bm X_m|S=0)$. Since $\mathbb P(S=0)\big[I(\bm\eta_t;\bm X_m|S=0)+I(\bm\eta_t^G;\bm X_m|S=0)\big] = o(\mathrm dt)$ due to~\eqref{seq:delta_I_outside_bound},~\eqref{seq:K_odt}, and~\eqref{seq:sum_Sk_odt} in step 5-6 of Section I.F, the last term in~\eqref{seq:deltaI_first_decom_trans} is also $o(\mathrm dt)$. For the remaining term $|\Delta {I_T}_1|$, following the exact same decomposition logic as in~\eqref{seq:MI_decomposition_bound}, we can bound it as
\begin{equation}
    |\Delta {I_T}_1| \leq 
    \underbrace{
    D_{\mathrm{KL}}\big(p_1^T(\bm u,\bm w)\|q_1^T(\bm u,\bm w)\big) }_{A_T}
    + \underbrace{
    \left|\mathbb E_{\bm X_m\in B(0,R)}\!\!\left[
    \int_{\mathbb R^{\dim(\bm u)}}\!\mathrm d\bm u\,\big(p_1^T(\bm u|\bm w)-q_1^T(\bm u|\bm w)\big)
    \ln\frac{q_1^T(\bm u|\bm w)}{q_1^T(\bm u)}\right]\right|}
    _{B_T}.
    \label{seq:MI_decomposition_bound_trans}
\end{equation}

Here, $p_1^T(\bm u, \bm w)$ and $p_1^T(\bm u | \bm w)$ serve as shorthand notations for the joint and conditional probability densities of the transformed actual dynamics, evaluated at $T_{\bm\eta}(\bm\eta_t)=\bm u$ and $T_{\bm X}(\bm X_m)=\bm w$ given $S=1$. The symbol $q_1^T$ denotes the analogous probability densities for the transformed Gaussian proxy $T_{\bm\eta}(\bm\eta_t^G)$. 
Because the marginal laws of $\bm X_m$ (and thus $T_{\bm X}(\bm X_m)$) are identical for both processes, the term $A$ in~\eqref{seq:MI_decomposition_bound} can be rewritten as the joint relative entropy $D_{\mathrm{KL}}\big(p_1(\bm z,\bm x)\|q_1(\bm z,\bm x)\big)$. Applying the DPI to this divergence directly implies $A_T \le A$. Given that $A=o(\mathrm dt)$ was established in~\eqref{seq:A_odt}, it straightforwardly follows that $A_T=o(\mathrm dt)$. Consolidating these results into~\eqref{seq:deltaI_first_decom_trans}, we obtain
\[
    |\Delta I_T|\leq B_T+o(\mathrm dt).
\]

Therefore what we need is bounding $B_T=o(\mathrm dt)$. To this end, let us focus on the region $S=1$. The key point is that, even after applying the transformation $T$, the same argument as in Step 4 of Section I.F still applies.

If $T_{\bm\eta}=0$, then $B_T=0$ trivially. Otherwise we restrict the codomain of $T_{\bm\eta}$ to its image space $\mathrm{Im}(T_{\bm\eta})$, so that the transformed Gaussian noise has covariance $\matbf\Sigma\coloneq T_{\bm\eta}T_{\bm\eta}^\intercal$
which is nondegenerate on this space. Then we obtain 
$
    \left\|\matbf\Sigma^{-\frac{1}{2}}T_{\bm\eta}\bm y\right\|\leq\|\bm y\|
$ for all $\bm y\in\mathbb R^d$ since
$
    \| \matbf\Sigma^{-1/2}T_{\bm\eta}\bm y\|^2
    =\bm y^\intercal T_{\bm\eta}^\intercal (T_{\bm\eta}T_{\bm\eta}^\intercal)^{-1}
    T_{\bm\eta}\bm y,
$
and the matrix $T_{\bm\eta}^\intercal(T_{\bm\eta}T_{\bm\eta}^\intercal)^{-1}T_{\bm\eta}$ is an orthogonal projection.

Now let $\phi_{\Sigma}$ denote the corresponding Gaussian density. The transformed Gaussian proxy then satisfies
\begin{equation}
    \bm U^G=\sqrt{\mathrm dt}\,T_{\bm\eta}\,\bm s(\bm X_m)+\bm\xi,
    \quad\bm\xi\sim\mathcal N(0,\matbf\Sigma).
    \label{seq:U_G_gaussian}
\end{equation}

Consequently,
\begin{equation*}
    \begin{aligned}
        q_1^T(\bm u|\bm w)&=\mathbb E\!\left[
        \phi_\Sigma\left(\bm u-\sqrt{\mathrm dt}\,T_{\bm\eta}\,\bm s(\bm X_m)\right)
        \middle|
        \bm W=\bm w,S=1\right],
        \\
        q_1^T(\bm u)&=\mathbb E\!\left[
        \phi_\Sigma\left(\bm u-\sqrt{\mathrm dt}\,T_{\bm\eta}\,\bm s(\bm X_m)\right)
        \middle|
        S=1\right].
    \end{aligned}
\end{equation*}

Using the bound $\left\|\sqrt{\mathrm dt}\,\bm s(\bm x)\right\|\leq
C_{\epsilon}\mathrm dt^{\frac{1}{2}-\epsilon}$ on $B(0,R)$ from~\eqref{seq:sqrt_dt_s_bound} with $\|\matbf\Sigma^{-1/2}T_{\bm\eta}\bm y\|\leq\|\bm y\|$ for both $\bm y=\bm s(\bm x)$ and $\bm y=\bm z$, the same argument as in~\eqref{seq:p1_sandwitch_bound}--\eqref{seq:phi_z_c_bound} in Step 4 of Section I.F yields
\begin{equation}
    \left|\ln\frac{q_1^T(\bm u|\bm w)}{q_1^T(\bm u)}\right|
    \leq C_\epsilon\mathrm dt^{\frac{1}{2}-\epsilon}
    \left\|\matbf\Sigma^{-\frac{1}{2}}\bm u\right\|
    +C_\epsilon\mathrm dt^{1-2\epsilon}
    \label{seq:ln_q1Tw_q1T_bound}
\end{equation}

Next, since $\bm W=T_{\bm X}(\bm X_m)$ is a deterministic function of $\bm X_m$,
\[
    p_1^T(\bm u|\bm w)-q_1^T(\bm u|\bm w)=\mathbb E\!\left[
    \int\mathrm d\bm z\,\delta\big(\bm u-T_{\bm\eta}\bm z\big)
    \big(p_1(\bm z|\bm x)-q_1(\bm z|\bm x)\big)
    \middle|\bm W=\bm w\right].
\]

Hence, by Jensen's inequality, for any nonnegative function $g$,
\begin{equation}
    \int \mathrm d\bm u\big|p_1^T(\bm u|\bm w)-q_1^T(\bm u|\bm w)\big|g(\bm u)
    \leq\mathbb E\!\left[\int\mathrm d\bm z\big|p_1(\bm z|\bm x)-q_1(\bm z|\bm x)\big|g(T_{\bm\eta}\bm z)
    \middle|\bm W=\bm w\right].
    \label{seq:p1_T-q1_T_g(u)_bound}
\end{equation}

Now one can choose $g(\bm u)=1+\|\matbf\Sigma^{-1/2}\bm u\|$. Since $1+\|\matbf\Sigma^{-1/2}\bm u\|=1+\|\matbf\Sigma^{-1/2}T_{\bm\eta}\bm z\|\leq1+\|\bm z\|$, substituting this into the bound~\eqref{seq:p1_T-q1_T_g(u)_bound} and taking expectation over $S=1$ yields
\begin{equation}
    B_T\leq C_\epsilon\mathrm dt^{\frac{1}{2}-\epsilon}\mathbb E\!\left[
    \int\mathrm d\bm z|p_1(\bm z|\bm x)-q_1(\bm z|\bm x)|\|\bm z\|
    \right]
    +C_\epsilon\mathrm dt^{1-2\epsilon}\mathbb E\!\left[
    \int\mathrm d\bm z|p_1(\bm z|\bm x)-q_1(\bm z|\bm x)|\right].
    \label{seq:B_T_bound_first}
\end{equation}

Finally, from the density expansion~\eqref{seq:p_dt_expansion_cancel_sqrt_dt} and the remainder bound~\eqref{seq:rem_dt_Rdt_choosen}, and by the same argument as in~\eqref{seq:ln_q1x_q1_bound}--\eqref{seq:B_odt} in Step 4 of section 1.F, the integrals on the RHS of~\eqref{seq:B_T_bound_first} are all $O(\mathrm dt)$. The technical details are identical and therefore omitted. Consequently $B_T=o(\mathrm dt)$, which proves~\eqref{seq:deltaI_T_odt}.

Next, let us evaluate the term $I(\bm U^G;\bm W)$. Normalizing the proxy defined in~\eqref{seq:U_G_gaussian} yields
\[
    \hat{\bm U}^G=\sqrt{\mathrm dt}\,\hat{\bm s}(\bm X_m)+\bm N,
    \quad\bm N\sim\mathcal N(0,\mathbf I),
\]
where $\hat{\bm U}^G\coloneq\mathbf\Sigma^{-1/2}\bm U^G$ and $\hat{\bm s}\coloneq\mathbf\Sigma^{-1/2}T_{\bm\eta}\,\bm s$. 

Since the linear transformation $\bm U^G\mapsto\hat{\bm U}^G$ is strictly invertible on the image space $\mathrm{Im}(T_{\bm\eta})$, it preserves mutual information, implying $I(\bm U^G;\bm W)=I(\hat{\bm U}^G;\bm W)$. Furthermore, because the channel signal is entirely determined by $\hat{\bm s}$, the variables form a Markov chain $\bm W \to \hat{\bm s} \to \hat{\bm U}^G$.
This Markov property ensures that the conditional mutual information vanishes ($I(\hat{\bm U}^G;\bm W|\hat{\bm s})=0$), which allows us to decompose the mutual information as follows:
\[
    I(\hat{\bm U}^G;\bm W)=
    I(\hat{\bm U}^G;\hat{\bm s})-I(\hat{\bm U}^G;\hat{\bm s}|\bm W).
\]

Then, by the same small-SNR I-MMSE expansion used in~\eqref{seq:I_G_I_MMSE} of Section I.G,
\[
    I(\hat{\bm U}^G;\hat{\bm s})=
    \frac{\mathrm dt}{2}\mathrm{Tr}\,\mathrm{Cov}(\hat{\bm s})+o(\mathrm dt),
\]
and, similarly, conditioning on $W$ and averaging,
\[
    I(\hat{\bm U}^G;\hat{\bm s}|\bm W)=
    \frac{\mathrm dt}{2}\mathrm{Tr}\,
    \mathbb E[\mathrm{Cov}(\hat{\bm s}|\bm W)]+o(\mathrm dt).
\]

Combining these results with the law of total covariance $\mathrm{Cov}(X)=\mathbb E[\mathrm{Cov}(X|Y)]+\mathrm{Cov}(\mathbb E[X|Y])$, we obtain
\begin{equation}
    I(\hat{\bm U}^G;\bm W)=
    \frac{\mathrm dt}{2}\mathrm{Tr}\,\mathrm{Cov}
    (\mathbb E[\hat{\bm s}|\bm W])+o(\mathrm dt)
    =
    \frac{\mathrm dt}{2}\mathrm{Tr}\,\matbf\Sigma^{-1}\mathrm{Cov}
    (\mathbb E[T_{\bm\eta}\bm s(\bm X_m,t_m)|T_{\bm X}(\bm X_m)])+o(\mathrm dt).
    \label{seq:I_U_hat_G_W_I_MMSE}
\end{equation}

Now, to justify replacing the midpoint $(\bm X_m,t_m)$ on RHS of~\eqref{seq:I_U_hat_G_W_I_MMSE} by the initial point $(\bm X_t,t)$, let $\bm W_u:=T_{\bm X}(\bm X_u)$ and $\bm m_u(\bm w):=\mathbb E[T_{\bm\eta}\,\bm s(\bm X_u,u)\mid \bm W_u=\bm w]$ for $u\in[t,t+\mathrm dt]$.
Since $T_{\bm X}$ is linear, after a fixed linear change of coordinates adapted to $T_{\bm X}$, $\bm m_u(\bm w)$ can be written as the ratio of fiber integrals of $T_{\bm\eta}\,\bm s(\bm x,u)p(\bm x,u)$ and $p(\bm x,u)$; hence, by the same continuity/moment assumptions used in Section I.G, $\bm m_u$ is continuous in $(\bm w,u)$ with at most polynomial growth. Since $\bm W_m=\bm W_t+o_{L^2}(1)$, we obtain $\bm m_{t_m}(\bm W_m)=\bm m_t(\bm W_t)+o_{L^2}(1)$
and therefore $\mathrm{Cov}\!\big(\bm m_{t_m}(\bm W_m)\big)
=
\mathrm{Cov}\!\big(\bm m_t(\bm W_t)\big)+o(1)$.
Combining all these results yields the exact analytic form of the information rate,
\begin{equation}
    \lim_{\mathrm dt\rightarrow0}\frac{
    I\big(T_{\bm\eta}(\bm\eta_t);T_{\bm X}(\bm X_m)\big)}{\mathrm dt}
    =\frac{1}{2}\mathrm{Tr}\big[(T_{\bm\eta}T_{\bm\eta}^\intercal)^{-1}\mathrm{Cov}
    (\mathbb E[\,T_{\bm\eta}\matbf B(\bm X_t,t)^{-1} \bm v(\bm X_t,t)\mid T_{\bm X}(\bm X_t)])\big],
    \label{seq:main_identity_T_extension}
\end{equation}
where $(T_{\bm\eta}T_{\bm\eta}^\intercal)^{-1}$ is inverse on $\mathrm{Im}(T_{\bm\eta})$. The expression~\eqref{seq:main_identity_T_extension} will serve as a crucial building block in subsequent proofs.

%As established in Section I.G , Assumptions~\hyperlink{assumption:A2}{A2}--\hyperlink{assumption:A3}{A3}  guarantee that $\bm s(\bm X_m,t_m)=\bm s(\bm X_t,t)+o(1)$ and $\bm X_m=\bm X_t+o(1)$. Consequently, the time evaluation point on the right-hand side of~\eqref{seq:I_U_hat_G_W_I_MMSE} can be asymptotically shifted from the midpoint $(\bm X_m,t_m)$ to the initial point $(\bm X_t,t)$. Substituting this result back into~\eqref{seq:deltaI_T_odt}, we finally arrive at the extended main identity:

\subsection{Proof of the local EP identity (Eq. 5)}
\vspace*{-0.6\baselineskip}

We now derive the subsystem identity directly from~\eqref{seq:main_identity_T_extension}. Choosing $T_{\bm\eta}=P_A$, $T_{\bm X}=\mathbf I$,~\eqref{seq:main_identity_T_extension} gives
\[
    \lim_{\mathrm dt\rightarrow0}\frac{I(P_A\,\bm\eta_t;\bm X_m)}{\mathrm dt}
    =\frac{1}{2}\mathrm{Tr}\,\mathrm{Cov}(\matbf B_A^{-1}\bm v_A).
\]
In the additive-noise case,
\[
    P_A\,\bm\eta_t=\frac{1}{\sqrt{\mathrm dt}}\matbf B_A^{-1}\mathrm d\bm x_A,
\]
and since $\matbf B_A$ is constant, this is an invertible linear transformation of $\mathrm d\bm x_A$. Hence mutual information is preserved: $I(P_A\,\bm\eta_t;\bm X_m)=I(\mathrm d\bm x_A;\bm x_m)$.
Now using $\matbf B_A\matbf B_A^\intercal=2\matbf D_A$, we obtain
\[
    \mathrm{Tr}\,\mathrm{Cov}(\matbf B_A^{-1}\bm v_A)
    =\frac{1}{2}\Big(\langle\bm v_A^\intercal\matbf D_A^{-1}\bm v_A\rangle-\langle\bm v_A\rangle^\intercal\matbf D_A^{-1}\langle\bm v_A\rangle\Big).
\]
Therefore
\[
    \lim_{\mathrm dt\rightarrow0}\frac{I(\mathrm d\bm x_A;\bm x_m)}{\mathrm dt}
    =\frac{1}{4}\Big(\langle\bm v_A^\intercal\matbf D_A^{-1}\bm v_A\rangle-\langle\bm v_A\rangle^\intercal\matbf D_A^{-1}\langle\bm v_A\rangle\Big).
\]
Recalling $\sigma_A=\langle\bm v_A^\intercal\matbf D_A^{-1}\bm v_A\rangle$ and $\sigma_A^{\rm mf}\coloneq\langle\bm v_A\rangle^\intercal\matbf D_A^{-1}\langle\bm v_A\rangle$, we immediately obtain
\begin{equation}
    \sigma_A=4\,\mathcal I(\mathrm d\bm x_A;\bm x_m)+\sigma_A^{\rm mf},
    \label{seq:local_EP_equality}
\end{equation}
which is Eq. (5) in the main text.

\subsection{Proof of current-space expression of the EP channels (Eq. 6b)}
\vspace*{-0.6\baselineskip}

Self channel contribution is defined by $
    \sigma_A^{\rm self}\coloneq4\,\mathcal I(\mathrm d\bm x_A;\bm x_m^A)+\sigma_A^{\rm mf}$.
To evaluate it, we again use~\eqref{seq:main_identity_T_extension}, now with $T_{\bm\eta}=T_{\bm X}=P_A$.
Then
\[
    \lim_{\mathrm dt\rightarrow0}\frac{I(P_A\,\bm\eta_t;P_A\bm X_m)}{\mathrm dt}
    =\frac{1}{2}\mathrm{Tr}\,\mathrm{Cov}\big(\mathbb E[\matbf B_A^{-1}\bm v_A\mid\bm x_A]\big).
\]
Let $\tilde{\bm v}_A\coloneq\mathbb E[\bm v_A|\bm x_A]$. Since $\matbf B_A$ is constant and $P_A\bm X_m=\bm x_m^A$, the previous relation becomes
\begin{equation*}
    \begin{aligned}
        \lim_{\mathrm dt\rightarrow0}\frac{I(\mathrm d\bm x_A;\bm x_m^A)}{\mathrm dt}
        &=\frac{1}{2}\mathrm{Tr}\,\mathrm{Cov}\big(\matbf B_A^{-1}\tilde{\bm v}_A\big)
        \\
        &=\frac{1}{4}\Big(\langle\tilde{\bm v}_A^\intercal\matbf D_A^{-1}\tilde{\bm v}_A\rangle
        -\langle\tilde{\bm v}_A\rangle^\intercal\matbf D_A^{-1}\langle\tilde{\bm v}_A\rangle\Big).
    \end{aligned}
\end{equation*}
Because $\langle\tilde{\bm v}_A\rangle=\langle\bm v_A\rangle$, adding the mean-flow term gives $
    \sigma_A^{\rm self}=\langle\tilde{\bm v}_A^\intercal\matbf D_A^{-1}\tilde{\bm v}_A\rangle.$

For the interaction channel contribution, the chain rule gives
\[
    I(\mathrm d\bm x_A;\bm x_m^{-A}|\bm x_m)
    =I(\mathrm d\bm x_A;\bm x_m)-I(\mathrm d\bm x_A;\bm x_m^A),
\]
where $\bm x_m^{-A}\coloneq(I-P_A)\bm X_m$. Applying~\eqref{seq:main_identity_T_extension} to the first term with $T_{\bm\eta}=P_A$, $T_{\bm X}=I$, and to the second with $T_{\bm\eta}=T_{\bm X}=P_A$, yields
\[
    \lim_{\mathrm dt\rightarrow0}\frac{I(\mathrm d\bm x_A;\bm x_m^{-A}|\bm x_m^A)}{\mathrm dt}
    =\frac{1}{4}\mathrm{Tr}\left[\matbf D_A^{-1}\left\{\mathrm{Cov}(\bm v_A)-\mathrm{Cov}(\widetilde{\bm v}_A)\right\}\right].
\]
By the law of total covariance and $\widetilde{\bm v}_A=\mathbb E[\bm v_A|\bm x_A]$,
\[
    \mathrm{Cov}(\bm v_A)-\mathrm{Cov}(\widetilde{\bm v}_A)
    =\mathbb E[\mathrm{Cov}(\bm v_A|\bm x_A)].
\]
Hence
\[
    \sigma_{A|-A}^{\rm int}=\langle(\bm v_A-\widetilde{\bm v}_A)^\intercal\matbf D_A^{-1}(\bm v_A-\widetilde{\bm v}_A)\rangle.
\]
Together with the expression for $\sigma_A^{\rm self}$ above, this gives Eq.~(6b) of the main text.

\section{Numerical illustration of the identity}
\vspace*{-0.6\baselineskip}

In this section, we numerically illustrate the multiplicative-noise identity~\eqref{seq:main_identity_multi} derived in Sec. I using a simple one-dimensional diffusion on a ring. This example is analytically transparent while retaining genuinely state-dependent noise, allowing us to compare the small-$\mathrm dt$ mutual information slope directly with the theoretical prediction. It also provides a useful contrast between midpoint and endpoint conditioning. In Sec. I.E, we showed that under midpoint conditioning the $\sqrt{\mathrm dt}$-order non-Gaussian corrections cancel, so that the conditional law is captured by a Gaussian channel up to $O(\mathrm dt)$ corrections; no analogous simplification is expected for endpoint conditioning. We therefore focus on the midpoint-conditioned quantity $I(\bm\eta_t;\bm X_m)$, and report $I(\bm\eta_t;\bm X_t)$ only as a control illustrating why the exact identity is formulated with the midpoint rather than with an endpoint. 

\vspace*{-0.6\baselineskip}
\subsection{Model and observables}
\vspace*{-0.6\baselineskip}

We consider a one-dimensional It\^{o} diffusion on a ring $x\in[0,2\pi)$,
\begin{equation}
\begin{aligned}
    &\mathrm d X_t
    =F(X_t)\,\mathrm dt
    +\sqrt{2D(X_t)}\,\mathrm dW_t
    \qquad(\mathrm{mod}\ 2\pi),
    \\
    \text{with}\quad&
    D(x)=2+\sin x,
    \quad F(x)=D'(x)+f=\cos x+f.
\end{aligned}
\label{seq:1D_ring_model}
\end{equation}

This choice yields a smooth uniformly elliptic diffusion coefficient, $1\leq D(x)\leq3$, on a compact state space. It therefore lies within the regular setting considered in Sec. I, where compact configuration spaces such as motion on a ring are natural examples covered by the assumption~\hyperlink{assumption:A5-A}{A5-A}. Also the model is analytically convenient because its steady state can be calculated exactly. Writing the stationary current in one dimension as
\[
    j_{ss}(x)=F(x)p_{ss}(x)-\partial_x[D(x)p_{ss}(x)],
\]
one finds that the choice $F=D'+f$ gives a uniform stationary density
\[
    p_{ss}=\frac{1}{2\pi},
\]
together with a constant current
\[
j_{ss}=\frac{f}{2\pi}.
\]

Accordingly, the current velocity is spatially uniform,
\[
    v_{ss}(x)=\frac{j_{ss}(x)}{p_{ss}(x)}=f.
\]

Thus $f=0$ corresponds to equilibrium, whereas $f\neq0$ generates a nonequilibrium steady state with a nonzero circulating current on the ring.

In the multiplicative-noise setting, the natural observable is not the raw increment itself but the midpoint-whitened increment
\[
    \eta_t\coloneq\frac{1}{\sqrt{\mathrm dt}}B(X_m)^{-1}
    (X_{t+\mathrm dt}-X_t),
    \qquad B(x)=\sqrt{2D(x)}=\sqrt{4+2\sin x},
\]
introduced in Sec. I. We therefore take the midpoint-conditioned mutual information $I(\eta_t;X_m)$ as the primary quantity of interest. For comparison, we also compute $I(\eta_t;X_t)$, which serves as a control rather than as the quantity appearing in the exact identity. This comparison is conceptually useful because Sec. I.E shows that midpoint conditioning removes the $\sqrt{\mathrm dt}$-order non-Gaussian correction, whereas endpoint conditioning does not share this cancellation.

In the present model, although the current velocity itself is spatially constant, $v_{ss}(x)=f$, the effective normalized signal entering the midpoint-conditioned channel is
\[
    B(x)^{-1}v_{ss}(x)=\frac{f}{\sqrt{2D(x)}}
    =\frac{f}{\sqrt{4+2\sin x}},
\]
which remains position dependent through the diffusion amplitude. The midpoint therefore remains informative about $\eta_t$ whenever $f\neq0$. Specializing the general multiplicative-noise identity~\eqref{seq:main_identity_multi} to this steady-state model yields
\[
    \sigma=\left\langle\frac{f^2}{D(X)}\right\rangle,
    \quad4\lim_{\mathrm dt\rightarrow 0}
    \frac{I(\eta_t;X_m)}{\mathrm dt}
    =\sigma-\left\langle\frac{f}{\sqrt{D(X)}}\right\rangle^2
\]

\subsection{Simulation and mutual information estimation}
\vspace*{-0.6\baselineskip}

To generate short-time trajectory pairs, we sampled the initial point $X_t$ directly from the exact steady state of the ring model. Since $p_{ss}(x)=1/(2\pi)$, this was done by drawing $X_t$ uniformly on $[0,2\pi)$, thereby avoiding any burn-in or equilibration procedure. For each time step $\mathrm dt$, we set $h=\mathrm dt/2$ and constructed the pair $(X_m,\eta_t)$ through two successive half-steps. The midpoint variable was defined as $X_m=X_{t+h}$, reduced modulo $2\pi$, while the total increment was taken from the corresponding unwrapped trajectory so that short-time displacement remained consistent with the local SDE dynamics.

Because the present example involves state-dependent diffusion, we did not use a plain Euler-Maruyama update. Instead, each half-step was propagated with the one-dimensional Milstein scheme~\cite{milstein1975strong,kloeden1992numerical}. Writing $\sigma(x)=\sqrt{2D(x)}$, with $D(x)=2+\sin x$ and $F(x)=D'(x)+f$, the first half-step was evolved as
\[
    X_m^{un}=X_t+F(X_t)h+\sigma(X_t)\Delta W_1
    +\frac{1}{2}D'(X_t)\big((\Delta W_1)^2-h\big),
\]
followed by $X_m=X_m^{un}\mod{2\pi}$. The second half-step was then evolved from $X_m^{un}$ using the same Milstein correction, but with coefficients evaluated at the midpoint:
\[
    X_{t+\mathrm dt}^{un}=X_m^{un}+F(X_m)h+\sigma(X_m)\Delta W_2
    +\frac{1}{2}D'(X_m)\big((\Delta W_2)^2-h\big).
\]

Here $\Delta W_1$ and $\Delta W_2$ are independent Gaussian increments with variance $h$. The normalized midpoint-whitened increment was finally constructed as
\[
    \eta_t=\frac{X_{t+\mathrm dt}^{un}-X_t}{\sqrt{\mathrm dt}\,B(X_m)},
    \qquad B(x)=\sqrt{2D(x)}=\sqrt{4+2\sin x}.
\]

This choice is consistent with the diffusion-metric normalization introduced in Sec. I and, for multiplicative noise, retains the $O(\mathrm dt)$ corrections relevant to mutual information estimation.

We estimated the mutual information using a Kraskov-Stögbauer-Grassberger (KSG) $k$-nearest-neighbor (knn) estimator with $k=10$. Since $X_m$ is an angular variable on the ring, the midpoint coordinate was treated as periodic with period $2\pi$, while $\eta_t$ was treated as nonperiodic. Operationally, we used the Chebyshev norm and implemented periodicity only in the $X_m$ coordinate through a periodic cKDTree representation. To reduce the small positive bias that is typical when the true mutual information is close to zero, we subtracted a permutation baseline,
\[
    \hat I_{\mathrm{deb}}(X;Y)=\hat I_{\mathrm{raw}}(X;Y)
    -\frac{1}{n_{\mathrm{perm}}}\sum_{l=1}^{n_{\mathrm{perm}}}
    \hat I_{\mathrm{raw}}(X;\pi_l (Y)),
\]
with $n_{\mathrm{perm}}=3$. We applied the same estimator both to the midpoint-conditioned quantity $I(\eta_t;X_m)$, which is the object entering the exact identity, and to the initial point-conditioned quantity $I(\eta_t;X_t)$, which we report only as a comparison quantity.

For each driving strength $f$ and each value of $\mathrm dt$, we used $N=10^6$ independently generated samples and repeated the calculation over five random seeds. The reported error bars correspond to the standard deviation across seeds. To quantify the small-$\mathrm dt$ slope, we considered two regression forms. First, we used an unconstrained linear fit,
\[
    \hat I(\mathrm dt)\approx m_{\mathrm{lin}}\,\mathrm dt +c,
\]
mainly as a visual guide and to expose the approximately $f$-independent baseline offset associated with finite-sample and estimator fluctuations. Second, we used a constrained quadratic fit,
\[
    \hat I(\mathrm dt)\approx m_{\mathrm{quad}}\,\mathrm dt+b\,\mathrm dt^2,
\]
which enforces $I(0)=0$ and was used to estimate the leading coefficient in the $\mathrm dt\rightarrow0$ limit. In the figures below, the linear fit is shown only for comparison, whereas the coefficient $m_{\mathrm{quad}}$ is taken as the numerical estimate of the asymptotic mutual information rate.

\subsection{Numerical results}
\vspace*{-0.6\baselineskip}

\begin{figure}[H]
    \centering
    \includegraphics[width=\textwidth]{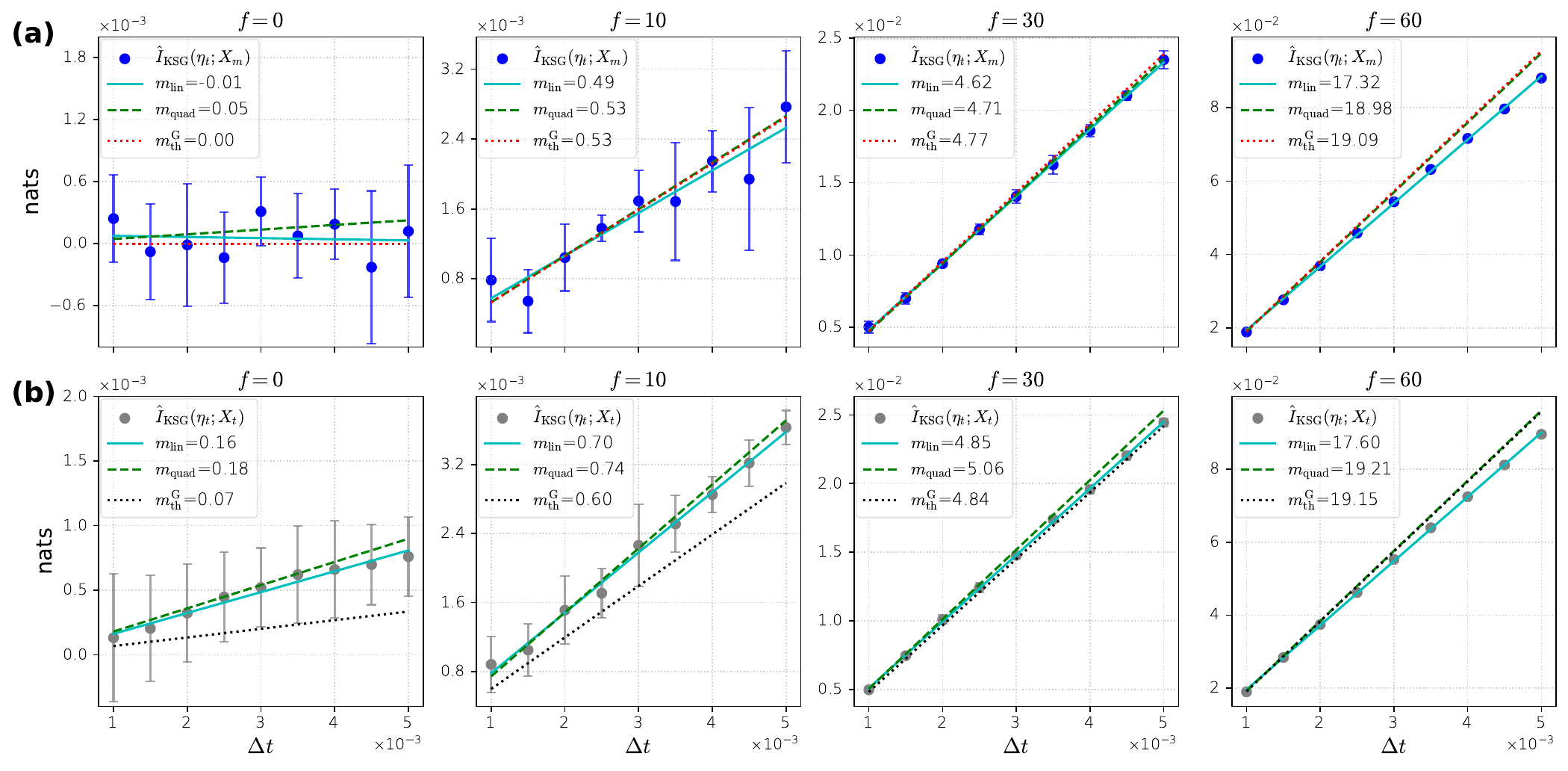}
    \vskip -0.02in
    \setlength{\abovecaptionskip}{0pt}
    \caption{Mutual information versus $\Delta t$ for the multiplicative-noise model of Sec. III.A--B. 
    \textbf{(a)} $I_{\rm KSG}(\eta_t;X_m)$ using the time midpoint $X_m$.
    \textbf{(b)} $I_{\rm KSG}(\eta_t;X_t)$ using the initial point $X_t$. Columns from left to right correspond to $f=0,10,30,60$. Error bars show the standard deviation over five independent runs. Cyan solid, green dashed, and red dotted lines denote the linear fit, the line from the first-order coefficient of the quadratic fit, and the theoretical slope from the Gaussian-channel approximation, respectively; the corresponding slope values are shown in each panel.
    }\label{sfig:multi_numerical_results}
    \vskip -0.05in
\end{figure}

The numerical results are shown in Fig.~\ref{sfig:multi_numerical_results}. Three main features are worth emphasizing. First, midpoint and initial point conditioning lead to different Gaussian-channel predictions, and the numerical data clearly reflect this distinction. Most notably, at $f=0$, where the ring diffusion is in equilibrium, the midpoint-conditioned theory predicts a vanishing slope and the measured $I_{\rm KSG}(\eta_t;X_m)$ is likewise consistent with zero within numerical uncertainty. By contrast, the initial point conditioned quantity $I_{\rm KSG}$ exhibits a positive slope already at $f=0$, both in the Gaussian approximation and in the direct numerical estimates. Second, the initial point conditioned data show that the Gaussian approximation is not uniformly reliable, especially in the weak-drive regime. For $f=0$ and $f=10$, the measured small-$\mathrm dt$ slope deviates visibly from the Gaussian-channel prediction, whereas the midpoint conditioned results remain in close agreement with the theoretical value. This is consistent with the analysis of Sec. I.E: under midpoint conditioning, the effect of the spatial variation of $D(x)$ is canceled at the relevant order, while no analogous cancellation occurs for conditioning on the initial point. Third, in the larger-drive regime, where finite-$\mathrm dt$ effects become more pronounced, the coefficient $m_{\rm quad}$ obtained from the constrained fit $I(\mathrm dt)\approx m_{\rm quad}\mathrm dt+b\,\mathrm dt^2$ provides a more faithful estimate of the asymptotic slope than a simple linear fit, most clearly for the $f=60$. This is consistent with the theoretical small-time form $I(\mathrm dt)=m\,\mathrm dt+o(\mathrm dt)$, for which the quadratic fit isolates the leading-order coefficient while partially absorbing subleading finite-time corrections.

\section{Simulation details for the main text Figs.2--3}
\vspace*{-0.6\baselineskip}

\subsection{Numerical procedure for Fig. 2}
\vspace*{-0.6\baselineskip}

The caption of Fig.~2 lists the model equations and parameter values; here we record only the additional details needed for reproduction. Panels (a,b) were not generated from trajectory data, but from the closed-form steady state of linear additive-noise model
\begin{equation}
    \dot{\bm X}=\matbf A\bm X+\sqrt{2\matbf D}\bm\xi,
    \quad
    \matbf A=
    \begin{pmatrix}
        -k & s+a \\
        s-a & -k
    \end{pmatrix},
    \quad
    \matbf D=\matbf I.
    \label{seq:linear_Langevin_fig2}
\end{equation}

For each parameter set, the stationary covariance $\matbf C$ was obtained from the Lyapunov equation
\begin{equation}
    \matbf A\matbf C+\matbf C\matbf A^\intercal+2\matbf D=\matbf 0.
    \label{seq:lyapunov_equation}
\end{equation}

The black arrows show the drift $\bm F(\bm x)=\matbf A\bm x$, the yellow arrows show the local mean velocity
\begin{equation}
    \bm v(\bm x)=\bm F(\bm x)-\matbf D\nabla\ln p_{ss}(\bm x)
    =(\matbf A+\matbf D\matbf C^{-1})\bm x,
    \label{seq:cal_v_linear}
\end{equation}
and the contours show the stationary Gaussian density $p_{ss}(\bm x)\propto\exp\left(-\frac{1}{2}\bm x^\intercal\matbf C^{-1}\bm x\right)$.
The crosses mark the midpoint locations used for the conditional displacement plots in panels (c,d).

For panels (c,d), conditional displacement clouds were generated from the same linear model~\eqref{seq:linear_Langevin_fig2} using a two-half-step Euler-Maruyama update with $\Delta t=10^{-3}$. In each case, $10^5$ trajectories were initialized at the origin and propagated for $10^4$ steps, and the first $10^3$ steps were discarded as burn-in. The midpoint $\bm X_m$ was recorded after the first half-step, and the full displacement was defined by $\Delta\bm X=\bm X_{t+\Delta t}-\bm X_t$. Conditional samples were then selected by restricting $\bm X_m$ to two square windows centered at $(-0.6,0.6)$ and $(0.6,-0.6)$, with half-width $0.02$ in each coordinate. The panels show scatter plots of the resulting conditional displacements together with their one-dimensional marginals. Dashed crosshairs indicate the sample means, solid black lines indicate the origin, and the dashed circles indicate the $1.96\sigma$ scale of each conditional cloud.

For panels (e,f), the linear model~\eqref{seq:linear_Langevin_fig2} and the nonlinear model
\begin{equation*}
    \begin{aligned}
        \dot x&=-x+y+\epsilon y^3+\sqrt 2\,\xi_x, \\
        \dot y&=-y+\sqrt 2\,\xi_y,
    \end{aligned}
\end{equation*}
with $\matbf D=\mathrm{I}$, were used. The information rate was estimated from midpoint-conditioned displacements. In the additive-noise setting,
\[
    I(\Delta\bm X;\bm X_m)=\frac{\Delta t}{4}
    \mathrm{Tr}\big(\matbf D^{-1}
    \mathrm{Cov}(\bm v(\bm X_m))\big)+o(\Delta t),
    \quad
    \bm v(\bm x_m)=\frac{\mathbb E[\Delta\bm X|\bm X_m=\bm x_m]}{\Delta t}.
\]
For the main-text panels, we therefore used the regression-based estimator
\begin{equation}
    \hat I_{\mathrm{MMSE}}(\Delta t)\coloneq\frac{\Delta t}{4}
    \mathrm{Tr}\left(\matbf D^{-1}\hat{\mathrm{Cov}}(\hat{\bm v}(\bm X_m))\right),
    \label{seq:I_hat_MMSE}
\end{equation}
where $\hat{\bm v}(\bm x_m)$ was obtained by regressing $\Delta\bm X$ on $\bm X_m$. The subscript MMSE refers to the leading-order small-SNR Gaussian/I-MMSE expression underlying this estimator~\cite{guo2005mutual}. In practice, polynomial features of the midpoint coordinates were used together with sample splitting (cross-fitting) to reduce overfitting bias. For each parameter value, $\hat I_{\mathrm{MMSE}}(\Delta t)$ was evaluated over a range of $\Delta t$ and fitted both by
\begin{equation}
    \hat I(\Delta t)\approx m_{\mathrm{lin}}\Delta t+c
    \label{seq:m_lin}
\end{equation}
and by the constrained quadratic form
\begin{equation}
    \hat I(\Delta t)\approx m_{\mathrm{quad}}\Delta t+b\Delta t^2,
    \quad \hat I(0)=0.
    \label{seq:m_quad}
\end{equation}

The main text reports $m_{\mathrm{quad}}$ as the representative estimate of the $\Delta t\rightarrow0$ slope.

As an independent cross-check, we also computed direct mutual information estimates with a debiased KSG $k$-nearest-neighbor estimator for $k=10$, denoted by $\hat I_{\mathrm{KSG}}$. These KSG calculations were not used for the main-text panels, but only to verify that the regression-based estimator $\hat{I}_{\rm MMSE}$ gives consistent small-$\Delta t$ slopes. At fixed sample size, $\hat{I}_{\rm KSG}$ exhibits substantially larger fluctuations than $\hat{I}_{\rm MMSE}$, especially in the nonlinear case, while remaining consistent with the same $\sigma/4$ trend. The corresponding summary for the linear model and comparison plots for the nonlinear model are shown in Fig.~\ref{sfig:main_lin_summary}--\ref{sfig:main_nonlin_MMSE}.

\begin{figure}[H]
    \centering
    \includegraphics[width=0.85\textwidth]{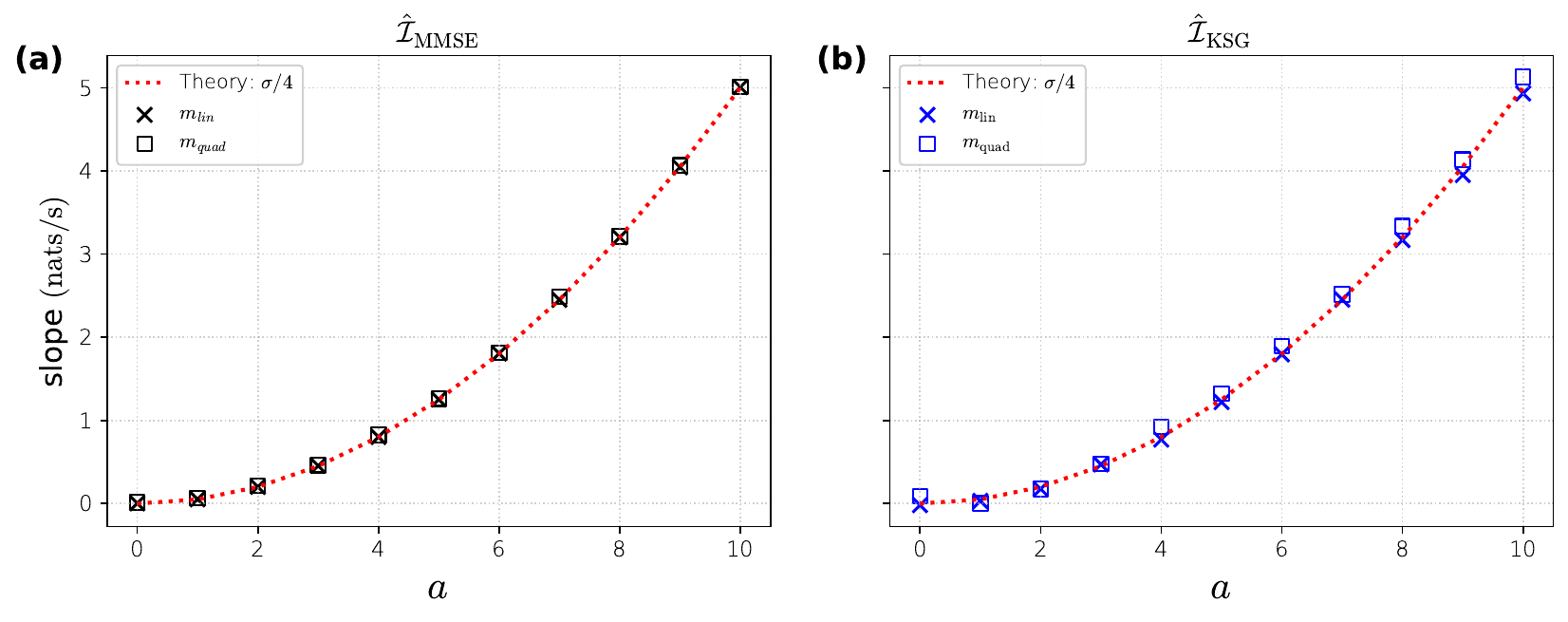}
    \vskip -0.0in
    \setlength{\abovecaptionskip}{0pt}
    \caption{Summary of slope estimates for the linear model used in Fig. 2e of the main text. Panel \textbf{(a)} shows $\hat{\mathcal I}_{\rm MMSE}$, and panel \textbf{(b)} shows $\hat{\mathcal I}_{\rm KSG}$, as a function of the nonequilibrium parameter $a$. Crosses denote the slope $m_{\rm lin}$ obtained from a linear fit in $\mathrm dt$, while squares denote the first-order coefficient $m_{\rm quad}$ extracted from a quadratic fit. The red dotted line indicates the theoretical prediction, $\sigma/4$, where $\sigma$ is the EP rate.
    }\label{sfig:main_lin_summary}
    \vskip -0.2in
\end{figure}

\begin{figure}[H]
    \centering
    \includegraphics[width=\textwidth]{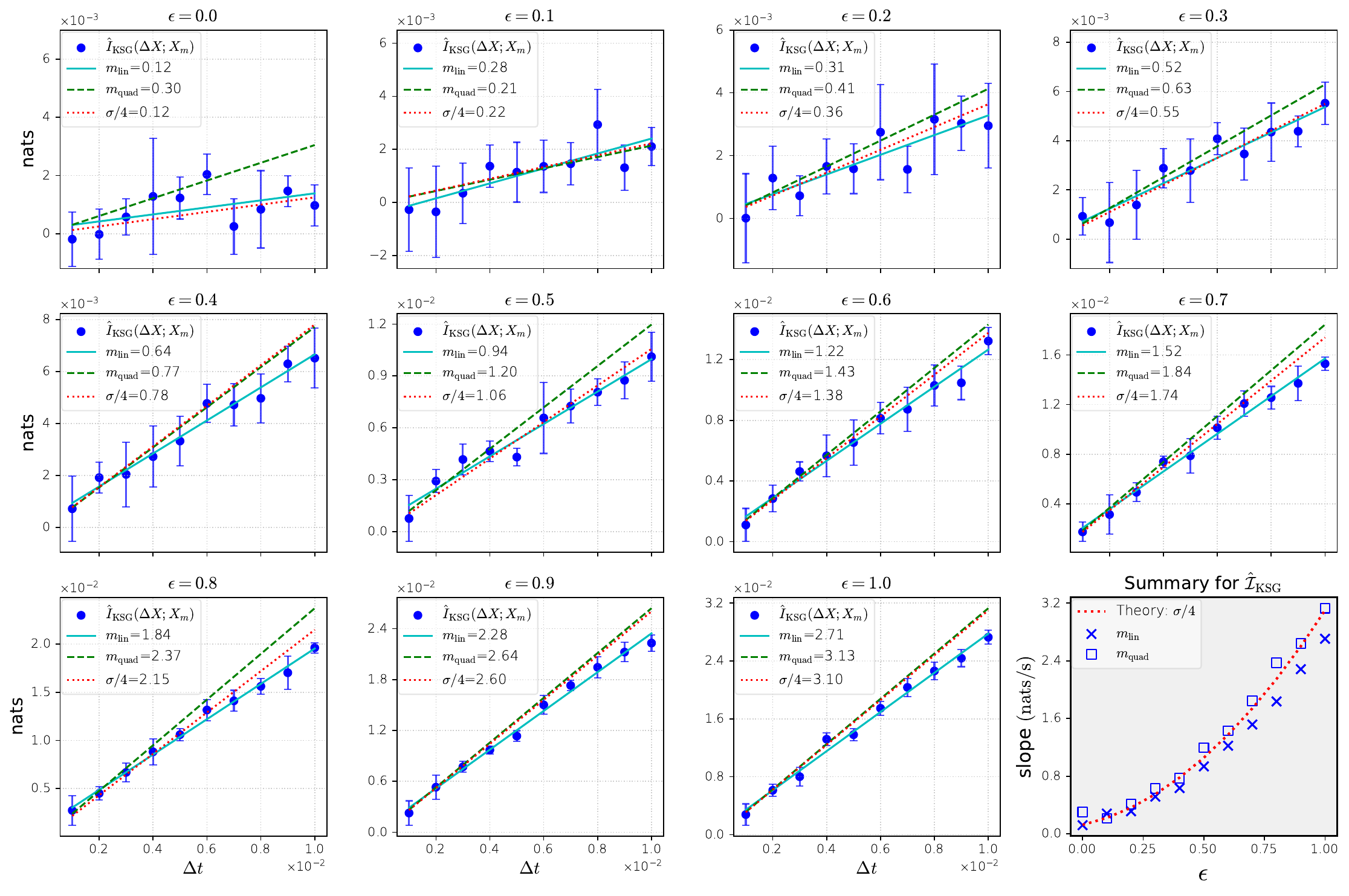}
    \vskip -0.0in
    \setlength{\abovecaptionskip}{0pt}
    \caption{Numerical results for the nonlinear model used in Fig. 2f of the main text, shown for $\hat I_{\rm KSG}(\Delta \bm X;\bm X_m)$. The first 11 panels display the mutual information as a function of $\mathrm dt$ for $\epsilon=0.0,0.1,\cdots,1.0$ and the last panel summarizes the extracted slopes over the full range of $\epsilon$. Blue circles show the mean $\hat I_{\rm KSG}$, and error bars denote the standard deviation over independent 5 runs. The cyan solid line is the linear fit, the green dashed line is the line determined by the first-order coefficient extracted from a quadratic fit in $\mathrm dt$, and the red dotted line indicates the simulated EP rate divided by four, $\sigma/4$. The corresponding slope values are listed in each panel.
    }\label{sfig:main_nonlin_KSG}
    %\vskip -0.2in
\end{figure}

\begin{figure}[H]
    \centering
    \includegraphics[width=\textwidth]{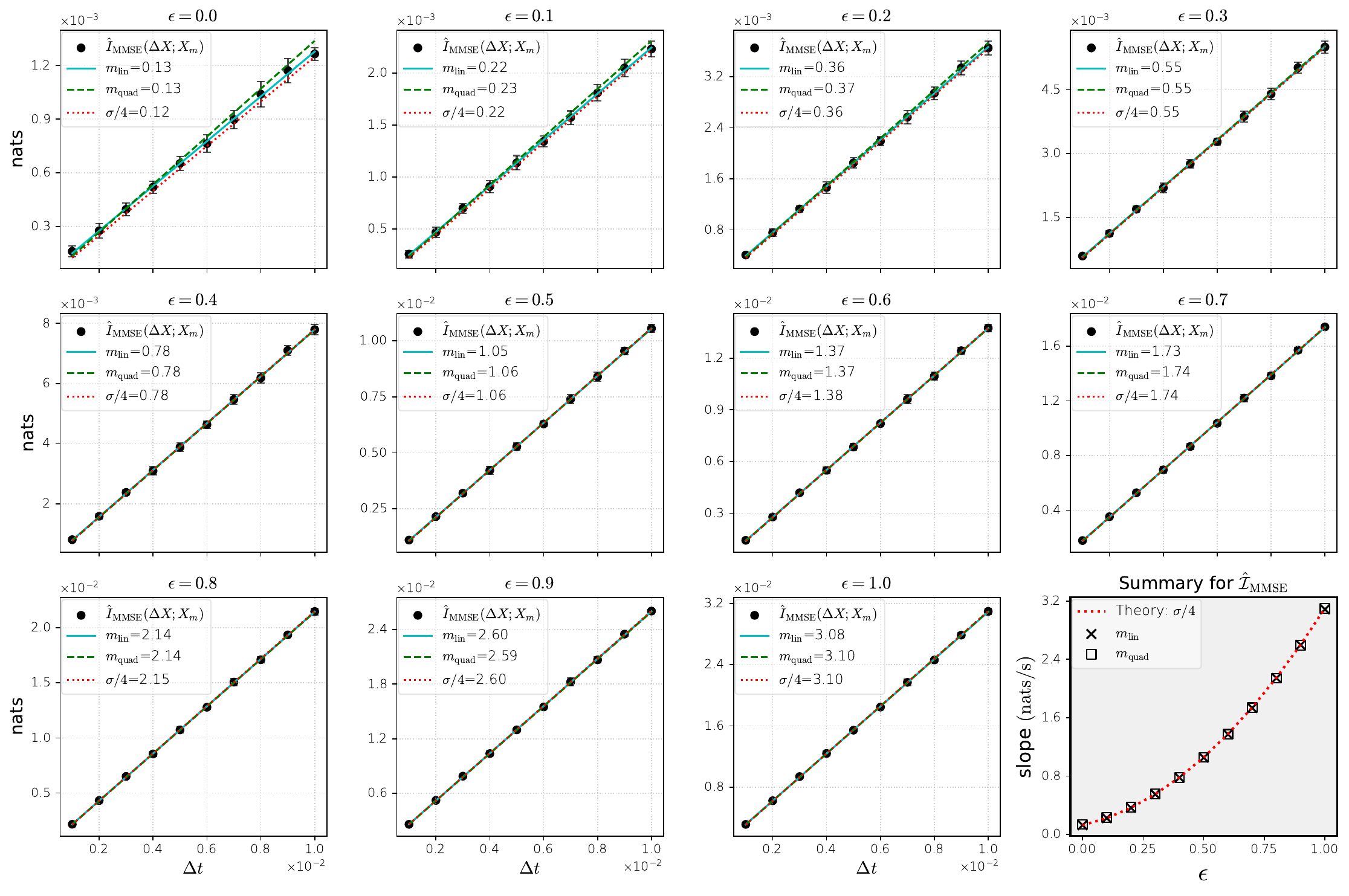}
    \vskip -0.0in
    \setlength{\abovecaptionskip}{0pt}
    \caption{Same as Fig.~\ref{sfig:main_nonlin_KSG}, but for $\hat I_{\rm MMSE}$. Black circles show the numerical estimates, the cyan solid line the linear fit, the green dashed line the first-order coefficient extracted from a quadratic fit in $\mathrm dt$, and the red dotted line the simulated value $\sigma/4$.
    }\label{sfig:main_nonlin_MMSE}
    %\vskip -0.2in
\end{figure}

For panel (e), the EP rate was evaluated analytically for the linear Ornstein-Uhlenbeck process. 
With stationary covariance $\matbf C$ and local mean velocity $\bm v$ in~\eqref{seq:lyapunov_equation} and~\eqref{seq:cal_v_linear}, the total EP rate is given by
\[
    \sigma=\langle\bm v^\intercal\matbf D^{-1}\bm v\rangle=
    \mathrm{Tr}\left[\matbf D^{-1}(\matbf A+\matbf D\matbf C^{-1})\matbf C(\matbf A+\matbf D\matbf C^{-1})^\intercal\right].
\]
Using the Lyapunov equation~\eqref{seq:lyapunov_equation}, this is equivalently written as
\begin{equation}
    \sigma=\mathrm{Tr}(\matbf D^{-1}\matbf A\matbf C\matbf A^\intercal)
    +\mathrm{Tr}(\matbf A).
    \label{seq:sigma_linear}
\end{equation}

For panel(f), the EP rate was estimated independently from the midpoint discretization of the Stratonovich expression
\[
    \sigma=\left\langle\bm F(\bm X_t)^\intercal\matbf D^{-1}\circ\dot{\bm X}_t\right\rangle,
\]
namely
\[
    \hat\sigma=\frac{1}{\Delta t}\left\langle
    \bm F\left(\frac{\bm X_t+\bm X_{t+\Delta t}}{2}\right)^\intercal
    \matbf D^{-1}\big(\bm X_{t+\Delta t}-\bm X_t\big)\right\rangle.
\]

For each value of $\epsilon$, $10^4$ trajectories of length $10^4$ steps were simulated with $\Delta t=10^{-3}$, starting from the origin, and the initial $5\times10^3$ steps were discarded as burn-in. This estimator follows from the steady-state identity
\[
    \sigma=\int\frac{\bm j_{ss}^\intercal\matbf D^{-1}\bm j_{ss}}{p_{ss}}\mathrm d\bm x
    =\int\bm F^\intercal\matbf D^{-1}\bm j_{ss}\mathrm d\bm x,
\]
where the second equality uses $\bm j_{ss}=\bm Fp_{ss}-\matbf D\nabla p_{ss}$ together with stationarity, so that the term involving $\nabla\ln p_{ss}$ integrates to zero under the adopted boundary conditions.

\subsection{Numerical procedure for Fig. 3}
\vspace*{-0.6\baselineskip}

For Fig. 3, we first collect the general steady-state expressions for the self and interaction EP rates in linear Langevin systems, together with the quantities entering the learning-rate bound, and then summarize the panel-specific calculations and results. Panel (a) is purely schematic and is therefore omitted.

\bigskip
\noindent\textbf{\textbullet\ B-1. Steady-state formulas for $\sigma_A^{\mathrm{self}},\sigma_{A|B}^{\mathrm{int}},\dot I_A$ and $\mathcal R_A$}

We consider the linear Langevin dynamics in steady-state,
\begin{equation*}
    \dot{\bm{x}}(t)=\matbf{A}\bm{x}(t)+\matbf{B}\,\bm{\xi}(t).
    %\label{seq:linear_langevin}
\end{equation*}

Let $\matbf C$ denote the steady-state covariance matrix of the system, which satisfies the Lyapunov equation~\eqref{seq:lyapunov_equation}.
To define the time midpoint variable $\bm x_m\coloneq \bm x(t+\mathrm{d}t/2)$, let $\bm y_1$ denote the Brownian increment $\matbf{B}\,\mathrm{d}\bm W_1\sim\mathcal{N}(0,\matbf{B}\matbf{B}^\intercal\mathrm dt/2)=\mathcal{N}(0,\matbf{D}\mathrm{d}t)$ over the first half-step $[t,t+\mathrm{d}t/2]$, and let $\bm y_2$ denote the increment over the second half-step $[t+\mathrm dt/2,t+\mathrm dt]$. Then, the relevant variables are given by
\[
    \mathrm{d}\bm x=\matbf{A}\bm x\,\mathrm{d}t+\bm y_1+\bm y_2,
    \quad
    \bm x_m=\bm x+\matbf{A}\bm x\frac{\mathrm{d}t}{2}+\bm y_1.
\]

Under the partition $\{A,B\}$, we write
\begin{equation*}
    \begin{aligned}
        \bm x=
        \begin{pmatrix}
            \bm x_A\\\bm x_B
        \end{pmatrix},
        \quad
        \mathrm{d}\bm x&=
        \begin{pmatrix}
            \mathrm{d}\bm x_A\\\mathrm{d}\bm x_B
        \end{pmatrix},
        \quad
        \bm x_m=
        \begin{pmatrix}
            \bm{x}_m^A\\\bm x_m^B
        \end{pmatrix},
        \\
        \matbf{A}=\begin{pmatrix}
            \matbf{A}_{AA} & \matbf{A}_{AB} \\
            \matbf{A}_{BA} & \matbf{A}_{BB}
        \end{pmatrix},
        \quad
        \matbf{C}&=\begin{pmatrix}
            \matbf{C}_{AA} & \matbf{C}_{AB} \\
            \matbf{C}_{BA} & \matbf{C}_{BB}
        \end{pmatrix},
        \quad
        \matbf{D}=\begin{pmatrix}
            \matbf{D}_{AA} & \matbf{D}_{AB} \\
            \matbf{D}_{BA} & \matbf{D}_{BB}
        \end{pmatrix}.
    \end{aligned}
    %\label{seq:def_partition_linear}
\end{equation*}
For Gaussian random variables $\bm X$ and $\bm Y$, the mutual information $I(\bm X;\bm Y)$ is given by
\[
    I(\bm X;\bm Y)
    =-\frac{1}{2}\log\det\left(
    I-\mathrm{Var}(\bm X)^{-1}
    \mathrm{Cov}(\bm X, \bm Y)
    \mathrm{Var}(\bm Y)^{-1}
    \mathrm{Cov}(\bm Y,\bm X)
    \right).
\]
Hence $I(\mathrm{d}\bm x_A;\bm x_m^A)$ can be evaluated by taking $\bm X=\mathrm d\bm x_A$ and $\bm Y=\bm x_m^A$,
\begin{equation*}
\begin{aligned}
    I(\mathrm d\bm x_A;\bm x_m^A)
    &=-\frac{1}{2}\log\det\left(
    I-\mathrm{Var}(\mathrm d\bm x_A)^{-1}
    \mathrm{Cov}(\mathrm d\bm x_A,\bm x_m^A)
    \mathrm{Var}(\bm x_m^A)^{-1}
    \mathrm{Cov}(\bm x_m^A,\mathrm d\bm x_A)
    \right)\\
    &=-\frac{1}{2}\log\det\left(
    I-(2\matbf{D}_{AA}\mathrm{d}t+O(\mathrm{d}t^2))^{-1}
    (\matbf{K}_{AA}\mathrm{d}t+O(\mathrm{d}t^2))
    (\matbf{C}_{AA}+O(\mathrm{d}t^2))^{-1}
    (\matbf{K}_{AA}\mathrm{d}t+O(\mathrm{d}t^2))^\intercal
    \right)\\
    &=-\frac{1}{2}\log\det\left(
    I-((2\matbf{D}_{AA})^{-1}\matbf{K}_{AA}\matbf{C}_{AA}^{-1}\matbf{K}_{AA}^\intercal)\mathrm{d}t+O(\mathrm{d}t^2)
    \right)\\
    &=\frac{1}{4}\mathrm{Tr}\left[
    \matbf{D}_{AA}^{-1}\matbf{K}_{AA}\matbf{C}_{AA}^{-1}\matbf{K}_{AA}^\intercal\right]\mathrm{d}t+O(\mathrm{d}t^2),
\end{aligned}
\end{equation*}
where $\matbf{K}\coloneq\matbf{A}\matbf{C}+\matbf{D}$. 
The last equality follows from $\log\det(I-\epsilon\matbf{M})=-\epsilon\mathrm{Tr}[\matbf{M}]+O(\epsilon^2)$.
Note that we used the Lyapunov equation~\eqref{seq:lyapunov_equation}, which cancels the $O(\mathrm dt)$ term in $\mathrm{Var}(\bm x_m^A)$.

Likewise, the conditional mutual information $I(\bm X;\bm Y\,|\,\bm Z)$ is given by
\[
    I(\bm X;\bm Y\,|\,\bm Z)
    =-\frac{1}{2}\log\det\left(
    I-\mathrm{Var}(\bm X|\bm Z)^{-1}
    \mathrm{Cov}(\bm X, \bm Y|\bm Z)
    \mathrm{Var}(\bm Y|\bm Z)^{-1}
    \mathrm{Cov}(\bm Y,\bm X|\bm Z)
    \right).
\]
Here, for jointly Gaussian variables $\bm X,\bm Y$ and $\bm Z$,
\[
    \mathrm{Cov}(\bm X,\bm Y \mid \bm Z)
    =\mathrm{Cov}(\bm X,\bm Y)
    -\mathrm{Cov}(\bm X,\bm Z)\,\mathrm{Var}(\bm Z)^{-1}\,\mathrm{Cov}(\bm Z,\bm Y),
\]
and
\[
    \mathrm{Var}(\bm X\mid\bm Z)
    =\mathrm{Var}(\bm X)-\mathrm{Cov}(\bm X,\bm Z)\mathrm{Var}(\bm Z)^{-1}\mathrm{Cov}(\bm Z,\bm X).
\]

Using these results, we obtain
\begin{equation*}
\begin{aligned}
    I(\mathrm{d}\bm{x}_A;\bm{x}_m^B\,|\,\bm{x}_m^A)
    &=-\frac{1}{2}\log\det\left(
    I-\mathrm{Var}(\mathrm{d}\bm{x}_A|\bm{x}_m^A)^{-1}
    \mathrm{Cov}(\mathrm{d}\bm{x}_A, \bm{x}_m^B|\bm{x}_m^A)
    \mathrm{Var}(\bm{x}_m^B|\bm{x}_m^A)^{-1}
    \mathrm{Cov}(\bm{x}_m^B,\mathrm{d}\bm{x}_A|\bm{x}_m^A)
    \right)\\
    &=-\frac{1}{2}\log\det\left(I-
    (2\matbf{D}_{AA}\mathrm{d}t+O(\mathrm{d}t^2))^{-1}
    (\matbf{L}_{AB|A}\mathrm{d}t+O(\mathrm{d}t^2))
    (\matbf{C}_{B|A}+O(\mathrm{d}t^2))^{-1}
    (\matbf{L}_{AB|A}\mathrm{d}t+O(\mathrm{d}t^2))^\intercal
    \right)\\
    &=-\frac{1}{2}\log\det\left(
    I-((2\matbf{D}_{AA})^{-1}
    \matbf{L}_{AB|A}\matbf{C}_{B|A}^{-1}
    \matbf{L}_{AB|A}^\intercal)\mathrm{d}t
    +O(\mathrm{d}t^2)\right)\\
    &=\frac{1}{4}\mathrm{Tr}\left[
    \matbf{D}_{AA}^{-1}\matbf{L}_{AB|A}
    \matbf{C}_{B|A}^{-1}\matbf{L}_{AB|A}^\intercal
    \right]\mathrm{d}t+O(\mathrm{d}t^2),
\end{aligned}
    %\label{seq:cal_cmi_dx_a_xm_b_xm_a}
\end{equation*}
where $\matbf{C}_{B|A}\coloneq\matbf{C}_{BB}-\matbf{C}_{BA}\matbf{C}_{AA}^{-1}\matbf{C}_{AB}$ and $\matbf{L}_{AB|A}\coloneq\matbf{K}_{AB}-\matbf{K}_{AA}\matbf{C}_{AA}^{-1}\matbf{C}_{AB}$.

Therefore, the analytic forms of the self and interaction EP rates are given by,
\begin{align}
    \sigma_A^{\mathrm{self}}\coloneq\lim_{\mathrm{d}t\rightarrow0}
    \frac{4I(\mathrm{d}\bm x_A;\bm x_m^A)}{\mathrm{d}t}
    &=\mathrm{Tr}\!\left[
    \matbf{D}_{AA}^{-1}\matbf{K}_{AA}
    \matbf{C}_{AA}^{-1}\matbf{K}_{AA}^\intercal\right],
    \label{seq:cal_self_EP_linear}
    \\
    \sigma_{A|B}^{\mathrm{int}}\coloneq\lim_{\mathrm{d}t\rightarrow0}
    \frac{4I(\mathrm{d}\bm x_A;\bm x_m^B\mid\bm x_m^A)}{\mathrm{d}t}
    &=\mathrm{Tr}\!\left[
    \matbf{D}_{AA}^{-1}\matbf{L}_{AB|A}
    \matbf{C}_{B|A}^{-1}\matbf{L}_{AB|A}^\intercal
    \right].
    \label{seq:cal_int_EP_linear}
\end{align}

The learning-rate in Matsumoto et al. (2025)~\cite{matsumoto2025learningrate}, or equivalently, the information flow in Horowitz (2015)~\cite{horowitz2015multipartite} is defined as
\[
    \dot I_A(\bm x_A;\bm x_B)
    =\int\bm J_A(\bm x)\cdot
    \nabla_{\bm x_A}\ln p(\bm x_B|\bm x_A)\,\mathrm d\bm x.
\]

Since $\bm x_B|\bm x_A\sim\mathcal N(\matbf C_{BA}\matbf C_{AA}^{-1}\bm x_A,\matbf C_{B|A})$, we obtain
\[
    \nabla_{\bm x_A}\!\ln p(\bm x_B|\bm x_A)
    =\matbf C_{AA}^{-1}\matbf C_{AB}\matbf C_{B|A}^{-1}
    \,\underbrace{\!
    \big(\bm x_B-\matbf C_{BA}\matbf C_{AA}^{-1}\bm x_A\big)}_{=:\bm r_B}
\]
while $\bm J_A=p(\bm x)\bm v_A$ where
\begin{equation*}
\begin{aligned}
    \bm v_A&=\big[(\matbf A+\matbf D\matbf C^{-1})\bm x\big]_A
    =(\matbf K\matbf C^{-1}\bm x)_A
    =(\matbf K\matbf C^{-1})_{AA}\bm x_A+(\matbf K\matbf C^{-1})_{AB}\bm x_B
    \\
    &=\big[(\matbf K\matbf C^{-1})_{AA}+(\matbf K\matbf C^{-1})_{AB}\matbf C_{BA}\matbf C_{AA}^{-1}\big]\bm x_A
    +(\matbf K\matbf C^{-1})_{AB}\bm r_B.
\end{aligned}
\end{equation*}
Note that $\mathrm{Cov}(\bm x_A,\bm r_B)=0$ and $\mathrm{Var}(\bm r_B)=\matbf C_{B|A}$.
Combining these results yields
\begin{equation}
    \dot I_A=
    \mathrm{Tr}\big[\matbf C_{AA}^{-1}\matbf C_{AB}\big((\matbf K\matbf C^{-1})_{AB}\big)^\intercal\big]
    =\mathrm{Tr}\big[\matbf C_{AA}^{-1}\matbf C_{AB}\matbf C_{B|A}^{-1}\matbf L_{AB|A}^\intercal\big].
    \label{seq:cal_IF_A_linear_ss}
\end{equation}

We emphasize that although \eqref{seq:cal_IF_A_linear_ss} is formally well-defined for any Gaussian diffusion, its interpretation as the information-flow term entering a subsystem-resolved second-law balance in the sense of Horowitz (2015) requires the partition ${A,B}$ to be bipartite, i.e. $\matbf D_{AB}=\matbf D_{BA}=\matbf 0$~\cite{horowitz2015multipartite}.

Lastly, to write $\mathcal R_A$ explicitly in the steady-state linear Gaussian case, we first calculate the conditional Fisher information matrix
\[
    \matbf F_A^{|A}\coloneq
    \Big\langle\big(\nabla_{\bm x_A}\!\ln p(\bm x_B|\bm x_A)\big)
    \big(\nabla_{\bm x_A}\!\ln p(\bm x_B|\bm x_A)\big)^\intercal\Big\rangle
    =\matbf C_{AA}^{-1}\matbf C_{AB}\matbf C_{B|A}^{-1}\matbf C_{BA}\matbf C_{AA}^{-1},
\]
and hence
\begin{equation}
    \mathcal R_A\coloneq
    \mathrm{Tr}\big[\matbf D_{AA}\matbf F_A^{|A}\big]^{-1}
    =\mathrm{Tr}\big[\matbf D_{AA}\matbf C_{AA}^{-1}\matbf C_{AB}\matbf C_{B|A}^{-1}\matbf C_{BA}\matbf C_{AA}^{-1}\big]^{-1}.
    \label{seq:cal_R_A}
\end{equation}

\bigskip
\noindent\textbf{\textbullet\ B-2. Details for models and calculations of Fig. 3}

For panels (b,c), the curves were not generated from trajectory data, but were evaluated analytically from the steady-state Gaussian statistics of three-dimensional linear cascade model shown in Fig.~3(a), with partition $A=\{x,y\}$ and $B=\{z\}$. For each parameter value, the total EP rate $\sigma_{AB}$, self/interaction EP rate $\sigma_A^{\mathrm{self}},\sigma_{A|B}^{\mathrm{int}}$, learning rate $\dot I_A$, and information resistance $\mathcal R_A$ were computed from~\eqref{seq:sigma_linear}--\eqref{seq:cal_R_A}, while the local EP rate $\sigma_A$ were obtained as $\sigma_A=\sigma_A^{\mathrm{self}}+\sigma_{A|B}^{\mathrm{int}}$.

For panel (b), we fixed $\beta=1$ and varied the internal coupling $\alpha$ in the drift matrix
\[
\matbf A=
\begin{pmatrix}
    -2 & \alpha & 0 \\
    0 & -2 & 1 \\
    0 & 0 & -2
\end{pmatrix},
\quad
\matbf D=\mathrm I.
\]
The analytic form of $\sigma_A^{\mathrm{self}}$ and $\sigma_{A|B}^{\mathrm{int}}$ is given by
\begin{align*}
    \sigma_A^{\mathrm{self}}(\alpha)
    &=\frac{289\alpha^2(19\alpha^2+272)}{64(323\alpha^2+4608)}
    =\frac{4913}{18432}\alpha^2+O(\alpha^4)
    \approx0.267\alpha^2,
    \\
    \sigma_{A|B}^{\mathrm{int}}(\alpha)
    &=\frac{23409\alpha^4+706048\alpha^2+5308416}{2(305\alpha^2+4352)(323\alpha^2+4608)}
    =\frac{9}{68}-\frac{5057}{5326848}\alpha^2+O(\alpha^4)
    \approx
    0.132-0.001\alpha^2.
\end{align*}
Therefore $\sigma_A^{\mathrm{self}}$ increases markedly with $\alpha$, whereas $\sigma_{A|B}^{\mathrm{int}}$ remains nearly unchanged.

For panel (c), we fixed $\alpha=1$ and varied the upstream coupling $\beta$ in
\[
\matbf A=
\begin{pmatrix}
    -2 & 1 & 0 \\
    0 & -2 & \beta \\
    0 & 0 & -2
\end{pmatrix},
\quad
\matbf D=\mathrm I.
\]
In this case,
\begin{align*}
    \sigma_A^{\mathrm{self}}(\beta)
    &=\frac{(\beta^2+16)^2(19\beta^2+272)}{64(3\beta^4+576\beta^2+4352)}
    =\frac{1}{4}+\frac{\beta^2}{64}+O(\beta^4)
    \approx
    0.25+0.0156\beta^2
    \\
    \sigma_{A|B}^{\mathrm{int}}(\beta)
    &=\frac{\beta^2(\beta^8+560\beta^6+83776\beta^4+1218560\beta^2+4734976)}{2(\beta^4+304\beta^2+4352)(3\beta^4+576\beta^2+4352)}
    =\frac{\beta^2}{8}+O(\beta^4)
    \approx0.125\beta^2.
\end{align*}
This shows that $\sigma_{A|B}^{\mathrm{int}}$ is nearly an order of magnitude more sensitive to $\beta$ than $\sigma_{A}^{\mathrm{self}}$.
Because $\alpha=1$, changes in $\beta$ also alter the steady-state distribution of partition $A=\{x,y\}$, and hence weakly affect $\sigma_A^{\mathrm{self}}$ as well. However, this effect is much smaller.

For panel (d), we used the same $\alpha$ sweep as in panel (b), with $\beta=1$. At $\alpha=1$, we obtain
\[
    \sigma_A^{\rm self}=0.26649,
    \quad
    \sigma_{A|B}^{\rm int}=0.13147,
    \quad
    \sigma_A=0.39795,
\]
and
\[
    \dot I_A=0.11767,
    \quad
    \mathcal R_A=9.4871,
    \quad
    |\dot I_A|^2\mathcal R_A=0.13137.
\]
Thus, $\sigma_{A|B}^{\rm int}/\sigma_A=0.33036$, while $|\dot I_A|^2\mathcal R_A/\sigma_{A|B}^{\rm int}=0.99924$. As $\alpha$ increase, the growing self channel contribution therefore makes the previous bound looser, whereas the sharpened interaction channel bound remains nearly saturated in this model.

\section{Sharpened learning-rate bound}

Beyond the single toy model considered in the main text, we examine the sharpened learning-rate bound for a broader set of randomly generated linear Ornstein-Uhlenbeck (OU) systems and for linear OU models fitted to red-blood-cell (RBC) data.

\subsection{Randomly generated Linear OU systems}
\vspace*{-0.6\baselineskip}

For each dimension $d=2,...,7$ we retained $10^4$ accepted samples from a proposal ensemble with full drift matrix $\matbf A$ and diagonal diffusion matrix $\matbf D$. Before rejection, the magnitudes of the entries of $\matbf A$ and the diagonal entries of $\matbf D$ were drawn independently from log-uniform distributions on $[1,10^3]$. Off-diagonal signs of $\matbf A$ were chosen at random, while each diagonal entry was taken negative with probability $0.9$. We then retained only physical samples for which $\matbf A$ was Hurwitz, the stationary covariance $\matbf C$ solving the Lyapunov equation~\eqref{seq:lyapunov_equation} was positive definite, and the total steady-state EP was positive.
For each accepted sample, we randomly partitioned the $d$ variables into two nonempty subsets $A$ and $B$, with $A\cup B=\{1,\cdots,d\}$. The values are evaluated the steady-state Gaussian quantities $\sigma_{A|B}^{\rm int},\mathcal R_A$, and $|\dot I_A|$ using the analytic formulas derived above. We then plotted the dimensionless ratio
\[
    \rho_{\mathrm{int}}\coloneq
    \frac{\left|\dot I_{A}\right|^2\mathcal R_{A}}{\sigma_{A|B}^{\mathrm{int}}}
\]
against $\sigma_{A|B}^{\mathrm{int}}$ on a logarithmic horizontal axis, with colors indicating the dimension $d$ in the left panel of Fig.~\ref{sfig:sharpened_learning}. The dashed horizontal line at unity marks saturation of Eq. (8). As expected, the sampled points remain below this line over a broad range of interaction EP rate values. The right panel additionally shows $\sigma_{A|B}^{\rm int}/\sigma_A$, the fraction of the previous local EP upper bound retained by the sharpened bound; smaller values indicate greater tightening.

\begin{figure}[!h]
    \centering
    \includegraphics[width=0.6\textwidth]{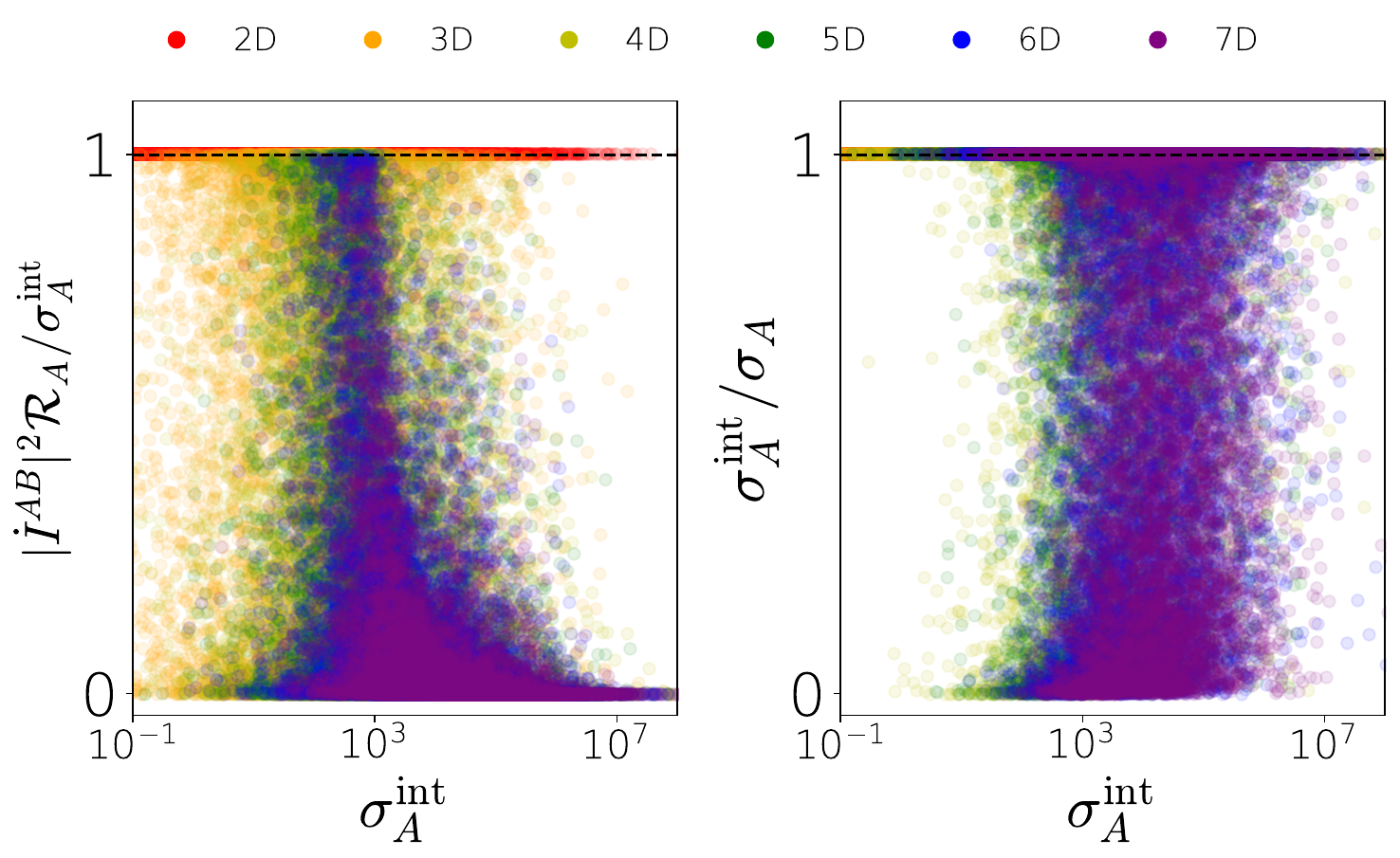}
    \vskip -0.0in
    \setlength{\abovecaptionskip}{0pt}
    \caption{Numerical assessment of the sharpened learning-rate bound in randomly generated stationary linear OU systems. For each dimension $d=2,\cdots,7$, $10^4$ accepted samples are shown, with colors indicating $d$. (Left) The saturation ratio $|\dot I_A|^2\mathcal R_A/\sigma_{A|B}^{\rm int}$. (Right) The ratio $\sigma_{A|B}^{\rm int}/\sigma_A$ of the sharpened to the previous upper bound. Both are plotted against the interaction EP rate $\sigma_{A|B}^{\rm int}$. The dashed lines mark unity. In the scalar-scalar case $(d=2)$, the sharpened bound is saturated and the self channel vanishes, so both ratios equal unity.
    }\label{sfig:sharpened_learning}
    %\vskip -0.2in
\end{figure}

It is worth noting that all $d=2$ samples lie exactly on the saturation line. This is not a numerical artifact. In a two-dimensional bipartite system each subsystem is one-dimensional, and at steady-state the self part vanishes identically. Indeed, with $\matbf K=\matbf A\matbf C+\matbf D$, the Lyapunov equation~\eqref{seq:lyapunov_equation} implies $\matbf K+\matbf K^\intercal=\matbf 0$, so any $1\times 1$ diagonal block must satisfy $\matbf K_{AA}=0$; hence $\sigma_A^{\mathrm{self}}=0$, and similarly $\sigma_B^{\mathrm{self}}=0$. More generally, any one-dimensional subsystem has zero self EP rate at steady-state. In the scalar-scalar case one can find $|\dot I_A|^2\mathcal R_A=\sigma_{A|B}^{\mathrm{int}}$.

\subsection{Fitted RBC models}
\vspace*{-0.6\baselineskip}

We use the fitted RBC parameter sets reported in Terlizzi et al.~(2024)~\cite{terlizzi2024variance}. The parameter values for the OT-sensing and OM datasets are listed in Tables~\ref{stab:RBC_parameter_OT_ss} and~\ref{stab:RBC_parameter_OM}, respectively.
These correspond to distinct measurement modalities: OT-sensing tracks probe-bead fluctuations of the membrane, whereas OM analyzes free-standing contour fluctuations. The OT-sensing table contains two passivated cells and five active cells, whereas the OM table contains one passivated cell and six active cells.

For each parameter set we construct the three-dimensional linear Langevin model,
\[
\dot{\bm X} = \matbf A \bm X + \sqrt{2\matbf D}\,\bm\xi, \qquad \bm X = (x,y,\eta)^\intercal,
\]
with
\[
\matbf A =
\begin{pmatrix}
-\mu_x k_x & \mu_x k_{\rm int} & 0 \\
\mu_y k_{\rm int} & -\mu_y k_y & \mu_y \\
0 & 0 & -1/\tau
\end{pmatrix},
\qquad
\matbf D =
\begin{pmatrix}
\mu_x k_B T & 0 & 0 \\
0 & \mu_y k_B T & 0 \\
0 & 0 & \epsilon^2/\tau
\end{pmatrix}.
\]
Here $x$ is the measured outer-membrane coordinate, $y$ the hidden membrane-cortex attachment, and $\eta$ the active force. The steady-state covariance matrix $\matbf C$ is obtained from the Lyapunov equation~\eqref{seq:lyapunov_equation}.
Throughout this section, we use the tabulated fitted parameter sets directly and without additional fitting or trajectory simulation. Entropies are measured in units of $k_B$, so the learning-rates and EP rates have units of $s^{-1}$.

\begin{table*}[h]
    \centering
    \small
    % --- 왼쪽: OT-stretching 표 (너비를 0.47로 소폭 축소) ---
    \begin{minipage}{0.47\textwidth}
        \centering
        \begin{tabular}{l rrrrrrr}
            \toprule
            \textbf{Parameters} & \textbf{P1} & \textbf{P2} & \textbf{A1} & \textbf{A2} & \textbf{A3} & \textbf{A4} & \textbf{A5} \\
            \midrule
            $k_x$ [$10^{-3}$ pN/nm] & 6.0 & 8.1 & 6.5 & 15.1 & 3.7 & 7.17 & 9.5 \\
            $\mu_x$ [$10^4$ nm/(pN s)] & 2.6 & 2.1 & 2.8 & 3.6 & 2.6 & 1.6 & 1.7 \\
            $k_y$ [$10^{-2}$ pN/nm] & 1.7 & 29.0 & 1.6 & 1.6 & 2.9 & 1.3 & 1.78 \\
            $\mu_y$ [$10^4$ nm/(pN s)] & 140 & 31.0 & 2.9 & 2.3 & 0.57 & 1.96 & 1.94 \\
            $k_{\text{int}}$ [$10^{-3}$ pN/nm] & 1.8 & 6.0 & 4.5 & 4.2 & 2.3 & 2.88 & 2.84 \\
            $\epsilon$ [pN] & 0.7 & 2.1 & 2.6 & 2.4 & 4.7 & 3.86 & 4.41 \\
            $\tau$ [$10^{-2}$ s] & 200 & 400 & 8.0 & 1.1 & 2.5 & 17.0 & 7.7 \\
            \bottomrule
        \end{tabular}
        \caption{Parameters for OT-sensing experiments.}
        \label{stab:RBC_parameter_OT_ss}
    \end{minipage}
    \hspace{0.002\textwidth} % <--- 여기서 간격을 명시적으로 줌 (원하는 만큼 조절 가능)
    % --- 오른쪽: OM 표 ---
    \begin{minipage}{0.48\textwidth}
        \centering
        \begin{tabular}{l rrrrrrr}
            \toprule
            \textbf{Parameters} & \textbf{P1} & \textbf{A1} & \textbf{A2} & \textbf{A3} & \textbf{A4} & \textbf{A5} & \textbf{A6} \\
            \midrule
            $k_x$ [$10^{-2}$ pN/nm] & 59.0 & 5.37 & 5.07 & 5.23 & 4.61 & 5.39 & 8.3 \\
            $\mu_x$ [$10^4$ nm/(pN s)] & 0.8 & 1.2 & 1.3 & 1.2 & 1.4 & 1.2 & 0.9 \\
            $k_y$ [$10^{-2}$ pN/nm] & 5.6 & 1.93 & 1.89 & 1.92 & 1.95 & 2.10 & 1.35 \\
            $\mu_y$ [$10^4$ nm/(pN s)] & 0.082 & 4.85 & 4.89 & 4.79 & 4.84 & 4.95 & 4.65 \\
            $k_{\text{int}}$ [$10^{-3}$ pN/nm] & 91.0 & 3.8 & 3.57 & 3.8 & 3.50 & 3.94 & 4.16 \\
            $\epsilon$ [pN] & 0.2 & 4.42 & 4.37 & 4.43 & 4.40 & 4.45 & 4.52 \\
            $\tau$ [$10^{-2}$ s] & 750 & 6.0 & 4.1 & 6.0 & 3.9 & 3.3 & 7.0 \\
            \bottomrule
        \end{tabular}
        \caption{Parameters for OM experiments.}
        \label{stab:RBC_parameter_OM}
    \end{minipage}
    \vspace{0.5ex}
    \begin{minipage}{0.95\textwidth}
        \footnotesize
        \textit{Common note.} Parameter sets are taken from Ref.~\cite{terlizzi2024variance}.
        We use $k_B T = 4.142\,\mathrm{pN\cdot nm}$.
    \end{minipage}
\end{table*}
\vspace*{-0.6\baselineskip}

\subsection{Observable-hidden partition and EP decomposition of the RBC}
\vspace*{-0.6\baselineskip}

For the RBC model, we partition the variables into the observable coordinate $\{x\}$ and the hidden block $\{y,\eta\}$. The total steady-state EP rate is evaluated from Eq.~\eqref{seq:sigma_linear}, while the self and interaction channels of the hidden block are evaluated from Eqs.~\eqref{seq:cal_self_EP_linear} and~\eqref{seq:cal_int_EP_linear}, respectively. This gives
\[
\sigma_{\mathrm{tot}}=\sigma_{x}+\sigma_{y\eta}, 
\qquad
\sigma_{y\eta}=\sigma^{\mathrm{self}}_{y\eta}+\sigma^{\mathrm{int}}_{y\eta|x}.
\]
In this setting, because the observable \(x\) sector is one-dimensional, its steady-state self channel contribution vanishes identically, and $\sigma_x=\sigma_{x|y\eta}^{\rm int}.$ 

For each fitted cell, we therefore compute the three normalized quantities
\[
\frac{\sigma_{x}}{\sigma_{\mathrm{tot}}},
\qquad
\frac{\sigma^{\mathrm{int}}_{y\eta|x}}{\sigma_{\mathrm{tot}}},
\qquad
\frac{\sigma^{\mathrm{self}}_{y\eta}}{\sigma_{\mathrm{tot}}},
\]
which sum to unity up to numerical precision. The bars in Fig.~\ref{sfig:RBC} display these fractions for each active fitted cell separately.

\begin{figure}[h]
    \centering
    \includegraphics[width=0.95\textwidth]{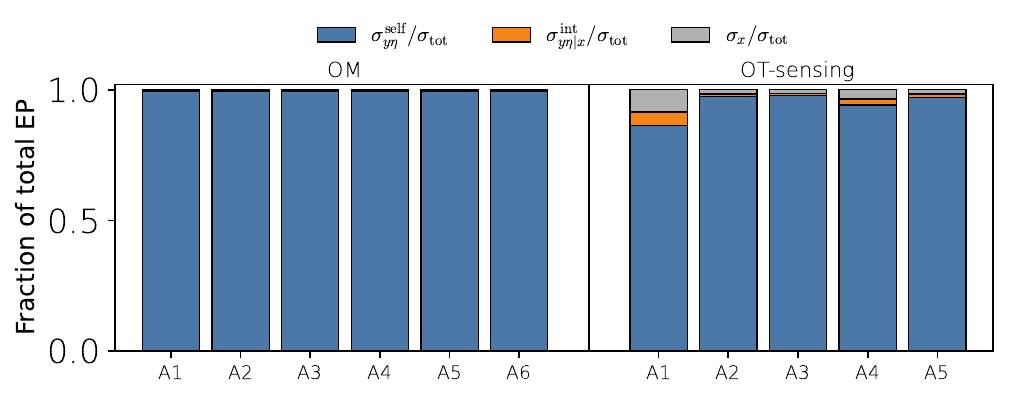}
    \vskip -0.0in
    \setlength{\abovecaptionskip}{0pt}
    \caption{Cell-resolved EP decomposition for the observable-hidden partition $\{x\}|\{y,\eta\}$ in the fitted RBC models of Ref.~\cite{terlizzi2024variance}. Each bar corresponds to one active-cell parameter set from Tables~\ref{stab:RBC_parameter_OT_ss} and ~\ref{stab:RBC_parameter_OM}, ordered as in the tables, and shows the fractions \(\sigma_{x}/\sigma_{\mathrm{tot}}\) (gray), \(\sigma^{\mathrm{int}}_{y\eta|x}/\sigma_{\mathrm{tot}}\) (orange), and \(\sigma^{\mathrm{self}}_{y\eta}/\sigma_{\mathrm{tot}}\) (blue). The left and right blocks show the OM $(n=6)$ and OT-sensing $(n=5)$ fits, respectively. All quantities are evaluated analytically from the stationary Gaussian model without additional fitting. For every active fitted cell, self EP within the hidden block is the dominant contribution. Passivated-cell results are discussed in the text. 
    }\label{sfig:RBC}
    %\vskip -0.2in
\end{figure}

For the active OM fits, the hidden-block self channel contribution accounts for $0.9928-0.9938$ of the total EP, with an arithmetic mean of $0.9932$. The active OT-sensing fits show broader cell-to-cell variation, but the same hierarchy persists: the self fraction ranges from $0.8645$ to $0.9773$, with a mean of $0.9459$. The mean hidden-block interaction fractions are only $0.00358$ and $0.02044$ for OM and OT-sensing, respectively. Thus, for this partition, most dissipation occurs within the hidden block rather than in its interaction channel with the observable coordinate. The cell-resolved presentation confirms that this hierarchy is not an artifact of averaging over heterogeneous fits.

\subsection{Sharpening of the learning-rate bound for RBC}
\vspace*{-0.6\baselineskip}

We apply the learning-rate bound to the hidden block, i.e., $A=\{y,\eta\}$ and $B=\{x\}$ in Eq. (8). Define
\[
    L_{y\eta}\coloneq\left|\dot I_{y\eta}\right|^2\mathcal R_{y\eta}.
\]
For the centered stationary OU models, the previous local-EP bound is $L_{y\eta}\le\sigma_{y\eta}$, whereas the sharpened bound is $L_{y\eta}\le\sigma_{y\eta|x}^{\rm int}$. Their saturation ratios and the tightening factor are
\[
    \rho_{\rm old}=\frac{L_{y\eta}}{\sigma_{y\eta}},
    \quad
    \rho_{\rm int}=\frac{L_{y\eta}}{\sigma_{y\eta|x}^{\rm int}},
    \quad
    \mathcal T=\frac{\sigma_{y\eta}}{\sigma_{y\eta|x}^{\rm int}}.
\]
The improvement is precisely the removal of $\sigma_{y\eta}^{\rm self}$ from the upper bound. In contrast, applying the bound to the scalar observable sector gives no such improvement because $\sigma_x^{\rm self}=0$.

\begin{figure}[h]
    \centering
    \includegraphics[width=0.7\textwidth]{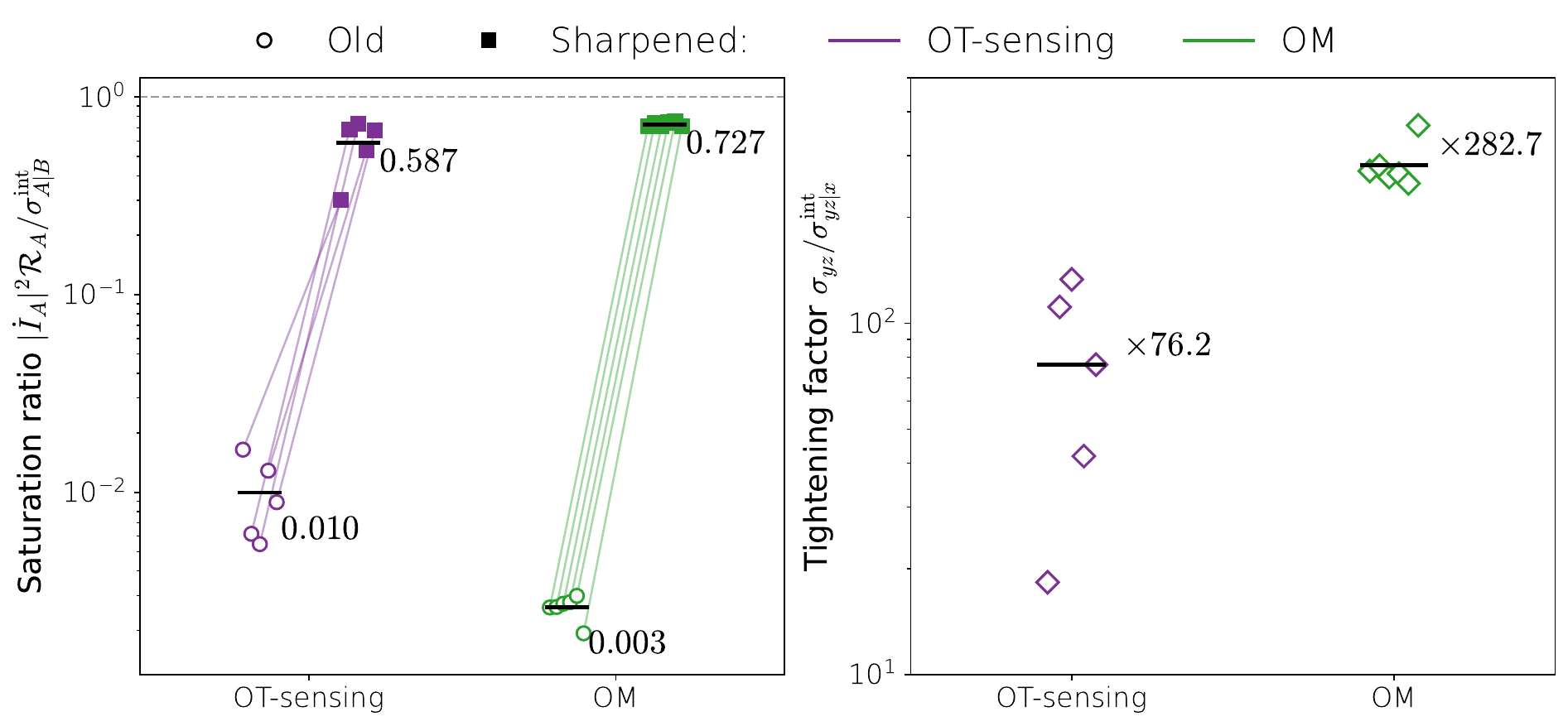}
    \vskip -0.0in
    \setlength{\abovecaptionskip}{0pt}
    \caption{Sharpening of the learning-rate bound for the hidden block $\{y,\eta\}$ relative to the observable coordinate $\{x\}$. (Left) saturation ratios for the previous bound (open circles) and the sharpened bound (filled squares); lines connect the two values for the same active-cell fit. The dashed horizontal line marks saturation. (Right) the tightening factor $\mathcal T$ for each active-cell fit. Purple and Green denote OT-sensing $(n=5)$ and OM $(n=6)$, respectively. Short black horizontal lines and their numerical labels indicate arithmetic means of the per-cell quantities, not ratios of ensemble-averaged rates. Passivated-cell results, whose saturation ratios involve very small numerators and denominators, are reported in the text only.
    }\label{sfig:RBC_sharpened}
    %\vskip -0.2in
\end{figure}

As shown in Fig.~\ref{sfig:RBC_sharpened}, the mean saturation ratio increases from $0.00998$ to $0.587$ for OT-sensing and from $0.00262$ to $0.727$ for OM. The mean per-cell tightening factors are $76.2$ and $282.7$, respectively. Individual factors range from $18.3$ to $133.3$ for OT-sensing and from $250.2$ to $366.2$ for OM. These model-based results illustrate that removing self dissipation can substantially sharpen the learning-rate bound while preserving a nontrivial degree of saturation.

For the passivated fits OT P1, OT P2, and OM P1, $L_{y\eta}$ is $1.67\times10^{-5}$, $1.47\times10^{-7}$, and $4.69\times10^{-10}\,s^{-1}$, respectively, much smaller than those of the active fits ($10.0$--$40.5\,\mathrm{s}^{-1}$). Their sharpened saturation ratios remain well defined and satisfy the bound, taking values of \(0.896\), \(0.806\), and \(0.975\). However, these relatively high ratios arise from dividing two very small rates and should not be interpreted as indicating appreciable absolute information flow. To avoid overstating their biophysical significance, we report the passivated-cell results in the text and retain their parameters in Tables I and II, while restricting the figures to the active fits.

%\newpage
\section{Hydrodynamic coarse-graining reversal: \\analytic calculation and finite-resolution recovery}
\subsection{Model and parameters}
\vspace*{-0.6\baselineskip}

We consider the three-dimensional stationary OU model introduced for two hydrodynamically coupled particles driven through a stochastic trap center~\cite{das2025irreversibility}. We write $\bm z\coloneq (x_1,x_2,\lambda)^\intercal$ and use the convention
\begin{equation}
    \mathrm d\bm z=-\matbf F\bm z\,\mathrm dt+\sqrt{2\matbf D}\,\mathrm dW,
\end{equation}
where
\begin{equation*}
    \matbf F\!=\!\!
    \begin{pmatrix}
    1/\tau_1 & \epsilon/{\tau_2} & -1/\tau_1 \\
    \epsilon/\tau_1 & 1/{\tau_2} & -\epsilon/\tau_1 \\
    0 & 0 & 1/\tau_\lambda
    \end{pmatrix}\!\!,
    \;\; 
    \matbf D\!=\!\!
    \begin{pmatrix}
    D_0 & \epsilon D_0 & 0 \\
    \epsilon D_0 & D_0 & 0 \\
    0 & 0 & \theta D_0
    \end{pmatrix}\!.   
    \label{eq:Das_model_details}
\end{equation*}
The parameters are $\tau_1=1.5\,ms$, $\tau_2=2.1\,ms$, $\tau_\lambda=4.0\,ms$, $D_0=0.16\,\mu m^2s^{-1}$, and $\theta=0.56$. The two conditions used in Fig.~4d are $\epsilon=0.29$ and $\epsilon=0.49$. $\epsilon$ and $\theta$ are dimensionless. All logarithms are natural, so rates in this section are reported in $k_B s^{-1}$ after applying the factor four in the midpoint identity.

\subsection{Analytic evaluation of the $\epsilon$ derivatives in Fig.~4c}
\vspace*{-0.6\baselineskip}

We use the closed-form expressions for linear systems in Eqs.~\eqref{seq:cal_self_EP_linear}--\eqref{seq:cal_int_EP_linear}, together with the Lyapunov equation~\eqref{seq:lyapunov_equation}. For the dynamics considered here, the drift matrix is $\matbf A=-\matbf F$. We define
\begin{equation}
    \matbf F_\epsilon \equiv \partial_\epsilon \matbf F,
    \qquad
    \matbf D_\epsilon \equiv \partial_\epsilon \matbf D,
    \qquad
    \matbf C_\epsilon \equiv \partial_\epsilon \matbf C.
\end{equation}
The derivatives $\matbf F_\epsilon$ and $\matbf D_\epsilon$ follow directly from the model matrices. Differentiating the Lyapunov equation $\matbf F\matbf C+\matbf C\matbf F^\intercal=2\matbf D$ with respect to $\epsilon$ gives a second continuous Lyapunov equation for the covariance derivative,
\begin{equation}
    \matbf F\matbf C_\epsilon
    +\matbf C_\epsilon\matbf F^\intercal
    =
    2\matbf D_\epsilon
    -\matbf F_\epsilon\matbf C
    -\matbf C\matbf F_\epsilon^\intercal.
    \label{seq:lyapunov_epsilon_derivative}
\end{equation}
After solving Eq.~\eqref{seq:lyapunov_epsilon_derivative} for $\matbf C_\epsilon$, we differentiated Eqs.~\eqref{seq:cal_self_EP_linear}--\eqref{seq:cal_int_EP_linear} with respect to $\epsilon$ and substituted $\matbf F_\epsilon$, $\matbf D_\epsilon$, and $\matbf C_\epsilon$. This yields the $\epsilon$ derivatives of the individual channel rates shown in Fig.~4(c).

The differentiation itself was performed analytically; however, we did not expand the resulting expressions into fully symbolic closed forms. Instead, at each value of $\epsilon$, Eq.~\eqref{seq:lyapunov_epsilon_derivative} was solved numerically using the continuous Lyapunov solver in SciPy. Thus, the curves in Fig.~4(c) do not rely on finite-difference approximations with respect to $\epsilon$.

\subsection{Synthesis of the trajectories used for Fig.~4d}
\vspace*{-0.6\baselineskip}

Synthetic trajectories were generated with the exact discrete transition of the stationary OU process. For observation interval $\delta t$ and stationary covariance $\matbf C$, the initial state is drawn from $\mathcal N(\bm 0,\matbf C)$, and subsequent states satisfy
\begin{equation}
    \bm z_{k+1}=\matbf M\bm z_k+\bm\eta_k,
    \quad
    \matbf M=\exp(-\matbf F\delta t),
    \quad
    \bm\eta_k\sim\mathcal N(\bm 0,\matbf C-\matbf M\matbf C\matbf M^\intercal).
    \label{seq:synthesis_trjs_equation}
\end{equation}

The Gaussian innovations are independent across steps and independent of the initial state. This construction initializes the process in stationarity and avoids time-integration bias. At each $\epsilon$, we generated 40 independent trajectories of duration $50\,s$ sampled at $\delta t=10^{-4} \,s$, giving 500001 recorded states per trajectory using~\eqref{seq:synthesis_trjs_equation}.

\subsection{Midpoint mutual information and zero-time extrapolation}
\vspace*{-0.6\baselineskip}

For this linear Gaussian model, the increment and midpoint are jointly Gaussian. We therefore estimate their (conditional) information directly from empirical covariances, rather than $\hat I_{MMSE}$ or $\hat I_{KSG}$ from Sec. IV.A. The Gaussian (conditional) information formula is exact at the population level for this model. For an integer half-lag $n$, we form
\[
    \tau_n=2n\,\delta t,
    \quad
    d\bm z_{k,n}=\bm z_{k+2n}-\bm z_k,
    \quad
    \bm {z_m}^{(k,n)}=\bm z_{k+n},
\]
where $\tau_n$ is the total lag, $d\bm z_{k,n}$ is the increment, and $\bm {z_m}^{(k,n)}$ is the recorded temporal midpoint (no interpolation is used). For each lag, all available overlapping windows within a trajectory enter the empirical covariance of $\left(d\bm z_{k,n},\bm z_m^{(k,n)}\right)$.

For Gaussian blocks $U,V,W$,
\begin{equation*}
    I(U;V|W)=\frac{1}{2}\log\left(\frac{\det\matbf\Sigma_{U|W}\det\matbf\Sigma_{V|W}}{\det\matbf\Sigma_{UV|W}}\right),
\end{equation*}
where for arbitrary $T$,
\begin{equation*}
    \matbf\Sigma_{T|W}=\matbf\Sigma_{TT}-\matbf\Sigma_{TW}\matbf\Sigma_{WW}^{-1}\matbf\Sigma_{WT}.
\end{equation*}
An empty conditioning block gives ordinary MI. Before the covariance is formed, each coordinate of the increment and midpoint blocks is divided by its sample standard deviation. This rescaling leaves Gaussian MI and CMI unchanged and improves numerical conditioning. 

For each channel we fit the five finite-lag values $n=1,\cdots,5$ to a polynomial constrained to pass through the origin,
\begin{equation}
    I_i(\tau_n)=a_{i1}(\tau_n/\delta t)+a_{i2}(\tau_n/\delta t)^2+a_{i3}(\tau_n/\delta t)^3,
    \quad
    \hat\sigma_i=4 a_{i1}/\delta t.
    \label{seq:W5-cubic}
\end{equation}

We call this the W5-cubic protocol. Figure~\ref{sfig:Das_midMI} displays the raw information values and the mean inferred slopes. For example, at $\epsilon=0.29$, the W5-linear, quadratic, and cubic total-rate estimates have means $200.665$, $259.069$, and $266.663\,k_{\rm B}s^{-1}$, respectively, compared with the exact value $267.261\,k_{\rm B}s^{-1}$.

\begin{figure}[h]
    \centering
    \includegraphics[width=0.8\textwidth]{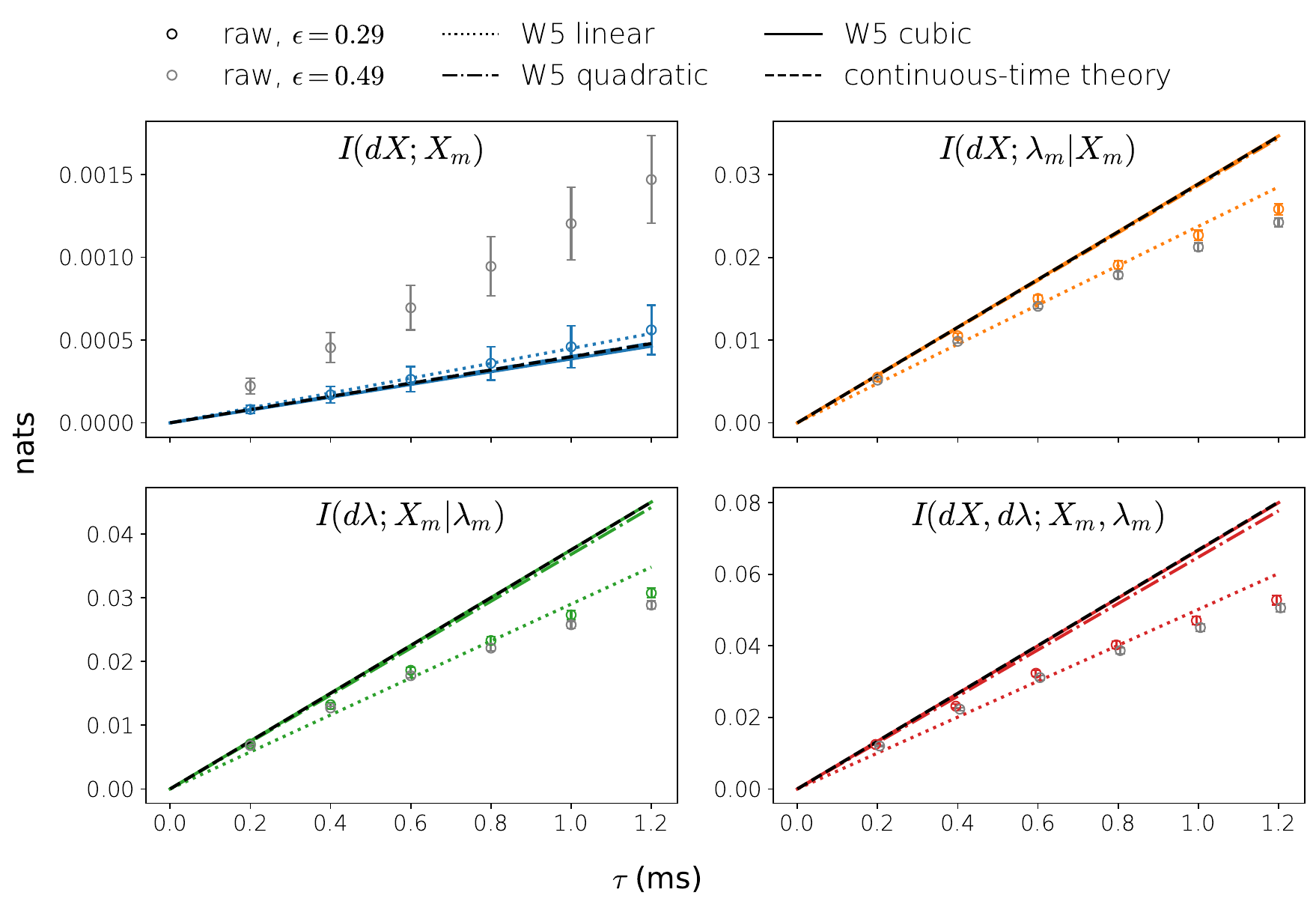}
    \vskip -0.0in
    \setlength{\abovecaptionskip}{0pt}
    \caption{Midpoint information and inferred zero-time slopes. The panels show $I(dX;X_m)$ (upper left, blue), $I(dX;\lambda_m|X_m)$ (upper right, orange), $I(d\lambda;X_m|\lambda_m)$ (lower left, green) and $I(d\bm z;\bm z_m)$ (lower right, red). Open symbols and error bars show the mean and SD across $40$ trajectories at each lag. For each panel, channel colors denote $\epsilon=0.29$ and gray denotes $\epsilon=0.49$. For $\epsilon=0.29$, dotted, dash-dotted, and solid lines are $\tau$ times the mean slope from the W5-linear, quadratic, and cubic fits, respectively. Black dashed lines are $\tau \sigma_i/4$. These lines display the inferred zero-time slopes. The sixth lag is displayed for context but excluded from the W5 fits.
    }\label{sfig:Das_midMI}
    %\vskip -0.2in
\end{figure}

\subsection{Protocol sensitivity and finite-resolution recovery}
\vspace*{-0.6\baselineskip}

We assess finite-resolution recovery using the same $40$ trajectories at each $\epsilon$. For channel $i$, with exact rate $\sigma_i$, the empirical bias and root-mean-square error (RMSE) are
\[
    \rm{bias}_i=\frac{1}{40}\sum_{r=1}^{40}\hat{\sigma}_{i,r}-\sigma_i,
    \quad
    \rm{RMSE}_i=\left[\frac{1}{40}\sum_{r=1}^{40}(\hat{\sigma}_{i,r}-\sigma_i)^2\right]^{\frac{1}{2}}.
\]

Figure~\ref{sfig:Das_W5_cubic_justification} compares quadratic and cubic fits over W4, W5, and W6, where W$m$ uses $n=1,\cdots,m$. We also evaluate the closure residual within each trajectory,
\[
    c_r=\hat{\sigma}_{X,r}^{\rm self}+\hat{\sigma}_{X|\lambda,\,r}^{\rm int}+\hat{\sigma}_{\lambda|X,\,r}^{\rm int}-\hat{\sigma}_{\rm tot,\, r}.
\]

\begin{figure}[h]
    \centering
    \includegraphics[width=\textwidth]{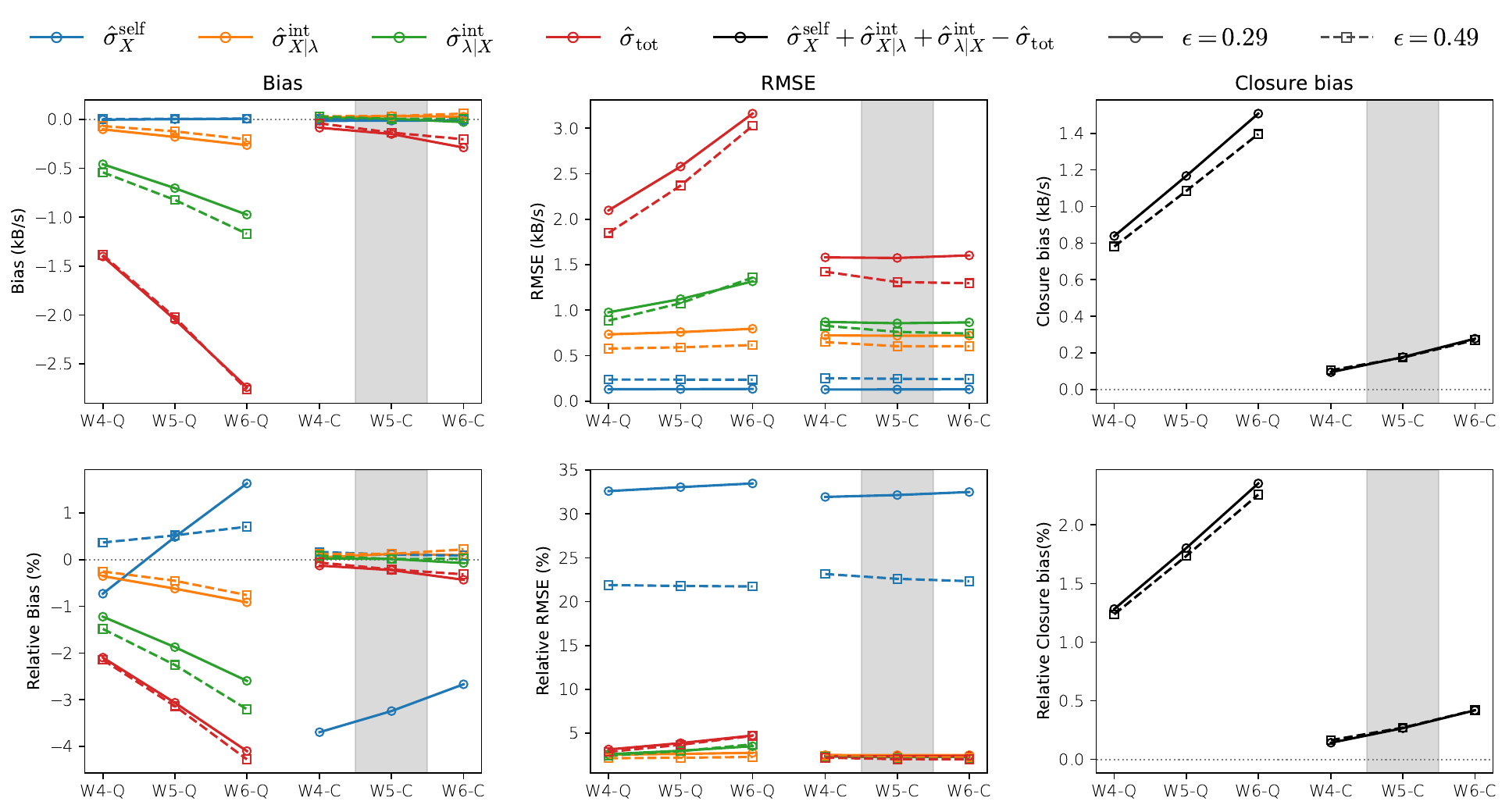}
    \vskip -0.0in
    \setlength{\abovecaptionskip}{0pt}
    \caption{Sensitivity to the lag window and polynomial order. Columns show channel bias, channel RMSE, and mean closure residual. The upper row reports dimensional quantities in $k_{\rm B} s^{-1}$. The lower row reports percentages. Relative channel bias and RMSE are normalized by the corresponding exact rate. Relative closure bias is $100\langle c_r/\hat\sigma_{\rm{tot},r}\rangle$. Colors identify the four channels, with black for closure. Solid lines with circles and dashed lines with squares denote $\epsilon=0.29$ and $\epsilon=0.49$, respectively. Q and C indicate quadratic and cubic fits. The shaded W5-cubic protocol is used in Fig. 4d in the main text.
    }\label{sfig:Das_W5_cubic_justification}
    %\vskip -0.2in
\end{figure}

Quadratic fits increasingly underestimate the interaction and total rates as the window widens, while cubic fits suppress most of this bias over W4-W6. At W5, the total-rate biases are $-0.598$ and $-0.541\,k_{\rm B}s^{-1}$ for the cubic fits, compared with $-8.192$ and $-8.101\,k_{\rm B}s^{-1}$ for the quadratic fit. The corresponding cubic RMSEs are $6.294$ and $5.233\,k_{\rm B}s^{-1}$, or $2.35\%$ and $2.03\%$ of the exact total rates.

W5-cubic provides a common protocol for all channels without channel-specific tuning. It uses five lag values for three coefficients, leaving two residual degrees of freedom, compared with one for W4-cubic. These lag values are correlated, so this count describes the fit rather than independent statistical replication. The small $X$ self channel retains larger relative uncertainty: its relative RMSEs are $32.2\%$ and $22.6\%$. We therefore use W5-cubic as a practical choice for this model and sampling resolution, without claiming that it is optimal for every channel or other systems.

The W5-cubic closure residual is small but detectable: its means are $0.709$ and $0.697\,k_{\rm B}s^{-1}$, with mean relative residuals $0.266\%$ and $0.271\%$. Thus, finite-data extrapolation does not enforce perfectly exact closure at either condition.

Figure~\ref{sfig:Das_channel_difference} resolves the W5-cubic results into individual trajectories. For each condition, shaded regions show the 95\% Student-t confidence interval for the ensemble mean. The theoretical rate, denoted by a black star, lies within each of the eight displayed intervals. The spread of the individual points is visibly larger than the uncertainty in the ensemble mean, particularly for the $X$ self channel. Overlapping windows within a trajectory are not treated as independent realizations for these intervals.

\begin{figure}[h]
    \centering
    \includegraphics[width=0.9\textwidth]{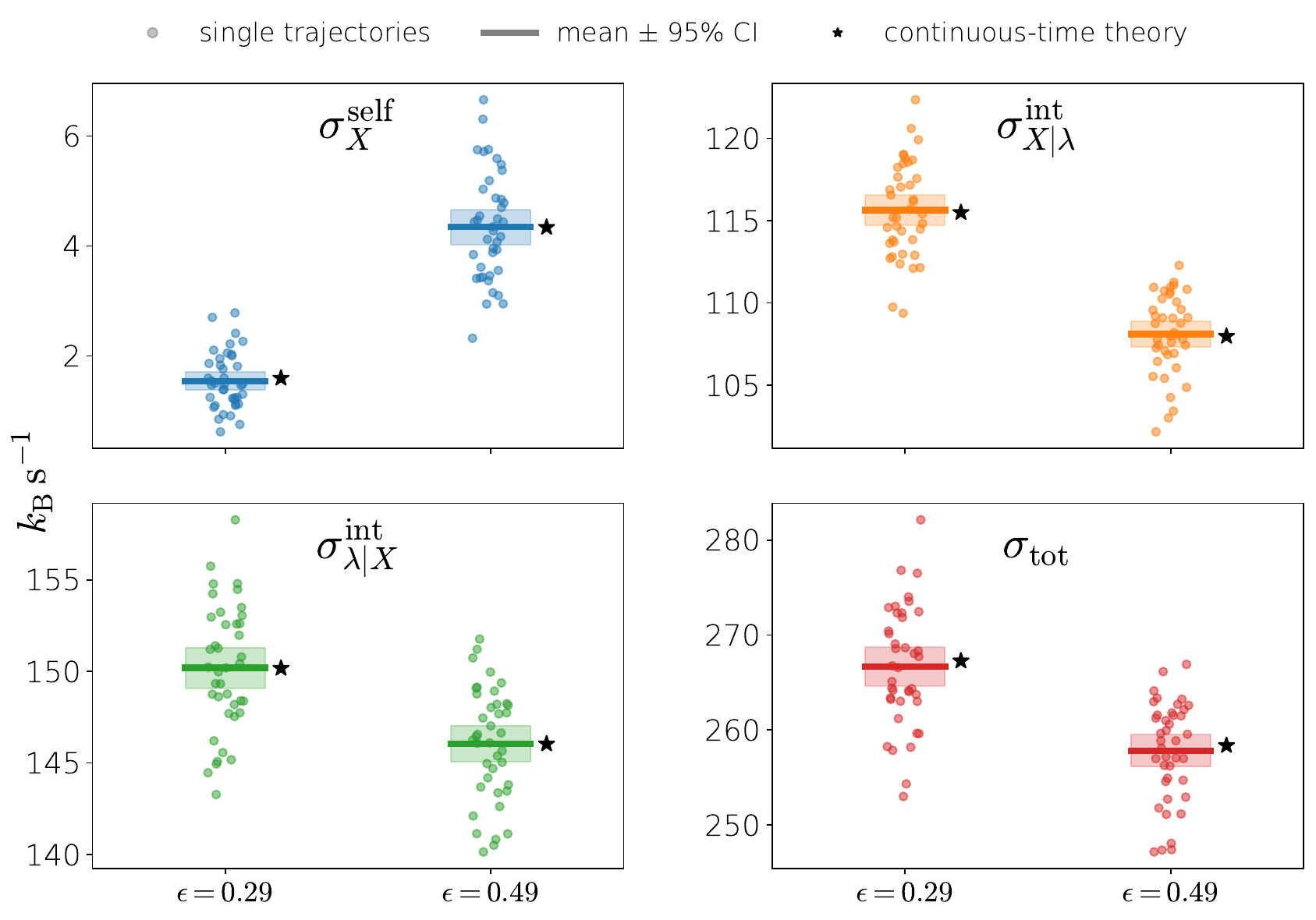}
    \vskip -0.0in
    \setlength{\abovecaptionskip}{0pt}
    \caption{W5-cubic channel rates at the two conditions used in Fig. 4d, with the same panel order as Fig.~\ref{sfig:Das_midMI}. Colored points show $40$ independent trajectory estimates at $\epsilon=0.29$ (left) and $0.49$ (right) in each panel. Black stars show the exact continuous-time rates. Colored horizontal bars mark the ensemble means and their $95\%$ Student-$t$ confidence intervals. These intervals describe uncertainty in the mean, while the point clouds show trajectory-to-trajectory variation. All eight theoretical rates lie within their respective intervals.
    }\label{sfig:Das_channel_difference}
    %\vskip -0.2in
\end{figure}

The changes in Fig. 4d are differences of the two ensemble means, $\Delta\bar{\hat{\sigma}}_i=\bar{\hat\sigma}_i(0.49)-\bar{\hat\sigma}_i(0.29)$. Their $95\%$ confidence intervals are computed using the Welch formula for two independent ensembles. The interaction losses exceed the self-channel gain, recovering the coarse-graining reversal as in the main text.

The change obtained from the three-channel sum also agrees closely with the directly estimated total change: their difference is $-0.012\,k_{\rm B}s^{-1}$, with $95\%$ CI $[-0.146,0.123]\,k_{\rm B}s^{-1}$. This interval is computed by first forming $c_r$ within each trajectory, preserving the channel covariances, and then taking the difference between the two independent ensemble means of $c_r$. This agreement of changes is distinct from the small nonzero closure residual at each individual condition.

\subsection{Cross-check against short-time TUR inference}
\vspace*{-0.6\baselineskip}

We cross-check the total EP obtained from midpoint information against the short-time thermodynamic-uncertainty-relation (TUR) inference of Manikandan, Gupta, and Krishnamurthy~\cite{manikandan2020inferring}, using the linear-current basis employed in subsequent applications~\cite{manikandan2021quantitative, das2022inferring} and the hydrodynamic studies~\cite{das2025irreversibility, das2025uraveling}. Both methods are evaluated on the same trajectories, so their agreement tests consistency between two inference routes for this application.

Let $m$ be the lag in recorded sampling steps, $h_m=m\delta t$, and let $\tilde{\bm z}_k$ denote the trajectory after subtracting its empirical
mean. The nine current features are
\[
    \bm\psi_{ab}^{(k,m)}=
    (\tilde{\bm z}_{k+m} -\tilde{\bm z}_k)_a\left(\frac{\tilde{\bm z}_{k+m} +\tilde{\bm z}_k}{2}\right)_b,
    \qquad
    a,b\in\{1,2,3\}.
\]

We form the empirical feature mean $\bm\mu_m$ and covariance $\matbf\Xi_m$, both using the normalization $1/N_m$, where $N_m$ is the number of available windows. The optimized quantity is
\begin{equation}
    Q_m=2\bm\mu_m^\intercal\matbf\Xi_m^{-1}\bm\mu_m.
    \label{seq:das_estimator}
\end{equation}

The short-time EP estimate is $Q_m/h_m$. The arithmetic mean of the two endpoints is part of this current construction; unlike the midpoint-MI identity, it does not require a recorded state at the temporal midpoint.

Finite-lag bias remains appreciable if this quantity is used without extrapolation. At the shortest lag $h_1=\delta t$, the mean estimates are $250.593$ and $242.346\,k_{\rm B}s^{-1}$, giving bias $-16.668$ and $-16.003\,k_{\rm B}s^{-1}$, respectively (about $-6.2\%$). At $2\delta t$, the biases are $-32.101$ and $-31.090\,k_{\rm B}s^{-1}$. We therefore also extrapolate the TUR quantity to the continuous-time target as in~\eqref{seq:W5-cubic}, using
\[
    Q_m\simeq b_{1}(h_m/\delta t)+b_{2}(h_m/\delta t)^2+b_{3}(h_m/\delta t)^3,
    \quad
    \hat\sigma_{\rm TUR}=b_{1}/\delta t.
\]

This is a zero-intercept cubic fit of $Q_m$; it carries no additional factor four.

We use two lag conventions in Fig.~\ref{sfig:Das_cross_check}. The matched-physical-lag check uses $m=2,4,6,8,10$ for both methods, corresponding to $0.2$--$1.0\,ms$. The native-lag check retains this midpoint-MI window but uses $m=1,2,3,4,5$ for TUR, or $0.1$--$0.5\,ms$. The latter gives each method its five shortest accessible lags; it does not involve an optimization over lag choices. Both fits are performed separately on every trajectory.

\begin{figure}[h]
    \centering
    \includegraphics[width=0.9\textwidth]{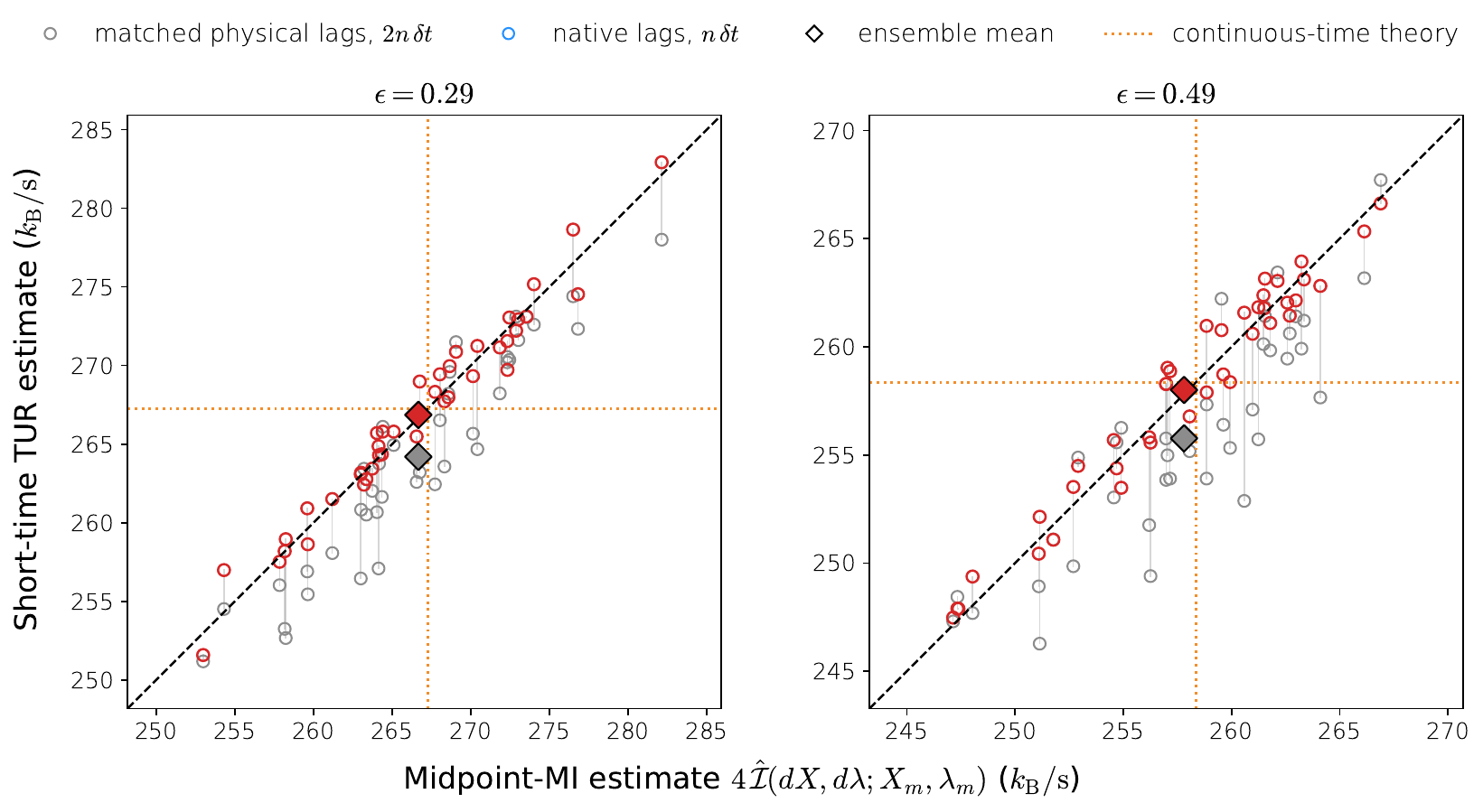}
    \vskip -0.0in
    \setlength{\abovecaptionskip}{0pt}
    \caption{Cross-check of the inferred total EP against short-time TUR inference at $\epsilon=0.29$ (left) and $0.49$ (right). Each open symbol corresponds to the same trajectory on both axes. The horizontal coordinate is the midpoint-MI W5-cubic estimate. Gray symbols use a TUR W5-cubic fit at matched physical lags $2n\delta t$; red symbols use the five native TUR lags $n\delta t$, with $n=1,2,3,4,5$. Faint vertical connectors join the two TUR results for each trajectory. Diamonds mark pairs of ensemble means. The black dashed diagonal denotes equality, and dotted orange horizontal and vertical lines mark the exact total EP. The annotated RMSEs are those of the two TUR protocols relative to the exact rate.
    }\label{sfig:Das_cross_check}
    %\vskip -0.2in
\end{figure}

At $\epsilon=0.29$, the midpoint-MI, matched-lag TUR, and native-lag TUR means are $266.663$, $264.197$, and $266.857\,k_{\rm B}s^{-1}$; at $\epsilon=0.49$, they are $257.808$, $255.768$, and $258.004\,k_{\rm B}s^{-1}$. The exact values are $267.261$ and $258.349\,k_{\rm B}s^{-1}$. For the native-lag check, the trajectory-level Pearson correlations are $0.983$ and $0.980$, and the RMS differences between the two methods are $1.179$ and $1.066\,k_{\rm B}s^{-1}$. At matched physical lags, the corresponding correlations are $0.944$ and $0.886$, with RMS differences $3.311$ and $3.201\,k_{\rm B}s^{-1}$ and mean TUR-minus-MI offsets $-2.466$ and $-2.401\,k_{\rm B}s^{-1}$. Thus, the two routes give closely agreeing trajectory-level EP values, with small protocol-dependent offsets. This cross-check supports the consistency of the finite-resolution midpoint-information analysis for the present model and sampling conditions; the result is not a claim of general estimator superiority.
\newpage
\end{widetext}

\end{document}